\documentclass{aa}  

\usepackage{graphicx}
\usepackage{txfonts}
\usepackage[colorlinks=true,linkcolor=blue,citecolor=blue]{hyperref}

\usepackage{url} 
\usepackage{xcolor} 
\usepackage{soul} 
\usepackage{siunitx} 
\usepackage{multirow}

\definecolor{bluehl}{rgb}{0.75,0.75,1}

\begin{document}

   \title{The COLIBRE-SKIRT pipeline}

   \subtitle{Calibration-free dust radiative transfer postprocessing \\ for cosmological simulations}

   \author{Andrea Gebek\inst{\ref{inst:1}}, Maarten Baes\inst{\ref{inst:1}}, Nick Andreadis\inst{\ref{inst:1}}, Joop Schaye\inst{\ref{inst:2}}, Anand Utsav Kapoor\inst{\ref{inst:1}}, Connor Bottrell\inst{\ref{inst:3}}, Shengdong~Lu\inst{\ref{inst:4}}, Cedric G. Lacey\inst{\ref{inst:4}}, Alejandro Benítez-Llambay\inst{\ref{inst:5}}, Peter Camps\inst{\ref{inst:1}}, Evgenii Chaikin\inst{\ref{inst:2},\ref{inst:4}}, Anna Durrant\inst{\ref{inst:6}}, Carlos S. Frenk\inst{\ref{inst:4}}, Filip Huško\inst{\ref{inst:2}}, Sylvia Ploeckinger\inst{\ref{inst:7}}, Alexander J. Richings\inst{\ref{inst:8},\ref{inst:9}}, Matthieu Schaller\inst{\ref{inst:2},\ref{inst:10}}, James~W.~Trayford\inst{\ref{inst:11}}, and Aswin P. Vijayan\inst{\ref{inst:12}}}

   \institute{\label{inst:1}Department of Physics and Astronomy, Universiteit Gent,               Proeftuinstraat 86 N3, B-9000 Ghent, Belgium\\
               \email{andrea.gebek@ugent.be}
               \and\label{inst:2}
             Leiden Observatory, Leiden University, PO Box 9513, 2300 RA Leiden, the Netherlands
             \and\label{inst:3}
             International Centre for Radio Astronomy Research, University of Western Australia, 35 Stirling Hwy, Crawley, 6009, WA, Australia
             \and\label{inst:4}
             Institute for Computational Cosmology, Department of Physics, University of Durham, South Road, Durham, DH1 3LE, UK
            \and\label{inst:5}
             Dipartimento di Fisica G. Occhialini, Università degli Studi di Milano Bicocca, Piazza della Scienza, 3 I-20126 Milano MI, Italy
             \and\label{inst:6}
             Astrophysics Research Institute, Liverpool John Moores University, Liverpool L3 5RF, UK
            \and\label{inst:7}
             Department of Astrophysics, University of Vienna, Türkenschanzstrasse 17, A-1180 Vienna, Austria
             \and\label{inst:8}
             Centre for Data Science, Artificial Intelligence and Modelling, University of Hull, Cottingham Road, Hull, HU6 7RX, UK
             \and\label{inst:9}
             E. A. Milne Centre for Astrophysics, University of Hull, Cottingham Road, Hull, HU6 7RX, UK
             \and\label{inst:10}
             Lorentz Institute for Theoretical Physics, Leiden University, PO Box 9506, 2300 RA Leiden, the Netherlands
             \and\label{inst:11}
             Institute of Cosmology and Gravitation, University of Portsmouth, Dennis Sciama Building, Burnaby Road, Portsmouth PO1 3FX, UK
             \and\label{inst:12}
             Astronomy Centre, University of Sussex, Falmer, Brighton BN1 9QH, UK
             }

   \date{Received XXX; accepted YYY}

 
  \abstract
   {Three-dimensional dust radiative transfer provides a powerful framework to connect cosmological galaxy simulations to multiwavelength observations. Until recently, in large-volume simulations, the formation of a cold ISM phase was prevented and dust was not evolved self-consistently. This required calibration of dust-to-metal ratios and extra subgrid dust attenuation in birth clouds, thereby reducing the predictive power.}
   {We present the COLIBRE-SKIRT pipeline, a calibration-free dust radiative transfer framework for the novel COLIBRE suite of large-volume cosmological simulations, which include a live dust model and directly simulate the multiphase ISM. Our primary aim is to establish a reference pipeline for generating multiwavelength mock observables from these simulations. As a first application, we produce far-ultraviolet (FUV) to far-infrared (FIR) spatially integrated spectra and assess them by comparison with the observed low-redshift cosmic spectral energy distribution (CSED).}
   {We apply the SKIRT dust radiative transfer code to the COLIBRE simulations. Dust masses and species fractions are taken directly from the simulation, and no birth cloud model is added in post-processing. We introduce a `split \& scale' approach that maps the simulated two-size, multi-species dust distribution onto continuous grain size distributions without introducing free parameters.}
   {We find that, for the first time, a large-volume cosmological simulation directly reproduces the local Universe CSED without calibrating the postprocessing routine \textit{a priori}. Residual tensions in the mid-infrared (of the order of $\approx0.2\,\mathrm{dex}$) point towards insufficient heating of the hottest dust components and uncertainties in the modelling of the PAH-emission carriers.}
   {We present a reference framework for generating synthetic observations including the effects of dust for the COLIBRE simulations using the SKIRT code. This framework can be readily applied at low and high redshift to create synthetic spectra and images from the FUV to the FIR.}

    \titlerunning{The COLIBRE-SKIRT pipeline}
    \authorrunning{Andrea Gebek et al.}
   \maketitle
%

\section{Introduction}

Large-volume ($L_\mathrm{box}\gtrsim100\,\mathrm{cMpc}$) cosmological hydrodynamical simulations evolve baryonic and cold dark matter in a $\Lambda$CDM Universe, predicting the properties of thousands to millions of galaxies from their formation at high redshift to the present day (see e.g. \citealt{Vogelsberger2020a}; \citealt{Crain2023} for recent reviews). The numerical techniques, physical realism, and number of resolution elements has improved substantially since the early simulation projects (e.g. TreeSPH, \citealt{Katz1996}; OWLS, \citealt{Schaye2010}; Illustris, \citealt{Vogelsberger2014}; Horizon-AGN, \citealt{Dubois2014}) to more modern simulations (e.g. EAGLE, \citealt{Schaye2015}; \citealt{Crain2015}; Magneticum Pathfinder, \citealt{Dolag2016}; IllustrisTNG, \citealt{TNG_Nelson}; \citealt{TNG_Pillepich}; SIMBA, \citealt{Dave2019}; FLAMINGO, \citealt{Schaye2023}; \citealt{Kugel2023}). Thanks to their realism and comparatively high resolution (baryonic mass resolution of $\sim10^5\,\mathrm{M}_\odot$ for the high-resolution runs), combined with relatively large volumes, EAGLE and IllustrisTNG have been particularly influential in the field of galaxy evolution. These simulations successfully reproduce many properties of the observed galaxy population, including a realistic morphological mix of galaxies at redshift zero (e.g. \citealt{Rodriguez-Gomez2019}; \citealt{Huertas-Company2019}; \citealt{Pfeffer2023}). This renders large-volume cosmological simulations invaluable to study galaxy evolution, given their ability to follow the time evolution of individual objects for statistical samples of galaxies.

To verify that the simulated galaxies are realistic, the simulation predictions need to be confronted with observational data. This comparison is hindered by the fact that almost all observational data trace light (in projection), while simulations trace mass (and other physical properties) in 3D. The conversion between light and physical properties is often complicated by dust attenuation, since dust grains in the interstellar medium (ISM) efficiently absorb and scatter starlight ($\approx30\,\%$ of the stellar luminosity is absorbed in Milky-Way-like galaxies, \citealt{Popescu2002}; \citealt{Viaene2016}; \citealt{Bianchi2018}). Hence, two approaches emerge: 

\begin{itemize}

\item \textit{Inverse modelling} retrieves the physical properties of observed galaxies, mostly through spectral energy distribution (SED) fitting (e.g. \citealt{Walcher2011}; \citealt{Conroy2013}). This approach requires assuming simplified star-formation histories and dust-to-star-geometries in the form of effective attenuation laws (see \citealt{Ramnichal2025} for a notable exception). Such strong assumptions are expected to introduce systematic uncertainties. For instance, \citet{Pacifici2023} found $\approx0.3\,\mathrm{dex}$ differences in star-formation rates when applying different SED fitting codes to the same ultraviolet - near-infrared (UV-NIR) photometric dataset at $z\approx1$.

\item \textit{Forward modelling} calculates observables from the simulated galaxies, mostly through 3D dust radiative transfer (\citealt{Steinacker2013}). This method uses the true star-formation history and dust-to-star geometry from the simulated galaxies, but requires knowledge on the distribution and physical/optical properties of dust grains in the ISM.

\end{itemize}

The forward modelling approach has been used to predict spatially integrated galaxy fluxes (e.g. \citealp{Trayford2015,Trayford2017}; \citealp{Camps2016,Camps2018}; \citealt{TNG_Nelson}; \citealt{Vogelsberger2020b}; \citealt{Trcka2022}; \citealp{Gebek2024,Gebek2025}), SEDs (e.g. \citealt{Torrey2015}; \citealt{Goz2017}; \citealt{Sharbaf2025}), images (e.g. \citealt{Granato2015}; \citealt{Rodriguez-Gomez2019}; \citealt{Cochrane2019}; \citealt{Baes2024a}; \citealt{Bottrell2024}; \citealt{Guzman-Ortega2025}; \citealt{Zhou2025}; \citealt{Eisert2026}), integral-field unit (IFU) spectra (e.g. \citealt{Harborne2020}; \citealt{Bottrell2022}; \citealt{Nanni2022}; \citealt{Barrientos2023}), and polarization maps (\citealt{Vandenbroucke2021}). These mock observables allow apples-to-apples comparisons with various observable properties of the galaxy population such as colours, luminosity functions, and image morphologies. The forward modelling methods for dust range from simple birth cloud prescriptions (\citealt{Harborne2020}; \citealt{Negri2022}) and geometrical dust attenuation models (where the dust screen optical depths scales with galaxy inclination, \citealt{Trayford2015}; \citealt{Zuckerman2021}) to line-of-sight dust extinction (\citealt{TNG_Nelson}; \citealt{Vijayan2021}; \citealt{Lovell2025}) and full three-dimensional dust radiative transfer with specialized software (GRASIL, \citealt{Silva1998}; \citealt{Granato2000}; \citealt{Dominguez-Tenreiro2014}; SUNRISE, \citealt{Jonsson2006}; RADMC-3D, \citealt{Dullemond2012}; SKIRT, \citealt{Camps2020}; RASCAS, \citealt{Michel-Dansac2020}; POWDERDAY, \citealt{Narayanan2021}). While the former methods are orders of magnitude faster than performing 3D radiative transfer calculations, they can model dust emission and scattering only approximately. Using such approximate methods severely limits the wavelength range (typically to the optical\footnote{However, we note that scattering approximations can induce the largest errors in the optical (\citealt{Baes2001}).}), as dust attenuation is strongest in the ultraviolet (UV) and dust emission becomes important for wavelengths $\gtrsim2\,\mu\mathrm{m}$ (\citealt{Camps2018}; \citealt{Gebek2025}).

In this work, we present a 3D dust radiative transfer pipeline for the recent COLIBRE simulation suite (\citealt{Schaye2026}; \citealt{Chaikin2026a}) to predict observables in the far-ultraviolet (FUV) to far-infrared (FIR) wavelength range, an endeavor that is not possible with approximate methods. Similar to EAGLE and IllustrisTNG, COLIBRE simulations exist in various resolutions and volumes\footnote{The L050m5 and L100m5 boxes have not reached $z=0$ at the time of writing.}, up to 100 cMpc at a mass resolution (for both baryons and dark matter) of $\sim10^5\,\mathrm{M}_\odot$ (L100m5), 200 cMpc at $\sim10^6\,\mathrm{M}_\odot$ resolution (L200m6), and 400 cMpc at $\sim10^7\,\mathrm{M}_\odot$ resolution (L400m7). For the purpose of forward modelling, the most important innovations of COLIBRE are the incorporation of a live dust model (\citealt{Trayford2026}) which is coupled to the thermochemistry of the ISM, and the direct modelling of the cold ISM phase (\citealt{Ploeckinger2025}).

In most large-volume cosmological simulations\footnote{SIMBA is a notable exception as it incorporates an on-the-fly dust model, but the fidelity of this model is limited by the coarse resolution (baryonic mass resolution of $\sim10^7\,\mathrm{M}_\odot$ for the main $100/h\,\mathrm{cMpc}$-box) and lack of a cold ISM phase. For simulations with fewer resolution elements, typically with box sizes of $L_\mathrm{box}\sim30\,\mathrm{cMpc}$, various cosmological simulations have also implemented live dust models (\citealt{McKinnon2017}; \citealt{Aoyama2018}; \citealt{Graziani2020}; and \citealt{Parente2022}).}, dust is not tracked on-the-fly. Hence, when building such radiative transfer pipelines for the EAGLE (\citealt{Camps2016}) and IllustrisTNG (\citealt{Trcka2022}) simulations, dust needed to be added in a postprocessing step through a fixed dust-to-metal ratio (DTM). The DTM is effectively a free parameter, which needs to be calibrated using observational data. Additionally, previous large-volume simulations lack a cold ISM phase, such that the attenuation from dusty birth clouds is implemented in a subgrid fashion (\citealt{Jonsson2010}). These shortcomings degrade the predictive power of the postprocessing framework and the simulation itself. Put another way, errors in the cosmological simulation or the radiative transfer postprocessing can be compensated to some degree by tuning the DTM and the subgrid dust attenuation. This is no longer the case for COLIBRE: The DTM is an outcome of the model and varies for each individual gas particle, and the direct modelling of the cold ISM phase circumvents the need to invoke an extra subgrid dust component to describe the attenuation from dusty birth clouds.

We use the Monte Carlo radiative transfer code SKIRT (\citealt{Baes2011}; \citealp{Camps2015,Camps2020}) to perform the dust radiative transfer calculations, and propose a fiducial COLIBRE-SKIRT pipeline. We emphasize that the output from this pipeline is calibration-free\footnote{By `calibration-free' we refer exclusively to the postprocessing pipeline. The underlying COLIBRE hydrodynamical simulations have been calibrated to reproduce the $z=0$ galaxy stellar mass function and size-mass relation (\citealt{Chaikin2026a}).}, i.e. it has not been tuned to reproduce any observables. We provide a first look at the predictions of this pipeline in terms of the $z\leq0.1$ cosmic spectral energy distribution (CSED). More detailed analyses of the COLIBRE-SKIRT predictions will be presented in \citet{Lu2026} and Lu et al. in prep. (luminosity functions), Andreadis et al. in prep. (attenuation curves), Baes et al. in prep. (optical color-magnitude diagrams), and Gebek et al. in prep. (flux-flux and color-color relations). 

Even though we do not calibrate the COLIBRE-SKIRT pipeline, a number of choices need to be made in the postprocessing. These include numerical choices (e.g. how the 3D dust distribution is discretized), spectral energy distribution (SED) templates for the stellar populations, and dedicated treatments for star-forming regions. Most importantly, the COLIBRE dust predictions need to be converted into optical properties. For each gas particle, COLIBRE predicts the dust mass in two size bins ($0.01\,\mu\mathrm{m}$ and $0.1\,\mu\mathrm{m}$) and three representative chemical species: graphite, forsterite (Mg$_2$SiO$_4$), and fayalite (Fe$_2$SiO$_4$). To perform dust radiative transfer, the two discrete sizes need to be expanded into continuous distributions, and the different chemical species need to be associated with optical properties.

The outline of this paper is as follows: We describe the COLIBRE simulations, the SKIRT code, and our fiducial COLIBRE-SKIRT pipeline in Section~\ref{sec:Methods}. The low-redshift CSED from COLIBRE is compared to observational data and results from the EAGLE and IllustrisTNG simulations in Section~\ref{sec:CSED}. In Section~\ref{sec:PostprocessingChoices} we discuss variations in our postprocessing choices (specifically, the dust model and the treatment of star-forming regions) and how these affect galaxy spectra. We discuss convergence with COLIBRE resolution and potential future improvements to our pipeline in Section~\ref{sec:Discussion}, and summarize our results in Section~\ref{sec:Summary}. We adopt a flat $\Lambda$CDM cosmology with $H_0=68.1\,\mathrm{km}\,\mathrm{s}^{-1}\,\mathrm{Mpc}^{-1}$ and $\Omega_\mathrm{m,0}=0.306$, with parameters measured by the Dark Energy Survey year three plus external constraints (\citealt{Abbott2022}), consistent with the COLIBRE cosmology.

\section{Simulation methods}\label{sec:Methods}

\subsection{The COLIBRE simulation suite}

COLIBRE\footnote{\url{https://colibre.strw.leidenuniv.nl/}} (COLd Ism and Better REsolution) is a state-of-the-art large-volume cosmological, hydrodynamical simulation suite. We briefly describe the aspects of COLIBRE that are most relevant to this work here, and refer the interested reader to \citet{Schaye2026} for a detailed description of the simulation methods and \citet{Chaikin2026a} for the calibration of the COLIBRE subgrid parameters.

COLIBRE employs the smoothed particle hydrodynamics (SPH) method to solve the equations of gravity (using \texttt{SWIFT}, \citealt{Schaller2024}) and hydrodynamics (using \texttt{SPHENIX}, \citealt{Borrow2022}) in an expanding $\Lambda$CDM Universe. The simulation runs from $z=63$ to $z=0$, following the evolution of baryonic and cold dark matter (CDM) particles. Unlike most large-volume cosmological simulations, four times more CDM than baryonic particles are used, which leads to comparable particle masses and suppresses spurious energy transfer from CDM to baryons (\citealp{Ludlow2019,Ludlow2021,Ludlow2023}). 

Similar to EAGLE and IllustrisTNG, COLIBRE features a range of resolutions with mean initial baryonic particle masses of $2.30\times10^5\,\mathrm{M}_\odot$ (`m5' resolution), $1.84\times10^6\,\mathrm{M}_\odot$ (`m6' resolution), and $1.47\times10^7\,\mathrm{M}_\odot$ (`m7' resolution). Various simulation box sizes (differing by factors of two in side length) exist for each resolution, starting at $25\,\mathrm{cMpc}$ for all resolutions and going up to $100\,\mathrm{cMpc}$ at m5 resolution (L100m5), $200\,\mathrm{cMpc}$ at m6 resolution (L200m6), and $400\,\mathrm{cMpc}$ at m7 resolution (L400m7). For each resolution, the largest volume uses more than an order of magnitude more resolution elements than EAGLE or IllustrisTNG.

In terms of physical models, the most important improvement of COLIBRE compared to other large-volume simulations is the direct modelling of a cold ISM phase. COLIBRE allows radiative cooling down to $\approx10\,\mathrm{K}$ (\citealt{Ploeckinger2025}) and models the formation of molecules as well as photo-ionization, photo-dissociation, and photo-heating from the interstellar\footnote{For the interstellar radiation field, the spectral shape is fixed to the \citet{Black1987} spectrum, and the intensity is scaled according to the local star formation rate surface density.} and metagalactic radiation field as well as cosmic rays. The heating and cooling rates and the species fractions are solved using the \texttt{CHIMES} (\citealp{Richings2014a,Richings2014b}) chemical reaction network.

Of particular relevance to dust radiative transfer postprocessing is the inclusion of a live dust model in COLIBRE, which is coupled to the gas chemistry through radiative cooling, molecule formation, and metal depletion onto (spherical) dust grains (\citealt{Trayford2026}). Two grain sizes ($0.01,\mu\mathrm{m}$ and $0.1,\mu\mathrm{m}$) and three dust species - graphite, forsterite (Mg$_2$SiO$_4$), and fayalite (Fe$_2$SiO$_4$) - are tracked individually, similar to the live dust modelling approach of \citet{Parente2022} and \citet{Dubois2024}; the latter is implemented in the NewCluster zoom simulation (\citealt{Han2025}). Dust grains are seeded from asymptotic giant branch stars and supernovae, and grow in the ISM through gas accretion. Thermal sputtering, feedback events, and astration destroy dust grains; while coagulation and shattering change the relative fractions of small and large grains. An important caveat of the COLIBRE simulation is that the internal structure of molecular clouds is not resolved. This requires the usage of a clumping factor to boost the rates of dust processes. The clumping factor transitions from 1 (for gas densities $n_\mathrm{H}<0.1\,\mathrm{cm}^{-3}$) to 100 (for $n_\mathrm{H}>10^2\,\mathrm{cm}^{-3}$). We refer the reader to \citet{Trayford2026} for a detailed discussion on how the clumping factor affects the dust properties in the simulations.

Gas particles are eligible for star formation according to a resolution-dependent gravitational instability criterion (\citealt{Nobels2024}). Their instantaneous star-formation rate (SFR) is computed using a Schmidt law (\citealt{Schmidt1959}):

\begin{equation}\label{eq:SFlaw}
    \dot{\rho}_\star=\epsilon\,\Bigl(\frac{32G}{3\pi}\Bigr)^{1/2}\rho_\mathrm{g}^{3/2},
\end{equation}
where $\dot{\rho}_\star$ is the star formation rate density, $\epsilon=0.01$ the efficiency per free-fall time, and $\rho_\mathrm{g}$ the gas mass density. Star particles represent simple stellar populations with a Chabrier initial mass function within $0.1-100\,\mathrm{M}_\odot$. These stellar populations enrich the ISM with metals (\citealt{Correa2026}) and dust grains. Furthermore, the star particles inject energy and momentum into the ISM through early stellar feedback (i.e. before the onset of core-collapse supernovae, \citealt{Benitez-Llambay2026}), core-collapse supernovae (\citealt{DallaVecchia2012}; \citealt{Chaikin2023}; \citealt{Schaye2026}), and type-Ia supernovae. Black holes are seeded and grow through gas accretion and mergers based on modified prescriptions from \citet{Nobels2022} and \citet{Bahe2022}. The COLIBRE suite contains simulations run with two different types of AGN feedback (`Thermal' and `Hybrid', \citealt{Husko2026}). In this work, we only consider simulations run with the fiducial thermal AGN feedback model (modified from \citealt{Booth2009}) as described in \citet{Schaye2026}. 

As in other cosmological simulations, COLIBRE contains a number of subgrid parameters that cannot be constrained from first principles. These resolution-dependent subgrid parameters need to be determined by calibrating the simulation predictions to observed properties of the galaxy population. In the case of COLIBRE, four parameters (the black hole seed mass and three stellar feedback parameters) are part of the main subgrid calibration. The values for these parameters are calibrated using machine-learning techniques to reproduce the $z=0$ galaxy stellar mass function and size-mass relation (\citealt{Chaikin2026a}). \citet{Chaikin2026b} show that for the calibrated COLIBRE simulations, the evolution of the stellar mass function and star-formation rates agree remarkably well with the data over the entire redshift range where observations are available ($0\leq z\leq12$). Furthermore, \citet{Schaye2026} find a good agreement between predicted and observed dust masses in the local Universe.

The simulation data is saved at 128 redshifts between $z=30$ and $z=0$, including 36 full snapshots and 92 `snipshots' with reduced data output. To identify collapsed structures such as galaxies, a friends-of-friends halo finder is first run on the dark matter particles. Subhalos are then identified using \texttt{HBT-HERONS} (\citealt{Han2018}; \citealt{ForouharMoreno2025}). Based on the output of these structure finding algorithms, many halo and subhalo properties are then calculated in a variety of apertures using the Spherical Overdensity and Aperture Processor (\texttt{SOAP}, \citealt{McGibbon2025}). We describe the boxes and galaxy sample that we use for this work in Section~\ref{sec:GalaxySample}.

\subsection{Galaxy sample}\label{sec:GalaxySample}

We select individual galaxies from various COLIBRE boxes for SKIRT postprocessing. We use four different galaxy samples: A main sample for the reference calculation of the low-redshift CSED (Section~\ref{sec:MainSample}), a sample to test convergence with COLIBRE resolution (Section~\ref{sec:ConvergenceSample}), a sample to explore variations in the COLIBRE-SKIRT pipeline (Section~\ref{sec:VariationsSample}), and a sample to optimize the SKIRT spatial grid (Section~\ref{sec:spatialGridSample}). An overview of the different galaxy samples is given in Table~\ref{tab:COLIBREboxes}.
 
\subsubsection{Main sample}\label{sec:MainSample}

The main application of the SKIRT pipeline in this work consists of the low-redshift CSED. We use the COLIBRE L100m6 box for our main galaxy sample. While the largest box at this resolution is L200m6, the L100m6 box provides an adequately representative sample of galaxies, without the excess cost of postprocessing a box as large as L200m6 (see Appendix~\ref{sec:subsampling}). The observational CSED data that we compare to (see Section~\ref{sec:observationCSED}) spans the redshift range $0.02<z<0.1$, hence we use both the $z=0$ and $z=0.1$ snapshots.

Following \citet{Baes2019}, who compute the EAGLE CSED using SKIRT postprocessing, we select all COLIBRE galaxies in the L100m6 box with $M_\star\geq10^{8.5}\,\mathrm{M}_\odot$, where $M_\star$ is the total gravitationally bound stellar mass available from the \texttt{SOAP} catalogues. This stellar mass limit ensures that each galaxy is resolved by at least $\approx200$ star particles, and is a compromise to have a sufficient number of well-resolved galaxies. With this selection criterion we have 23\,490 galaxies at $z=0$ and 23\,729 galaxies at $z=0.1$ in our main sample. 

\subsubsection{Convergence test sample}\label{sec:ConvergenceSample}

To test how our results depend on COLIBRE resolution, we postprocess galaxies from the largest simulation volume available at $z=0$ in all three resolutions. These are the L025m5, L025m6, and L025m7 runs. Since we want to test what the minimum stellar mass is above which our results converge, we use a very low stellar mass threshold of $10^7\,\mathrm{M}_\odot$ for all three boxes. This amounts to 1\,502 galaxies for L025m5, 1\,633 galaxies for L025m6, and 2\,077 galaxies for L025m7.

\subsubsection{Model variations sample}\label{sec:VariationsSample}

All of the previously mentioned galaxy samples are postprocessed with our fiducial SKIRT pipeline. To test the impact of variations in the pipeline, we use a small galaxy sample drawn from the largest COLIBRE box at m6 resolution (L200m6) and compute their SEDs with various alterations in the SKIRT postprocessing. Specifically, we select galaxies from the L200m6 simulation with a stellar mass of $M_\star\geq10^{8.5}\,\mathrm{M}_\odot$, in subsamples of 50 galaxies per $0.5\,\mathrm{dex}$ in $\log_{10}(M_\star)$. Our highest-mass bin is $M_\star\geq10^{11}\,\mathrm{M}_\odot$, which means we have six mass bins and a total of 300 galaxies in the L200m6 sample. We find that the $z=0$ CSED converges to within $\pm0.05\,\mathrm{dex}$, with the largest variations in the FIR (see Appendix~\ref{sec:subsampling}). Since we test $\approx30$ alternative postprocessing versions, it would be too computationally expensive to re-run the main L100m6 sample that many times. Given that we adopt a fixed number of galaxies here (300 galaxies), we use the larger L200m6 box instead of L100m6 to sample a larger cosmic volume and hence rarer modes of structure formation.

\subsubsection{Spatial grid sample}\label{sec:spatialGridSample}

To optimize the spatial grid for our SKIRT runs, we test various grid parameters using a small dedicated galaxy sample. We use the same subsampling strategy as outlined in Section~\ref{sec:VariationsSample}, with two adjustments: We only draw 25 galaxies per $0.5\,\mathrm{dex}$ in $\log_{10}(M_\star)$, and we only include dusty galaxies (see footnote~\ref{footnote:dustyGalaxies}). Since the COLIBRE resolution impacts the convergence behaviour of the SKIRT spatial grid, we construct a spatial grid sample for both the L025m5 simulation (using a minimum stellar mass of $10^{7.5}\,\mathrm{M}_\odot$) and for the L200m6 simulation (using a minimum stellar mass of $10^{8.5}\,\mathrm{M}_\odot$).

\begin{table}
    \centering
    \begin{tabular}{ccccc}
         SKIRT run & Box & $M_\star^\mathrm{min}\,[\mathrm{M}_\odot]$ & \shortstack{Sub- \\ sample} & $N_\mathrm{galaxies}$\\ \hline
        Main run & L100m6 & $10^{8.5}$ & no & 47\,219$^1$\\ \hline
        \multirow{3}{*}{Convergence} & L025m5 & & & 1\,502\\
        & L025m6 & $10^7$& no & 1\,633\\
        & L025m7 & & & 2\,077\\ \hline
        Model variations & L200m6 & $10^{8.5}$& yes & 300\\ \hline
        \multirow{2}{*}{Spatial grid} & L025m5 & $10^{7.5}$ & yes & 176\\
        & L200m6 & $10^{8.5}$ & yes & 150\\ \hline        
    \end{tabular}
    \caption{Overview of the COLIBRE galaxy samples that we use for SKIRT postprocessing. We use the $z=0$ snapshot for all COLIBRE boxes, except for L100m6 which we postprocess both at $z=0$ and $z=0.1$. For most boxes we use our fiducial SKIRT pipeline, except for the L200m6 box which we use to test variations in the COLIBRE-SKIRT pipeline (`Model variations'). $M_\star^\mathrm{min}$ indicates the minimum stellar mass which we use to select galaxies. For the `Model variations' and `Spatial grid' runs we use a subsampling strategy, see Section~\ref{sec:GalaxySample} for more details. The number of galaxies for each of our samples is given in the last column. \newline $^1$\textit{23\,490 galaxies at $z=0$ and 23\,729 galaxies at $z=0.1$.}}
    \label{tab:COLIBREboxes}
\end{table}

\subsection{The SKIRT radiative transfer code}

To perform the dust radiative transfer postprocessing, we use the open-source SKIRT\footnote{Version 9, \url{https://skirt.ugent.be}} code (\citealt{Baes2011}; \citealt{Camps2015, Camps2020}). SKIRT is widely used to compute mock observables for galaxies from cosmological simulations such as EAGLE (e.g. \citealp{Camps2016,Camps2018}; \citealt{Trayford2017}), FIRE-2 (e.g. \citealt{Ma2019}; \citealt{Parsotan2021}; \citealt{Cochrane2023}), IllustrisTNG (e.g. \citealt{Vogelsberger2020b}; \citealt{Shen2020}; \citealt{Trcka2022}; \citealt{Bottrell2024}; \citealt{Gebek2024}), AURIGA (\citealt{Kapoor2021}), ARTEMIS (\citealt{Camps2022}), NewHorizon (\citealt{Jang2023}), NIHAO (\citealt{Faucher2023}), and NewCluster (\citealt{Byun2025}). SKIRT provides a flexible environment to read in the distributions and properties of stars and dust for various simulation formats such as smoothed particles, structured adaptive grids, or unstructured Voronoi meshes (\citealt{Camps2013}).

The SKIRT code is modular and allows a variety of radiative transfer applications in dusty astrophysical systems. SKIRT employs the Monte Carlo method, launching a large number of photon packets from primary sources and tracing them through a discretized medium. The code uses diverse speed-up mechanisms such as peel-off, path-length stretching, composite biasing, and forced scattering.

In the context of galaxies, primary emission typically corresponds to light from stellar particles, which are treated as simple stellar populations (SSPs) and modelled with SED template libraries. The transfer medium is the dusty ISM, which scatters and absorbs the stellar light. The absorbed energy heats dust grains, giving rise to secondary emission from dust grains in the infrared. The code also accounts for stochastic heating of dust grains (\citealt{Camps2015b}), which is necessary to produce the emission features of polycyclic aromatic hydrocarbons (PAHs) in the mid-infrared (MIR). Lastly, the photon packets are recorded in synthetic instruments that emulate various apertures, orientations, spectral resolutions, and observational configurations (integrated fluxes/images/IFU cubes).

To solve the radiative transfer equation, the transfer medium (the dusty ISM) needs to be discretized, which is achieved through a variety of regular and adaptive grids (\citealp{Saftly2013,Saftly2014}). To track photon packets through the grid cells, SKIRT requires the optical and calorimetric properties of the dust to compute optical depths and emissivities. These dust grain properties are tabulated depending on the grain sizes and chemical compositions, which means that the user needs to specify the amount of dust as well as its composition and grain size distribution.

We describe our fiducial pipeline to postprocess COLIBRE galaxies with SKIRT in Section~\ref{sec:fiducialPostprocessing}. This pipeline largely follows the developments made for SKIRT data releases for the EAGLE (\citealp{Camps2016,Camps2018}) and IllustrisTNG (\citealt{Trcka2022}) simulations. However, it differs significantly from the aforementioned works in the treatment of dust due to COLIBRE's live dust model, which circumvents the need for calibration and allows the extraction of more detailed dust-related properties directly from the simulation. Additionally, COLIBRE's direct modelling of the cold ISM phase removes the need to add a model for subgrid dust attenuation in birth clouds.

\subsection{The fiducial COLIBRE-SKIRT pipeline}\label{sec:fiducialPostprocessing}

In this section, we describe and motivate our choices made for the fiducial COLIBRE-SKIRT pipeline. An overview of our SKIRT settings is given in Table~\ref{tab:SKIRTsettings}. We present variations of these choices and analyze their impact on synthetic spectra in Section~\ref{sec:PostprocessingChoices} and Appendix~\ref{sec:PipelineVariationsAppendix}.

\begin{table*}
\centering
\begin{tabular}{ccc}
\hline
\multirow{4}{*}{Evolved stellar populations} & COLIBRE particles & Star particles with ages $\geq10\,\mathrm{Myr}$ \\
& SED templates & BPASS v2.2.1 \\
& IMF & \citealt{Chabrier2003}, $0.1-100\,\mathrm{M}_\odot$ \\
& Smoothing & 32-nearest grav. bound star particle, quartic spline \\ \hline

\multirow{4}{*}{Star-forming regions} & COLIBRE particles & Star particles with ages $<10\,\mathrm{Myr}$ \& star-forming gas particles \\
& SED templates & TODDLERS (10-Myr-averaged, dust-free, with emission lines) \\
& IMF & \citealt{Chabrier2003}, $0.1-100\,\mathrm{M}_\odot$ \\
& Smoothing & gas smoothing, quartic spline \\ \hline

\multirow{5}{*}{Dust model} & COLIBRE particles & Gas particles \\
& Optical properties & \citet{Draine2007} \\
& Grain size distributions & `Split \& scale' \\ 
& Number of size bins & 10 (same for each component) \\
& Smoothing & gas smoothing, quartic spline \\ \hline

\multirow{5}{*}{Spatial grid} & SKIRT box size & 100\,kpc \\
& Adaptive grid algorithm & $k$-d tree \\
& Minimum refinement level & 15 (cell size $3.13\,\mathrm{kpc}$) \\
& Maximum refinement level & 36 (cell size $24.4\,\mathrm{pc}$) \\
& Refinement criterion & Scaled to $\Sigma_\mathrm{dust}$ \\ \hline

\multirow{4}{*}{Other SKIRT settings} & Number of photon packets & $10^{7.5}$ \\
& Primary source wavelength range & $0.09-2000\,\mu\mathrm{m}$ \\
& Grid to store primary radiation field & $0.09-20\,\mu\mathrm{m}$, 25 points (log. grid) \\
& Grid to calculate dust emission & $0.09-2000\,\mu\mathrm{m}$, 200 points (nested log. grid) \\ \hline
\end{tabular}
\caption{Summary of the fiducial settings used for our SKIRT simulations.}
\label{tab:SKIRTsettings}
\end{table*}

\subsubsection{Particle extraction from COLIBRE}

Since we postprocess the COLIBRE data on a galaxy-by-galaxy basis, we first extract the necessary data individually for each galaxy using all bound star and gas particles\footnote{Specifically, we use both \texttt{SWIFTGalaxy} (\citealt{Oman2025}) and the \texttt{SWIFTmask} facility from \texttt{swiftsimio} (\citealt{Borrow2020}) to loop over galaxies and access bound particles.} (as identified by \texttt{HBT-HERONS}). This means that all particles that belong to neighbouring galaxies are removed, even if they fall within the field-of-view of the synthetic observations. Following \citet{Trcka2022}, we treat all star particles with ages above 10\,Myr as evolved stellar populations (described in Section~\ref{sec:EvolvedStellarPopulations}).

For young stellar populations (or star-forming regions), the numerical resolution limits the number of star particles with ages below 10\,Myr - for a galaxy with a constant star-formation rate over the last 10\,Gyr, only one in a thousand star particles formed will fall within the category of star-forming regions. This leads to poor sampling, even though these particles can dominate the total UV luminosity. A solution for this problem consists of invoking the star-forming gas particles as sources for star-forming regions (\citealt{Trayford2015}; \citealt{Camps2016}), as those particles are significantly more numerous (by approximately two orders of magnitude). Hence, we use the star-forming gas particles (supplemented by evoking the parent gas particles of young stars) as star-forming regions (described in Section~\ref{sec:StarFormingRegions}).

Lastly, we store all gravitationally bound gas particles for the dust medium. Most earlier postprocessing efforts needed to assign an amount of dust for each gas particle/cell, typically by calibrating a fixed dust-to-metal ratio and using cuts in temperature, density, and/or SFR (e.g. \citealt{Camps2016}; \citealt{Kapoor2021}; \citealt{Trcka2022}). A big advantage of COLIBRE's live dust model is that we can directly take the dust masses (as well as the fractions of various dust species) from the particle data. We describe the treatment of dust in Section~\ref{sec:Dust}.

\subsubsection{Evolved stellar populations}\label{sec:EvolvedStellarPopulations}

The most important choices related to evolved stellar populations consist of the SED template library and the initial mass function (IMF). To be consistent with the early stellar feedback of COLIBRE (\citealt{Benitez-Llambay2026}) and to account for the effect of binaries, we use the BPASS v2.2.1 library\footnote{Natively, the BPASS SEDs are tabulated at a resolution of $1\textup{~\AA}$ up to a wavelength of $10\,\mu\mathrm{m}$. We have modified these templates for usage within SKIRT, as this high resolution is not needed in the infrared and the very large number of wavelength points slows down the postprocessing (the effect is significant for massive galaxies at m5 resolution which have many star particles). Hence, we keep the original BPASS resolution of $1\textup{~\AA}$ for wavelengths shorter than $0.6\,\mu\mathrm{m}$, and cover longer wavelengths with 2\,000 log-spaced grid points (corresponding to a fixed spectral resolving power of $R=710.88$). This brings down the number of wavelength grid points from the original $10^5$ to $8\times10^3$. Additionally, to avoid artifacts in images around $10\,\mu\mathrm{m}$, we extrapolate the SED templates to an upper wavelength of $160\,\mu\mathrm{m}$ assuming a fixed slope in log-log space measured at $8-10\,\mu\mathrm{m}$.} (\citealt{Eldridge2017}; \citealt{Stanway2018}) with a Chabrier IMF with the mass range $0.1-100\,\mathrm{M}_\odot$.

Most of the information that is required to model the emission of evolved stellar populations with BPASS is directly available from the COLIBRE star particle data (coordinates, initial mass, metallicity, age). Stellar particles represent large numbers of stars, which would in reality be distributed over some region of space. While for SPH simulations there exists a natural choice for the gas particle smoothing length, this is not the case for the length over which stellar properties, including luminosities, should be smoothed (e.g. \citealt{Torrey2015}). We follow the convention of IllustrisTNG postprocessing (\citealt{Trcka2022}) and take the distance to the 32nd nearest star particle\footnote{We find that this choice has a negligible impact on the CSED: varying the smoothing lengths by a factor of two (corresponding to approximately 4 or 256 neighbours) changes the CSED by $\lesssim 0.06\,\mathrm{dex}$.} (counting all gravitationally bound star particles) as the smoothing length. Unlike most previous SKIRT postprocessing efforts which use a cubic spline kernel, we adopt a quartic spline kernel here to be consistent with COLIBRE's gas particle smoothing kernel.

\subsubsection{Star-forming regions}\label{sec:StarFormingRegions}

Modelling the young stellar populations (with ages $\lesssim10\,\mathrm{Myr}$) is notoriously challenging in the context of dust radiative transfer, for two reasons:
\begin{itemize}
    \item Because star particles represent SSPs consisting of a very large number of stars, there are few young star particles in simulated galaxies. However, they can dominate the UV, leading to sampling problems. This is particularly problematic when creating spatially resolved observables like images.
    \item Young stellar populations are typically enshrouded by their dusty birth clouds, which dissolve on timescales of a few Myr (e.g. \citealt{Charlot2000}; \citealt{Kawamura2009}; \citealt{Murray2011}). Large-volume cosmological simulations cannot yet spatially resolve these birth clouds. For simulations that do not model the multiphase ISM this leads to an underestimation of the birth cloud dust attenuation. Hence, the reprocessing of starlight from the youngest stars is typically treated with subgrid models (\citealt{Jonsson2010}).
\end{itemize}
Regarding the sampling problem, we find that galaxies at m6 resolution and $10^{9.5}\,\mathrm{M}_\odot\lesssim M_\star\lesssim10^{11}\,\mathrm{M}_\odot$ have $\lesssim10$ young star particles (at $z=0$). Most lower-mass galaxies do not have a single young star particle (at m6 resolution), even though many of them have star-forming gas. The only way to overcome this is to increase the simulation resolution: At m6 resolution, the mean initial star particle mass is $1.84\times10^6\,\mathrm{M}_\odot$, requiring an SFR of $0.184\,\mathrm{M}_\odot\mathrm{yr}^{-1}$ to spawn (on average) a single star particle over $10\,\mathrm{Myr}$.

To combat this sampling issue for EAGLE postprocessing (at a similar resolution as COLIBRE m6), \citet{Trayford2015,Trayford2017} and \citet{Camps2016} sample the young stellar populations from the star-forming gas. We adopt a similar strategy and use all star-forming gas particles (and parent gas particles of young stars) as sources. This leads to a factor $\sim100$ increase in particle number of young stellar populations. With this approach, we effectively average the SFH of the galaxy over the last $10\,\mathrm{Myr}$ (for EAGLE, an averaging timescale of $100\,\mathrm{Myr}$ was used).

To resolve the second problem (sub-resolution dust and gas around young stars), dust radiative transfer postprocessing campaigns always invoke a dedicated set of SED templates that include this subgrid reprocessing of starlight. Exceptions include simulations with relatively high resolution (particle masses $\lesssim10^4\,\mathrm{M}_\odot$, \citealt{Behrens2018}; \citealt{Ma2019}), both with and without a cold ISM: \citet{Behrens2018} use the simulation from \citet{Pallottini2017}, which does not model the cold ISM, while \citet{Ma2019} use the FIRE-2 simulations (\citealt{Hopkins2018}) that do model the cold ISM. However, as we discuss in Section~\ref{sec:Variations-YoungStars}, for large-volume simulations that have lower resolution such as COLIBRE we find that the presence of a multiphase ISM can also reproduce the effect of birth-cloud dust attenuation without subgrid treatments.

For SKIRT postprocessing studies, the canonical choice for such a template library has long been MAPPINGS-III (\citealt{Groves2008}), but due to inconsistencies in the UV/MIR (e.g. \citealt{Trcka2020}; \citealt{Kapoor2021}), \citet{Kapoor2023} recently developed the TODDLERS templates to mitigate these issues. TODDLERS assumes spherical symmetry and models the time evolution of a newly formed stellar population within its birth cloud under the influence of gravity, radiation pressure, stellar winds, and pressure gradients in the gas. To calculate the emergent radiation, a stellar SED template library is used for the young stellar population. In our case, we adopt the same SED template library that we use for the evolved stellar populations (BPASS v2.2.1, Chabrier IMF with mass limits $0.1-100\,\mathrm{M}_\odot$). This light is then reprocessed by the dusty birth cloud using the photoionization code \texttt{Cloudy} (C23, \citealt{Chatzikos2023}). Since the COLIBRE multiphase ISM reproduces the effect of birth-cloud dust attenuation, we use a variation of the TODDLERS templates which omits dust attenuation from the birth cloud. We do include nebular emission lines predicted by TODDLERS using the high-resolution version of its library (see Section~\ref{sec:NebularEmission} for a discussion on the modelling of the nebular component). 

The main TODDLERS SED templates describe the emission from individual star-forming gas clouds, parametrized by cloud gas mass, cloud density, star-formation efficiency, metallicity, and age. Since star-forming regions typically host a range of cloud masses that follows a power-law distribution in masses (\citealt{Heyer2015}), TODDLERS also provides a time-averaged version of the library which integrates over the cloud masses. This version of the library is parametrized by cloud density, star-formation efficiency, metallicity, and scaled by star-formation rate. Since we use star-forming gas particles to model the emission from young stellar populations, we use the time-averaged version of TODDLERS, which assumes a constant SFR over a specific timescale ($10\,\mathrm{Myr}$ in our case).

By using the star-forming gas particles as sources, we effectively replace the few stochastically formed star particles with the typically much larger number of star-forming gas particles. This approach captures unresolved star formation, but misses the contribution from gas particles that have been converted into star particles during the last $10\,\mathrm{Myr}$. We correct for this effect by replacing all star particles with ages below $10\,\mathrm{Myr}$ by their parent gas particles (see below for details). We note that this is a very small effect: The parent gas particles and star-forming gas particles have similar luminosities, but there are typically $\sim100$ times more star-forming gas particles than young star particles.

The following properties are required for the TODDLERS SED template library: We take the coordinates and metallicity directly from the COLIBRE particle data. We want to use the gas neighborhood for the star-forming regions, hence we directly use the `smoothing length' property of the SPH particle data. We multiply\footnote{\label{footnote:smoothing}We adopt the convention that the smoothing length is the compact support radius of the smoothing kernel, unlike COLIBRE where the `smoothing length' property corresponds to two times the standard deviation $\sigma$ of the kernel. For the quartic spline kernel that we use here, the ratio between compact support radius and $2\sigma$ is 2.018932 (\citealt{Dehnen2012}).} this quantity to convert the `smoothing length' property of the particle data to the compact support radius of the smoothing kernel. We adopt $320\,\mathrm{cm}^{-3}$ for the cloud density and $2.5\,\%$ for the star-formation efficiency (i.e. the ratio of the mass of newly formed stars to gas mass in a star formation region) following \citet{Kapoor2024}, who calibrated these values by comparing FUV-FIR fluxes from the AURIGA simulations (\citealt{Grand2017}) to DustPedia observations (\citealt{Davies2017}). These parameters affect only the nebular emission lines from the template library (not the underlying continuum from the young stellar populations), which are not the primary interest of this study.

Lastly, TODDLERS requires a star-formation rate (which is assumed to be constant over $10\,\mathrm{Myr}$) to scale the SEDs. For the star-forming gas particles, we directly use the 10-Myr-averaged SFRs that are available from the particle data. For the young star particles that are replaced with their parent gas particles, we instead use the instantaneous SFR at birth since there is no time-averaged SFR available for the star particles. We compute the instantaneous SFR from the birth gas density and initial stellar mass using Eq.~\ref{eq:SFlaw}. This SFR is further multiplied by $(10\,\mathrm{Myr}-t_\star)/10\,\mathrm{Myr}$ ($t_\star$ denotes the age of the star particle) to account for the finite period over which the parent gas particle was eligible for star formation.

To summarize, we use the high-resolution TODDLERS templates without birth-cloud dust attenuation to describe the light from young stellar populations (using star-forming gas particles and the parent gas particles of young star particles). The resulting spectra are equivalent to the BPASS SED templates for a composite stellar population forming stars at a constant rate over $10\,\mathrm{Myr}$, with nebular emission lines added to the stellar continuum (but without nebular continuum). In our setup, the shape of the spectra only depends on metallicity, and the spectra are scaled according to the star-formation rate of the particle.

\subsubsection{Dust}\label{sec:Dust}

With respect to earlier SKIRT postprocessing efforts for galaxies in other simulations, the dust treatment is the aspect that differs the most in this work. This is because we no longer assume a constant dust-to-metal ratio and fixed dust grain size distributions and composition. Instead, we directly use the dust properties predicted by COLIBRE.

SKIRT supports a variety of dust models which describe the composition and properties of interstellar dust (e.g. \citealt{Zubko2004}; \citealt{Draine2007}; or THEMIS, \citealt{Jones2017}). These dust models combine multiple dust grain types (e.g. graphite, amorphous silicates, amorphous hydrocarbons, PAHs) with optical and calorimetric properties (known from theoretical calculations or lab measurements) as a function of grain size and wavelength (for each grain type). The grain size distributions in the dust models are determined (for each grain type) by fitting to a range of Milky Way observations, typically including extinction curves as well as emission spectra, polarization spectra, and metal depletions. Crucially, this means that dust models can start from very different dust grain types (with significant differences in the optical and calorimetric properties), but still end up reproducing the same set of Milky Way observations by tuning grain size distributions. This variety in dust models is ultimately due to our incomplete understanding of dust in the local ISM, an issue that is exacerbated at higher redshift.

These considerations are not a major concern when using a dust model `out-of-the-box', as the final dust model predictions (e.g. total dust mass extinction coefficient or emissivities) are somewhat comparable between dust models (because they are tuned to similar sets of observations). However, the live dust model of COLIBRE predicts six different grain species (small/large grains for three different grain compositions). This requires `splitting up' a dust model into its components and modifying the grain size distributions\footnote{We note that expanding the two discrete sizes into continuous size distributions is only strictly necessary when modelling dust emission. When considering only dust absorption and scattering, a single effective dust grain type and size with average optical properties can be used (\citealt{Steinacker2013}).}. This procedure can introduce large differences between dust models as the underlying dust components significantly differ in their optical and calorimetric properties.

Other recent works that employ radiative transfer for simulated galaxies with live dust models (\citealt{Dubois2024}; \citealt{Byun2025}) also use two size bins and employ the two-size approximation from \citet{Hirashita2015}. In these works, two discrete grain sizes are replaced by two continuous modified log-normal grain size distributions. We find this approach unsuitable for our purpose due to the following disadvantages: 

\begin{itemize}
    \item The two continuous size distributions overlap with each other and cover both discrete sizes. For example, this means that even if COLIBRE predicts that no small grains ($a=0.01\,\mu\mathrm{m}$) are present, this dust model would still have a small fraction of grains with sizes $a\leq0.01\,\mu\mathrm{m}$ from the continuous size distribution of the large grains.
    \item It is not clear a priori what value to choose for the width $\sigma$ of the modified log-normal grain size distribution (\citealt{Hirashita2015} adopts $\sigma=0.75$), leading to the introduction of free parameters.
    \item It is also not clear what optical and calorimetric properties (i.e. which grain types) should be chosen. Since the size distributions in full dust models like THEMIS do not resemble a combination of two modified log-normal size distributions (see Fig.~\ref{fig:dustModels_sizeDistributions}), it is difficult to use full dust models as starting points.
    \item Most importantly, the two-size approximation was developed to model dust extinction curves (\citealt{Hirashita2015}). PAH emission features in the MIR arise from small carbonaceous grains with sizes below $\approx8\textup{~\AA}$ (\citealt{Draine2007}) that are stochastically heated. These very small grains are absent in the two-size approximation, leading to a very significant underestimate of MIR dust emission. PAHs are not explicitly modelled in COLIBRE, hence it would be conceivable to add PAHs in a postprocessing step (Rodriguez Montero et al. in prep., J.K. Jang et al. in prep.). However, this approach introduces significant complexity and additional modelling uncertainties.
\end{itemize}

To simultaneously overcome all these limitations we propose a novel `split \& scale' approach: Starting from any dust model, we split the model`s size distributions at $a=0.03\,\mu\mathrm{m}$, and then scale the lower and upper parts according to the mass fractions of small and large grains in COLIBRE. This scaling is done separately for carbonaceous and silicate\footnote{We combine the Mg-rich and Fe-rich silicate grains from COLIBRE into a single silicate component. This is because the Milky Way silicate dust is very Mg-rich ($\mathrm{Mg}/(\mathrm{Mg}+\mathrm{Fe})\geq0.9$, \citealt{Min2007}), and hence optical and calorimetric properties for iron-rich silicate grains are scarce.} grains. The grain size at which we split the size distributions ($a=0.03\,\mu\mathrm{m}$) exactly corresponds to the boundary between small and large grains in the two-size approximation of \citet{Hirashita2015}, and is close to the geometric mean of the two COLIBRE grain sizes ($a=0.0316\,\mu\mathrm{m}$). This approach has the advantages that it can be applied to any dust model, it works for both dust attenuation and dust emission, and it does not introduce additional free parameters.

\begin{table}
    \centering
    \begin{tabular}{lccc}
         Dust grains & COLIBRE & DL07 & THEMIS \\ \hline
         Large silicates & 44.0\,\% & 61.5\,\% & 65.7\,\% \\
         Small silicates & 30.2\,\% & 11.2\,\% & 3.2\,\% \\
         Large carbonaceous & 15.4\,\% & 19.5\,\% & 7.7\,\% \\
         Small carbonaceous & 10.4\,\% & 7.8\,\% & 23.4\,\%
    \end{tabular}
    \caption{Breakdown of various dust models into their constituents in terms of mass fractions. For COLIBRE, we compute the species dust masses by summing over all galaxies in our L100m6 sample at $z=0$, and then quote the ratio relative to the total dust mass. For the DL07 (\citealt{Draine2007}) and THEMIS (\citealt{Jones2017}) dust models, separation between small and large grains occurs at $a=0.03\,\mu\mathrm{m}$.}
    \label{tab:sizeDistributions}
\end{table}

We still need to choose a dust model. Ideally, this dust model has relative proportions of grain types and sizes that align with the COLIBRE dust predictions at low redshift. We note that the COLIBRE dust properties, particularly the grain sizes, vary between and within galaxies depending on the ISM conditions (\citealt{Trayford2026}; \citealt{Schaye2026}). We calculate the total dust masses in small/large and carbonaceous/silicate grains in our main COLIBRE L100m6 sample at $z=0$. We find that $\approx75\,\%$ of the dust mass is composed of silicate grains, while the remaining $25\,\%$ consists of carbonaceous grains. For both grain types, the mass ratio between large and small grains is approximately 3:2. The exact breakdown of the COLIBRE dust grains is listed in Table~\ref{tab:sizeDistributions}, where we also compare to the relative grain abundances in two other commonly used dust models (\citealt{Draine2007}and THEMIS). We caution that the COLIBRE sample includes all galaxies above a stellar mass limit in the simulation box, while the observational dust models are calibrated to Milky Way data.

While both dust models broadly reproduce the carbonaceous-to-silicate ratio in COLIBRE, the \citet{Draine2007} (hereafter DL07) model yields small-to-large grain size ratios that are more consistent with COLIBRE than those of THEMIS. In contrast, THEMIS adopts markedly different grain size distributions relative to COLIBRE. Variations in the dust model and the treatment of grain size distributions are presented in Section~\ref{sec:Variations-DustModel}.

The dust particle coordinates as well as the dust masses in the four species that we consider (small silicate grains, large silicate grains, small carbonaceous grains, and large carbonaceous grains) are directly taken from the gas particle data. The smoothing lengths that are stored in the gas particle data are based on gas-gas neighborhood, hence we use those values (multiplied by 2.018932, see footnote~\ref{footnote:smoothing}) as smoothing lengths for the quartic spline smoothing kernels in SKIRT.

\subsubsection{SKIRT settings}\label{sec:SKIRTsettings}

In this section we describe the settings that we use for our SKIRT runs. While most of them generalise to other COLIBRE-SKIRT applications, we note that some settings (e.g. the SKIRT simulation mode, the primary source wavelength range, or the number of photon packets) might require adjustments depending on the application. We use the standard settings for biasing (\citealt{Baes2016}) and the photon packet life cycle, i.e. forced scattering is enabled but explicit absorption is turned off as we found no performance improvements when enabling explicit absorption (consistent with the results shown in figure 6 of \citealt{Baes2022}). Given the low spectral resolution of our synthetic observations (see Section~\ref{sec:Output}), we do not consider Doppler shifts due to the velocities from the particles.

As we are interested in the full FUV-FIR wavelength range, we use the `DustEmission' simulation mode. For galaxies that do not contain dust\footnote{\label{footnote:dustyGalaxies}We treat all galaxies that have zero dust mass within a 50-kpc exclusive sphere aperture (available from the \texttt{SOAP} catalogues) as dust-free, even though these galaxies could have dust outside this aperture (or in satellite galaxies inside the aperture). Of the 47\,219 galaxies in our main L100m6 sample, 7\,876 galaxies are dust-free. Almost all ($99.7\,\%$) of these dust-free galaxies have no current star formation, and they tend to have lower stellar masses compared to the main sample (the median stellar mass is lower by $\approx0.2\,\mathrm{dex}$).}, we instead postprocess them in the `NoMedium' mode. We use $N_\mathrm{pp}=10^{7.5}$ photon packets per galaxy, comparable to the number used by \citet{Trcka2022} for the TNG50 simulation ($N_\mathrm{pp}=5\times10^7$).

The primary source wavelength range is set to $0.09-2000\,\mu\mathrm{m}$. We store the radiation field from the primary sources from $0.09-20\,\mu\mathrm{m}$ using a log-spaced wavelength grid with 25 points. The dust emission wavelength grid is a nested logarithmic grid with 100 points between $0.09-2000\,\mu\mathrm{m}$ and 100 additional points between $1-30\,\mu\mathrm{m}$ to resolve the MIR PAH emission features. We enable stochastic heating of dust grains which is required to produce the PAH emission features. We do not include dust heating from the cosmic microwave background or dust self-absorption as our tests indicate a negligible impact of these effects (which is expected at low redshift, \citealt{Camps2018}). We performed convergence tests related to $N_\mathrm{pp}$ and the aforementioned wavelength grids, finding that fluxes converge to within $\lesssim3\,\%$ in the MIR and $\lesssim1\,\%$ for other wavelength ranges (see Appendix~\ref{sec:PipelineVariationsAppendix}).

For the dust medium, we use the DL07 dust model, the original version of which has five components: astronomical silicates, graphite 1 (modified power-law grain size distribution), graphite 2 (log-normal grain size distribution), neutral PAHs, and ionized PAHs (see Fig.~\ref{fig:dustModels_sizeDistributions}). While the graphite 2 and PAH components only contribute to the small carbonaceous grains, the silicates and graphite 1 components have to be split into small and large size parts by cutting their size distributions at $a=0.03\,\mu\mathrm{m}$. This gives a total of seven dust media in the SKIRT configuration file. While the silicates and the large carbonaceous grains can each be directly matched to a single component from the DL07 dust model, the situation is more complicated for the small carbonaceous grains as four different components contribute to it. To achieve this, we break the dust mass in small carbonaceous grains predicted by COLIBRE into graphite 1 ($19.46\,\%$), graphite 2 ($20.17\,\%$), neutral PAHs ($30.19\,\%$), and ionized PAHs ($30.19\,\%$). These values are chosen to preserve the relative contributions to small carbonaceous dust (integrating the size distributions up to $a=0.03\,\mu\mathrm{m}$) in the DL07 dust model. This means that $60.38\,\%$ of the dust mass in small carbonaceous grains predicted by COLIBRE will be assigned to PAHs. Since COLIBRE predicts an average of $\approx10\,\%$ of the total dust mass to be in small carbonaceous grains (at $z=0$), $\approx6\,\%$ of the total dust mass will be in PAHs. We use 10 size bins for each of the seven components to discretize the grain size distributions.

To discretize the dust distribution, we use a spatial grid with a box size of $100\,\mathrm{kpc}$ (proper) which is large enough to encompass most of the ISM dust. When constructing the dust grid, all gas particles that are outside of this box are discarded\footnote{Technically, gas particles that have their centers outside of the SKIRT box but overlap with it within the extent of their smoothing kernels would still contribute to the dust radiative transfer. However, according to our tests this effect is completely negligible. Hence, we directly discard gas particles that are located outside the SKIRT box to reduce the I/O load when performing SKIRT simulations.}. This introduces an inconsistency between the treatment of source particles inside and outside the 100-kpc box: primary sources located outside the box are not subject to (local) dust attenuation. This effect is particularly relevant for apertures with radii larger than $50\,\mathrm{kpc}$, but can also arise for smaller apertures, since SKIRT apertures are cylindrical and therefore extend to infinity along the line of sight. We test the impact of this approximation using 30-kpc apertures and find that the galaxy spectra are insensitive to whether primary sources outside the 100-kpc box are included. However, for aperture radii $\geq50\,\mathrm{kpc}$ (i.e. extending beyond our adopted SKIRT grid) the UV fluxes of massive galaxies ($M_\star\gtrsim10^{10.5}\,\mathrm{M}_\odot$) are significantly affected, indicating that larger dust grids should be adopted for such systems when studying extended emission.

The spatial grid parameters are motivated in Appendix~\ref{sec:spatialGrid}. Here we summarise the adopted setup. We use an adaptive $k$-d tree (`binary tree') to grid the dust distribution.  We find that a $k$-d tree is slightly more efficient than the more commonly used octree (e.g. \citealt{Camps2016}; \citealt{Trcka2022}), in agreement with \citet{Saftly2014}. The minimum (maximum) grid refinement level is 15 (36), corresponding to a maximum (minimum) cell size of $3.125\,\mathrm{kpc}$ ($24.41\,\mathrm{pc}$). This maximum refinement level is consistent with level 12 for the octree grids used by \citet{Kapoor2021} and \citet{Trcka2022}. The most important parameter for the construction of the adaptive grid is the refinement criterion $\delta_\mathrm{max}$: If the ratio of cell dust mass to total dust mass exceeds $\delta_\mathrm{max}$, then the grid cell is refined. We scale $\delta_\mathrm{max}$ with the dust surface density $\Sigma_\mathrm{dust}$. For the m6 and m7 COLIBRE resolutions, we use the following relation:
\begin{equation}\label{eq:refinement_m6}
\begin{aligned}
    &\log_{10}(\delta_\mathrm{max})=\\
    &\begin{cases}
      -6 & \text{if $\Sigma_\mathrm{dust}\geq10^6\,\mathrm{M}_\odot\mathrm{kpc}^{-2}$}\\
      -1.5-0.75\log_{10}(\Sigma_\mathrm{dust}/(\mathrm{M}_\odot\mathrm{kpc}^{-2})) & \text{else}\\
      -4.5 & \text{if $\Sigma_\mathrm{dust}\leq10^4\,\mathrm{M}_\odot\mathrm{kpc}^{-2}$}
    \end{cases}
\end{aligned}
\end{equation}
At m5 resolution, we use a more stringent refinement criterion:
\begin{equation}\label{eq:refinement_m5}
\begin{aligned}
    &\log_{10}(\delta_\mathrm{max})=\\
    &\begin{cases}
      -6.5 & \text{if $\Sigma_\mathrm{dust}\geq10^6\,\mathrm{M}_\odot\mathrm{kpc}^{-2}$}\\
      -0.5-\log_{10}(\Sigma_\mathrm{dust}/(\mathrm{M}_\odot\mathrm{kpc}^{-2})) & \text{else}\\
      -4.5 & \text{if $\Sigma_\mathrm{dust}\leq10^4\,\mathrm{M}_\odot\mathrm{kpc}^{-2}$}
    \end{cases}  
\end{aligned}
\end{equation}
The dust surface density is determined as $\Sigma_\mathrm{dust}=M_\mathrm{d}/(\pi R_{\star,1/2}^2)$, where $M_\mathrm{d}$ denotes the dust mass of the galaxy in a 50-kpc exclusive-sphere 3D aperture and $R_{\star,1/2}$ is the 3D stellar half-mass radius (both quantities are taken from the \texttt{SOAP} catalogues).

\subsubsection{Output}\label{sec:Output}

The SKIRT pipeline described in Section~\ref{sec:fiducialPostprocessing} can be used to compute a wide range of observables and other output that SKIRT offers (so-called `probes'). For this work, we always compute low-resolution SEDs in the wavelength range $0.09-2000\,\mu\mathrm{m}$ with 1\,000 log-spaced wavelength bins. This corresponds to a fixed resolving power of $R=99.91$. The SEDs are stored in eight different circular apertures (with radii of infinity, $10\,\mathrm{kpc}$, $30\,\mathrm{kpc}$, $50\,\mathrm{kpc}$, and 1/2/3/5 stellar half-mass radii). Throughout this work, we adopt an aperture radius of $30\,\mathrm{kpc}$ for the SKIRT results. We test the impact of different aperture choices on the CSED in Appendix~\ref{sec:aperture}. We observe all galaxies along the $z$-axis of the COLIBRE simulation box, i.e. effectively along a random orientation. Various components to the total SED are recorded by enabling the corresponding property (`recordComponents') in the SKIRT instruments. Lastly, we also store the luminosities of the primary sources (`LuminosityProbe') on the same wavelength grid as the instrument and information about the convergence of the spatial grid (`ConvergenceInfoProbe').

\section{The low-redshift CSED}\label{sec:CSED}

As a first benchmark for the COLIBRE-SKIRT predictions, we consider the cosmic SED at low redshift. The CSED measures the integrated energy output from the galaxy population in a specific redshift interval. We focus on the CSED for this work as it is relatively cheap to compute from the simulations (small galaxy samples are sufficient) and the comparison to observational data is straightforward (for instance, there is no need to mimic observational selection effects because those have already been accounted for in the observational measurement). For other applications, we caution that the low-redshift CSED is an insufficient metric to establish that the COLIBRE-SKIRT synthetic observations are realistic (particularly at higher redshifts). We defer the analysis of other COLIBRE observables (e.g. luminosity functions, attenuation curves, flux/color relations, galaxy images, IFU spectra) at various redshifts to future work. We show the low-redshift CSED in Fig.~\ref{fig:fiducialCSED} for the COLIBRE/IllustrisTNG/EAGLE simulations and observational data from GAMA/HerMES. We describe these datasets and how to obtain CSEDs in Sections~\ref{sec:simulationCSED} and~\ref{sec:observationCSED}, respectively, and discuss the results in~\ref{sec:CSEDresults}.

\subsection{The simulation CSED}\label{sec:simulationCSED}

\begin{figure*}
    \centering
    \includegraphics[width=\textwidth]{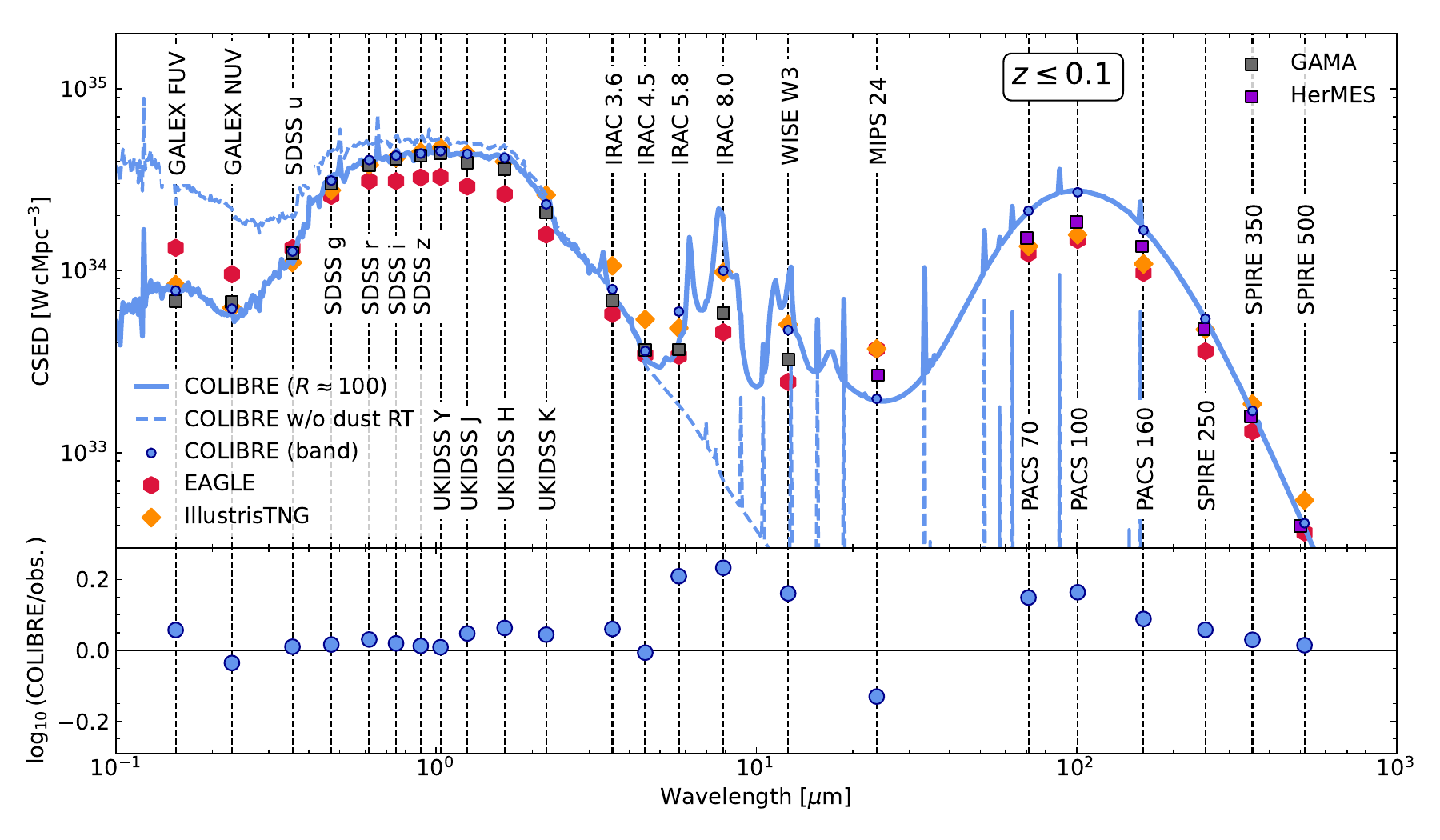}
    \caption{The rest-frame cosmic SED of the COLIBRE L100m6 box averaged over the $z=0$ and $z=0.1$ snapshots, obtained with our fiducial SKIRT setup (solid blue line) and without dust medium in SKIRT (i.e. without dust radiative transfer, dashed blue line). In addition to these CSEDs at the native SKIRT resolution in our setup ($R\approx100$), we also show the filter-convolved COLIBRE CSED as blue markers. The agreement with low-redshift observational data from GAMA ($0.02\leq z\leq0.08$, \citealt{Andrews2017}) and HerMES ($0.02\leq z\leq0.1$, \citealt{Marchetti2016}) is very good, as shown in the bottom panel where we compare the COLIBRE filter-convolved fluxes to the observations. For comparison, we also show SKIRT-derived CSEDs for the IllustrisTNG (TNG100-1) and EAGLE (RecalL0025N0752) simulations. We emphasize that, unlike was the case for EAGLE and IllustrisTNG, we did not calibrate the COLIBRE-SKIRT dust radiative transfer pipeline to reproduce any observations.}
    \label{fig:fiducialCSED}
\end{figure*}

To compute the CSED from a large-volume cosmological simulation, we follow \citet{Baes2019} and sum the rest-frame SEDs of all galaxies in our samples for the $z=0$ and $z=0.1$ snapshots. This cumulative SED is then divided by two times the simulation box co-moving volume. This averaging\footnote{Rather than simply averaging the CSEDs from the two snapshots, one could alternatively interpolate between them and evaluate the CSED at the median redshift of the observed galaxies. The evolution of the CSED in COLIBRE is small but non-negligible, increasing by $\approx10\,\%$ in the UV–NIR wavelength range and by up to $\approx24\,\%$ in the FIR between $z=0$ and $z=0.1$.} over the $z=0$ and $z=0.1$ snapshots is necessary because the observational data that we compare to approximately spans this redshift range (see Section~\ref{sec:observationCSED}). For all simulations, we use SKIRT-derived fluxes obtained by post-processing individual galaxies, where only particles gravitationally bound to each galaxy are considered.

For COLIBRE, we use our main galaxy sample, which comprises all galaxies in the L100m6 box (at $z=0$ and $z=0.1$) with a total stellar mass of $M_\star\geq10^{8.5}\,\mathrm{M}_\odot$. The SEDs are computed with our fiducial COLIBRE-SKIRT pipeline as detailed in Section~\ref{sec:fiducialPostprocessing}. In Figure~\ref{fig:fiducialCSED}, we show the rest-frame COLIBRE CSED at its native spectral resolution ($R\approx100$, solid blue line) as well as the band-convolved version for commonly used broadband filters (blue markers). In the full spectrum, strong nebular emission lines arising from the HII regions (modelled with TODDLERS) are visible. To be consistent with the other simulation datasets, we use the SEDs in a 30-kpc aperture.

For COLIBRE, we use our main galaxy sample, which comprises all galaxies in the L100m6 box (at $z=0$ and $z=0.1$) with a total stellar mass of $M_\star\geq10^{8.5}\,\mathrm{M}_\odot$. The SEDs are computed with our fiducial COLIBRE-SKIRT pipeline as detailed in Section~\ref{sec:fiducialPostprocessing}. In Figure~\ref{fig:fiducialCSED}, we show the rest-frame COLIBRE CSED at its native spectral resolution ($R\approx100$, solid blue line) as well as the band-convolved version for commonly used broadband filters (blue markers). In the full spectrum, strong nebular emission lines arising from the HII regions (modelled with TODDLERS) are visible. To be consistent with the other simulation datasets, we use the SEDs in a 30-kpc aperture.

We also show the COLIBRE CSED without dust radiative transfer (dashed blue line in Fig.~\ref{fig:fiducialCSED}) using the `transparent' SEDs predicted by SKIRT. These SEDs contain only the emission from evolved stellar populations and star-forming regions, without any impact of the dust medium. The impact of dust attenuation on the low-redshift CSED is strong in the UV, as it reduces the flux in the GALEX FUV band by a factor of $\approx4$. Dust attenuation stays relevant into the near-infrared UKIDSS bands, while dust emission becomes important for the MIR and FIR starting at the IRAC 5.8 band. When integrating the COLIBRE CSED to obtain the total energy output of the low-redshift galaxy population, we obtain energy densities that agree to within $\approx1.3\,\%$ between the dust-attenuated and dust-free CSEDs, confirming that energy balance holds on average for the galaxy population. On the other hand, when limiting the wavelength range to $0.09\,\mu\mathrm{m}<\lambda<3\,\mu\mathrm{m}$, the energy density computed from the dust-free CSED is $52.8\,\%$ higher. Put another way, one third of the emitted starlight of the low-redshift galaxy population is absorbed by dust and re-emitted in the infrared. This value is similar to the values inferred through SED fitting from observations of nearby galaxies (\citealt{Popescu2002}; \citealt{Viaene2016}; \citealt{Bianchi2018}).

The CSED for the EAGLE simulation has been analyzed in \citet{Baes2019}, based on the SKIRT-derived observables from \citet{Camps2018}. We use the same rest-frame absolute magnitudes at $z=0$ and $z=0.1$ in random orientation from the public EAGLE database\footnote{\url{https://icc.dur.ac.uk/Eagle/database.php}} (\citealt{McAlpine2016}). These magnitudes are computed in apertures of 30 kpc for various bands. The EAGLE-SKIRT database includes various simulation runs and snapshots, comprising all galaxies with a total stellar mass of $M_\star\geq10^{8.5}\,\mathrm{M}_\odot$ and at least 250 dust particles\footnote{We note that EAGLE does not include dust as a separate species. See section 3.1 of \citet{Camps2018} for more details.}. The latter criterion means that the EAGLE-SKIRT catalogue is biased against galaxies that have a low absolute dust content: For the reference 100-Mpc EAGLE box (RefL0100N1504), only 63.6\,\% of galaxies with $M_\star\geq10^{8.5}\,\mathrm{M}_\odot$ are included in the EAGLE-SKIRT catalogue. Hence, we follow \citet{Baes2019} and analyze the high-resolution recalibrated 25-Mpc EAGLE box (RecalL0025N0752) as our fiducial EAGLE CSED. The completeness is significantly increased to 95.7\,\% for this higher-resolution simulation.

For IllustrisTNG, we use the SKIRT-derived rest-frame SEDs from \citet{Gebek2024} for the 100-Mpc IllustrisTNG box (TNG100-1). These SEDs are available for all galaxies at redshifts $z=0$ and $z=0.1$ with a stellar mass of $M_\star\geq10^{8.5}\,\mathrm{M}_\odot$, where the stellar mass is measured within two stellar half-mass radii. This differs from the stellar mass convention for COLIBRE and EAGLE, where total stellar mass is used. However, we do not find any significant difference in the CSED when restricting the COLIBRE sample to galaxies that have a stellar mass within two stellar half-mass radii above $10^{8.5}\,\mathrm{M}_\odot$. We use the SEDs in random orientation and 30-kpc aperture from the public IllustrisTNG database\footnote{\url{https://www.tng-project.org/data/docs/specifications/\#sec5w}} (\citealt{Nelson2019a}). We convolve the IllustrisTNG CSED with various broadband filters, shown as orange diamonds in Fig.~\ref{fig:fiducialCSED}.

\subsection{The observed CSED}\label{sec:observationCSED}

We use the same observational data as \citet{Baes2019} for their analysis of the EAGLE CSED, namely GAMA data for $0.15\,\mu\mathrm{m}\lesssim\lambda\lesssim15\,\mu\mathrm{m}$ (\citealt{Andrews2017}) and data from the HerMES survey for $20\,\mu\mathrm{m}\lesssim\lambda\lesssim500\,\mu\mathrm{m}$ (\citealt{Marchetti2016}). Both datasets were corrected to the COLIBRE Hubble parameter ($H_0=68.1\,\mathrm{km}\,\mathrm{s}^{-1}\mathrm{Mpc}^{-1}$). Although the observational data include AGN contributions, which are not modelled in our pipeline (see Section~\ref{sec:AGN}), their impact on the low-redshift CSED is expected to be negligible given the low AGN number density at $z<0.5$ (\citealt{Andrews2017}; \citealt{Driver2018}).

GAMA is a spectroscopic survey which obtained redshifts for $\sim200\,000$ galaxies with an r-band magnitude below $19.8\,\mathrm{mag}$, with most galaxies at $z\lesssim0.5$ (\citealt{Driver2011}; \citealt{Liske2015}). Of particular relevance to this work is the panchromatic complementary data\footnote{Specifically, the GAMA database includes photometry from the XMM-
XXL, GALEX, SDSS, KiDS, VIKING, WISE, and Herschel-ATLAS surveys.} that is incorporated into the GAMA database (\citealt{Driver2016}), which enables the observational determination of the FUV-FIR CSED. This procedure is described in detail in \citet{Andrews2017}, we summarize the most important steps here. 

\citet{Andrews2017} use the three equatorial GAMA fields, covering a total area of $180\,\mathrm{deg}^2$. Fluxes are derived from matched-aperture photometry using the \texttt{LAMBDAR} code (\citealt{Wright2016}). To obtain the rest-frame SEDs, \citet{Driver2018} perform SED fitting for each object using \texttt{MAGPHYS} (\citealt{daCunha2008}). The CSED is then obtained in various redshift bins accounting for incompleteness effects due to the limited depth of the survey. For this work, we only consider the CSED\footnote{These data are publicly available at \url{https://academic.oup.com/mnras/article/470/2/1342/3854789\#supplementary-data}.} in the lowest-redshift bin of $0.02\leq z\leq0.08$.

The error budget of the GAMA CSED is dominated by uncertainties in the SED fitting and completeness corrections, as shown in figure 6 of \citet{Andrews2017}. Cosmic variance contributes another 20\,\% uncertainty on the CSED. \citet{Andrews2017} note that their CSED is potentially unreliable for wavelengths $\lambda\gtrsim25\,\mu\mathrm{m}$ due to the underlying data (from \textit{Herschel}-Atlas, \citealt{Eales2010}) being missing or of too low sensitivity and resolution.

Because of the uncertainties in the GAMA IR CSED, we instead
use the determination of the $24-500\,\mu\mathrm{m}$ CSED from \citet{Marchetti2016} for this wavelength range. Their results are based on the HerMES survey (\citealt{Oliver2012}), the largest single project with the
Herschel space telescope. The HerMES wide fields that are used (covering
an area of 38.9 square degrees) are part of the Spitzer Data Fusion
(\citealt{Vaccari2010}), with ancillary Spitzer seven-band IR imaging, 2MASS
3-band NIR imaging and SDSS five-band optical imaging and spectroscopy.

To identify sources, \citet{Marchetti2016} use the SPIRE 250 channel as this
is the most sensitive of the three SPIRE bands. To mitigate source confusion,
MIPS 24 source positions are used as priors. In total, $7\,087$ galaxies
are detected within $0.02<z<0.5$, with redshifts obtained from the SDSS
catalogues. The monochromatic rest-frame luminosities are then obtained
through template-based LE PHARE (\citealt{Arnouts1999}; \citealt{Ilbert2006}) SED fitting, using all available bands (SDSS, 2MASS, IRAC, MIPS, SPIRE). This procedure means that there are effectively three sensitivity limits that affect the sample: The limit in r-band magnitude which guides the photometric redshift estimates in SDSS ($m_r < 22.2\,\mathrm{mag}$), the MIPS 24 flux limit which is used as a prior for the source extraction ($S_{24}>300\,\mu\mathrm{Jy}$),
and the flux density limit in the SPIRE 250 band ($S_{250}>30\,\mathrm{mJy}$).

\citet{Marchetti2016} derive monochromatic\footnote{We caution that the HerMES data use monochromatic fluxes, whereas all other datasets consist of band-integrated fluxes. As a result, the comparison between COLIBRE and HerMES in Fig.~\ref{fig:fiducialCSED} is not fully consistent. However, by comparing monochromatic and band-integrated fluxes within COLIBRE, we verify that this effect is minor (see also \citealt{Lu2026}). The largest difference occurs in the SPIRE 500 band, where the monochromatic flux at $500\,\mu\mathrm{m}$ exceeds the band-integrated value by $\approx 6.7\,\%$.} rest-frame luminosity functions based on the $1/V_\mathrm{max}$ method (\citealt{Schmidt1968}) to correct for sample incompleteness. The infrared luminosity densities are then computed from modified Schechter function fits to the luminosity functions over the redshift range $0.02<z<0.1$. \citet{Marchetti2016} find good agreement between their IR CSED and earlier studies, including results from an earlier GAMA-based study (\citealt{Driver2012}). On the other hand, we found that between
$20\,\mu\mathrm{m}<\lambda<100\,\mu\mathrm{m}$ the GAMA-based CSED (\citealt{Andrews2017}) and the HerMES-based CSED (\citealt{Marchetti2016}) CSEDs differ significantly: At the lowest wavelength available available in \citealt{Marchetti2016} ($24\,\mu\mathrm{m}$), the monochromatic GAMA CSED is higher by $0.29\,\mathrm{dex}$. This difference reduces at larger wavelengths, e.g. $0.09\,\mathrm{dex}$ at $100\,\mu\mathrm{m}$ and $0.04\,\mathrm{dex}$ at $500\,\mu\mathrm{m}$.

When comparing CSEDs between simulations and observations, two caveats should be noted. The first concerns the choice of apertures. We adopt a fixed 30-kpc aperture for all simulation datasets. As shown in Fig.~\ref{fig:apertures}, the impact of aperture choice is strongest in the optical and UV, where we compare to GAMA data. In GAMA, photometry is measured using \texttt{LAMBDAR} apertures, which yield circularized radii of $\approx5-30\,\mathrm{kpc}$ depending on stellar mass, and can be approximated as $3\times R_{\star,1/2}$ (\citealt{Gebek2024}). We find that using a fixed 30-kpc aperture instead of $3\times R_{\star,1/2}$ yields a slightly higher CSED, with the biggest impact in the GALEX FUV band ($\approx24\,\%$, see Fig.~\ref{fig:apertures}). This suggests that our use of 30-kpc apertures slightly overestimates the UV CSED relative to the observations.

Second, we consider the impact of sample selection and completeness. The observational datasets that we consider are corrected for incompleteness and therefore aim to represent the full low-redshift galaxy population down to low stellar masses. Because the CSED is dominated by galaxies near the knee of the stellar mass function (see Fig.~2 in \citealt{Baes2019}), it is relatively insensitive to low-mass galaxies. On the other hand, our simulation samples are limited by a stellar mass threshold. We assess the impact of this limitation using the galaxy sample from the L025m6 simulation, which extends to lower stellar masses ($M_\star\geq10^7\,\mathrm{M}\odot$ rather than $M_\star\geq10^{8.5}\,\mathrm{M}_\odot$). Including these lower-mass galaxies increases the CSED most noticeably in the UV, with the GALEX FUV flux rising by $\approx11.8\,\%$. However, this remains a relatively small correction, and we therefore do not adjust the simulation CSEDs to account for sample incompleteness. We note that these two effects, aperture choice and sample completeness, act in opposite directions and therefore partially offset each other.

\subsection{CSED results}\label{sec:CSEDresults}

The low-redshift COLIBRE CSED from the L100m6 box shows excellent agreement with the observations over the UV–NIR wavelength range, while the match in the MIR–FIR regime is somewhat less accurate. As illustrated by the residuals in the lower panel of Fig.~\ref{fig:fiducialCSED}, the COLIBRE predictions are generally consistent with the observational data to within $\lesssim0.2\,\mathrm{dex}$. In the following, we discuss the wavelength ranges separately, comparing our COLIBRE results with observational data and with the EAGLE and IllustrisTNG simulations.

We emphasize that, contrary to EAGLE and IllustrisTNG, the COLIBRE-SKIRT pipeline as described in Section~\ref{sec:fiducialPostprocessing} has not been calibrated to any observational data. Hence, the synthetic observables generated with this framework are more directly tied to the underlying cosmological simulation compared to previous postprocessing efforts for large-volume simulations. In that sense, the good match of the COLIBRE CSED with low-redshift GAMA data is not necessarily expected, and is an important test for the realism of COLIBRE. Since the CSED is an integrated quantity, more fine-grained tests (e.g. in terms of luminosity functions or galaxy image morphologies) will be required to confirm the realism of the simulated COLIBRE galaxies. Using the COLIBRE-SKIRT pipeline, \citet{Lu2026} already find very good agreement between simulation predictions and observational data across most wavelengths, with the exception of the $8\,\mu\mathrm{m}<\lambda<24\,\mu\mathrm{m}$ regime.

\subsubsection{Ultraviolet (GALEX bands)}

While the UV slope in COLIBRE is slightly bluer than in the GAMA data, the agreement in the GALEX bands is nevertheless very good. In the UV, we find that COLIBRE predicts a similar CSED as IllustrisTNG, while EAGLE is significantly brighter. At first sight, this is a surprising result because both for EAGLE and IllustrisTNG a subgrid model for dust attenuation is used, which attenuates light from the youngest stellar populations locally and hence reduces the UV flux. Even though we did not use any subgrid dust attenuation for COLIBRE, the UV flux in COLIBRE is lower than for EAGLE and comparable to IllustrisTNG.

We stress that this is a remarkable result: for the first time, a large-volume simulation reproduces the UV CSED of the low-redshift galaxy population without the need to invoke extra subgrid dust attenuation. We attribute this to the presence of a multiphase ISM in COLIBRE which is clumpier compared to previous large-volume simulations, which imposed a temperature or entropy floor on the ISM, thus preventing the formation of a cold gas phase. Compared with these earlier simulations, this leads to higher dust densities\footnote{We find that, at $z=0$, the gas densities at star formation (i.e. when a gas particle is converted into a star particle) are approximately two orders of magnitude higher in COLIBRE than in EAGLE (at comparable resolutions).} around the sites of star formation and hence more effective attenuation of UV light from the diffuse (resolved) ISM. Therefore, the diffuse ISM dust attenuation is sufficiently strong in COLIBRE to bring down the intrinsic UV emission to observed levels, unlike IllustrisTNG and EAGLE which required a subgrid dust attenuation model.

\subsubsection{Optical-NIR (SDSS and UKIDSS bands)}

COLIBRE shows excellent agreement with the observational GAMA data, similar to IllustrisTNG. In the optical–NIR range, \citet{Baes2019} note that the EAGLE CSED is too low due to incompleteness in the galaxy sample (arising from the requirement that galaxies contain at least 250 dust particles) and the relatively small simulation volume ($L_\mathrm{box}=25\,\mathrm{cMpc}$). However, EAGLE also appears to underestimate dust attenuation more generally, as indicated by an excess in the UV and a deficit in the FIR. This discrepancy between EAGLE and the observational data cannot be explained by sample incompleteness alone, since the bias primarily affects dust-poor, quiescent galaxies (see Fig.~4 in \citealt{Camps2018}), which contribute only minimally to the UV and FIR emission.

\begin{figure*}
    \centering
    \includegraphics[width=\textwidth]{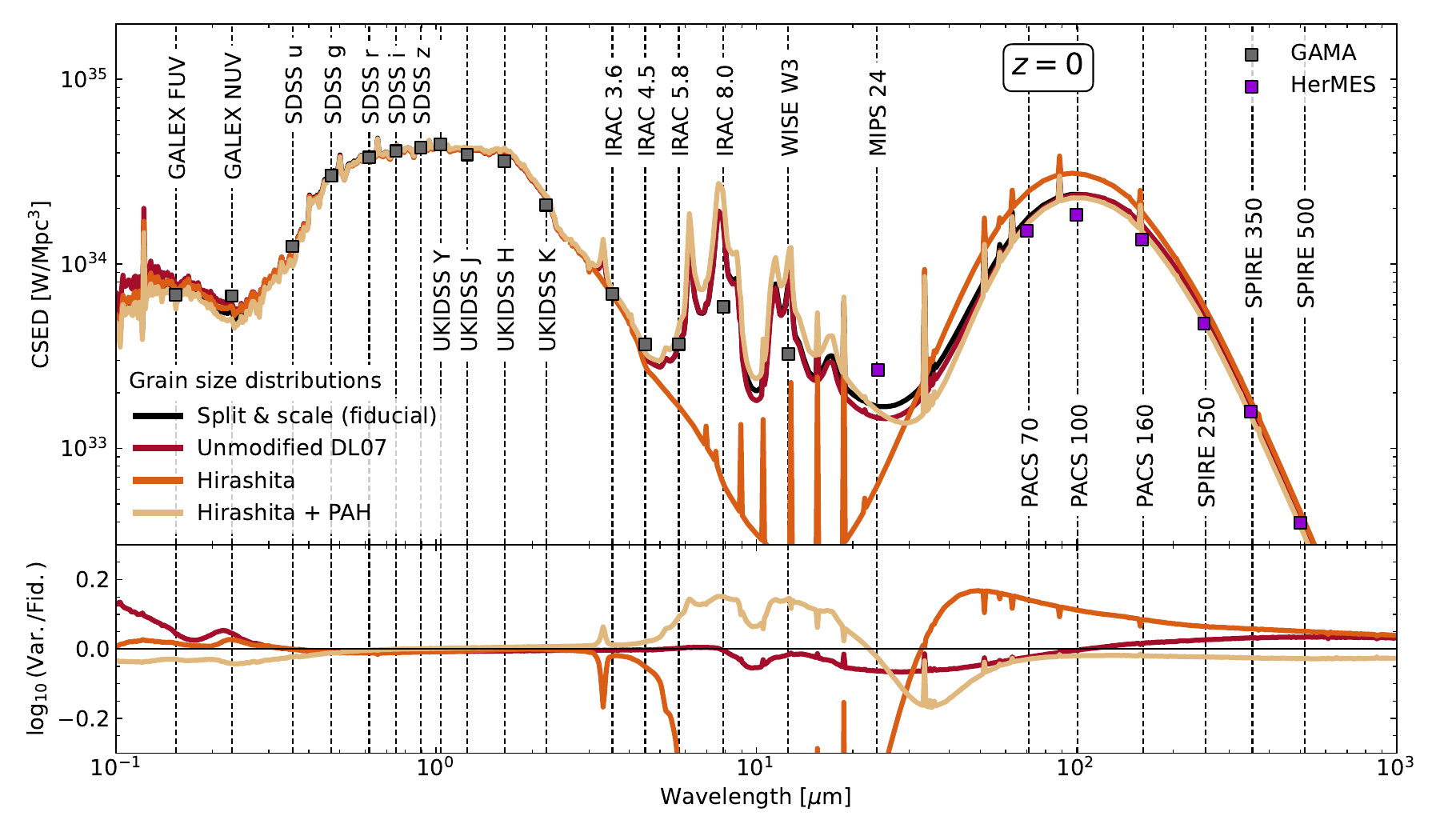}
    \caption{The COLIBRE L200m6 redshift-zero CSED, varying the treatment of the dust grain size distribution. The fiducial model is shown in black, where the original DL07 size distributions are split at $a=0.03\,\mu\mathrm{m}$ and the four dust components (small silicates, large silicates, small carbonaceous grains, and large carbonaceous grains) are scaled depending on the COLIBRE grain species abundances. For the `Unmodified DL07' CSED (red line), we did not take the dust species fractions predicted by COLIBRE into account, and instead used the original DL07 dust model for all gas particles. The `Hirashita' CSED (dark orange line) employs log-normal size distributions similar to \citet{Hirashita2015}, which lack small carbonaceous grains and hence the MIR PAH emission features are absent in this model. For the `Hirashita + PAH' CSED (light brown line) we treated all small carbonaceous grains in COLIBRE as PAHs, while the remaining COLIBRE dust grains follow log-normal size distributions. The observational data is shown by the grey (GAMA) and purple (HerMES) markers.}
    \label{fig:CSEDvariation_sizeDistribution}
\end{figure*}

\subsubsection{Mid-infrared (IRAC and WISE W3 bands)}

Significant discrepancies between all considered simulations and GAMA exist in the MIR, particularly for COLIBRE and IllustrisTNG. In the case of COLIBRE, the CSED begins to deviate from the observations in the IRAC 5.8 band (maximum overestimation of up to $\approx0.23\,\mathrm{dex}$ in the IRAC 8.0 band). As we show in Section~\ref{sec:Variations-DustModel}, the MIR is significantly affected by choices related to the dust model and grain size distributions, as MIR emission up to $\lambda\approx20\,\mu\mathrm{m}$ is completely dominated by small carbonaceous grains (see Fig.~\ref{fig:dustModels_opticalProperties}). While alterations to our fiducial postprocessing choices could mitigate some of the tensions in the MIR, we note that all of the variations that we consider that lower the MIR fluxes simultaneously increase the FIR fluxes, thereby creating new tensions. Hence, an entirely new dust model with different PAH emission spectra might be required to match the observed MIR CSED.

\subsubsection{Far-infrared (MIPS 24, PACS, and SPIRE bands)}

\begin{figure*}
    \centering
    \includegraphics[width=\textwidth]{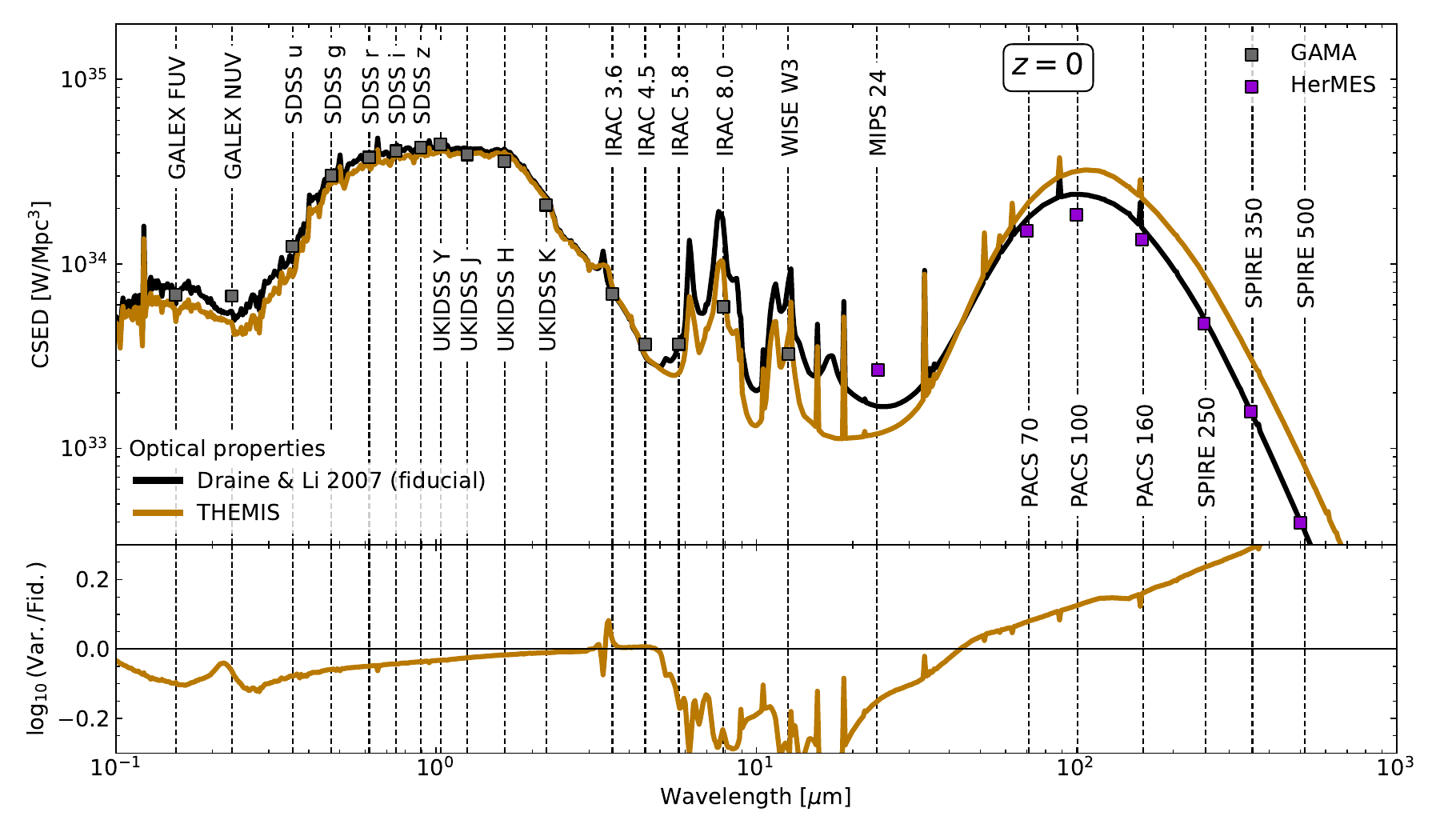}
    \caption{Same as Fig.~\ref{fig:CSEDvariation_sizeDistribution}, but varying the optical properties of the dust grains. For both the DL07 and THEMIS dust models we use our fiducial `split \& scale' method to treat the grain size distributions. We find that the COLIBRE CSED using THEMIS exhibits too high FIR fluxes.}
    \label{fig:CSEDvariation_opticalProperties}
\end{figure*}

As described in Section~\ref{sec:observationCSED}, we use HerMES data instead of GAMA in this wavelength regime. The cold dust emission traced by the SPIRE bands is well reproduced by COLIBRE, while EAGLE slightly underpredicts the SPIRE 250 flux and IllustrisTNG slightly overpredicts the SPIRE 500 flux. The slope across the SPIRE bands is directly related to the spectral index $\beta$, which parametrizes the frequency dependence of the dust emissivity ($\kappa_\nu\propto\nu^\beta$ where $\kappa_\nu$ is the dust mass absorption coefficient). The THEMIS dust model used in IllustrisTNG (\citealt{Gebek2024}) employs relatively low $\beta$ values, resulting in flatter FIR slopes (see also Fig.~\ref{fig:CSEDvariation_opticalProperties}). We note that the more recent THEMIS v2 model (\citealt{Ysard2024}) features higher $\beta$, which may alleviate this discrepancy.

In contrast, for the MIPS 24 band (tracing the hottest dust) and the PACS bands around the FIR peak, COLIBRE shows the opposite trend compared to EAGLE and IllustrisTNG: the MIPS 24 flux is underpredicted, while the PACS bands are overpredicted. The discrepancy is more pronounced in COLIBRE than in the other simulations. Improving the agreement in this regime requires higher dust temperatures. This may indicate that the star–dust geometry in COLIBRE is not sufficiently realistic and/or that the numerical resolution is insufficient. Mitigating this discrepancy may require alternative treatments of the smoothing lengths or the inclusion of subgrid dust, as discussed in Section~\ref{sec:Variations-YoungStars}. We also note that \citet{Lu2026} find that the largest discrepancy between COLIBRE-SKIRT predictions and observed luminosity functions at $z=0$ occurs in the $24\,\mu\mathrm{m}$ band, where the bright end is significantly underpredicted by the COLIBRE-SKIRT pipeline.

\section{COLIBRE-SKIRT pipeline variations}\label{sec:PostprocessingChoices}

While the COLIBRE-SKIRT pipeline is not calibrated to reproduce a specific set of observations, there are still a fair number of postprocessing choices that need to be made. In this section, we analyze how alterations to our fiducial pipeline affect the FUV-FIR SEDs. We use the subsampled L200m6 box at $z=0$ for these pipeline variations. This sample consists of 300 galaxies in six stellar mass bins of width $0.5\,\mathrm{dex}$, starting at $M_\star=10^{8.5}\,\mathrm{M}_\odot$. Since the CSED is effectively a stack of all galaxy SEDs, it is not necessary to postprocess all simulated galaxies - particularly low-mass galaxies can be subsampled and the CSED is still converged (see Appendix~\ref{sec:subsampling}).

We discuss the two most important pipeline variations in this section: choices related to the dust model (Section~\ref{sec:Variations-DustModel}) and variations in the treatment of young stellar populations (Section~\ref{sec:Variations-YoungStars}). Other pipeline variations (grain size variations when using dust optical properties from THEMIS instead of DL07, different SED templates for the evolved stellar populations) are discussed in Appendix~\ref{sec:PipelineVariationsAppendix}.

\subsection{Pipeline variations: Dust model}\label{sec:Variations-DustModel}

\begin{figure*}
    \centering
    \includegraphics[width=\textwidth]{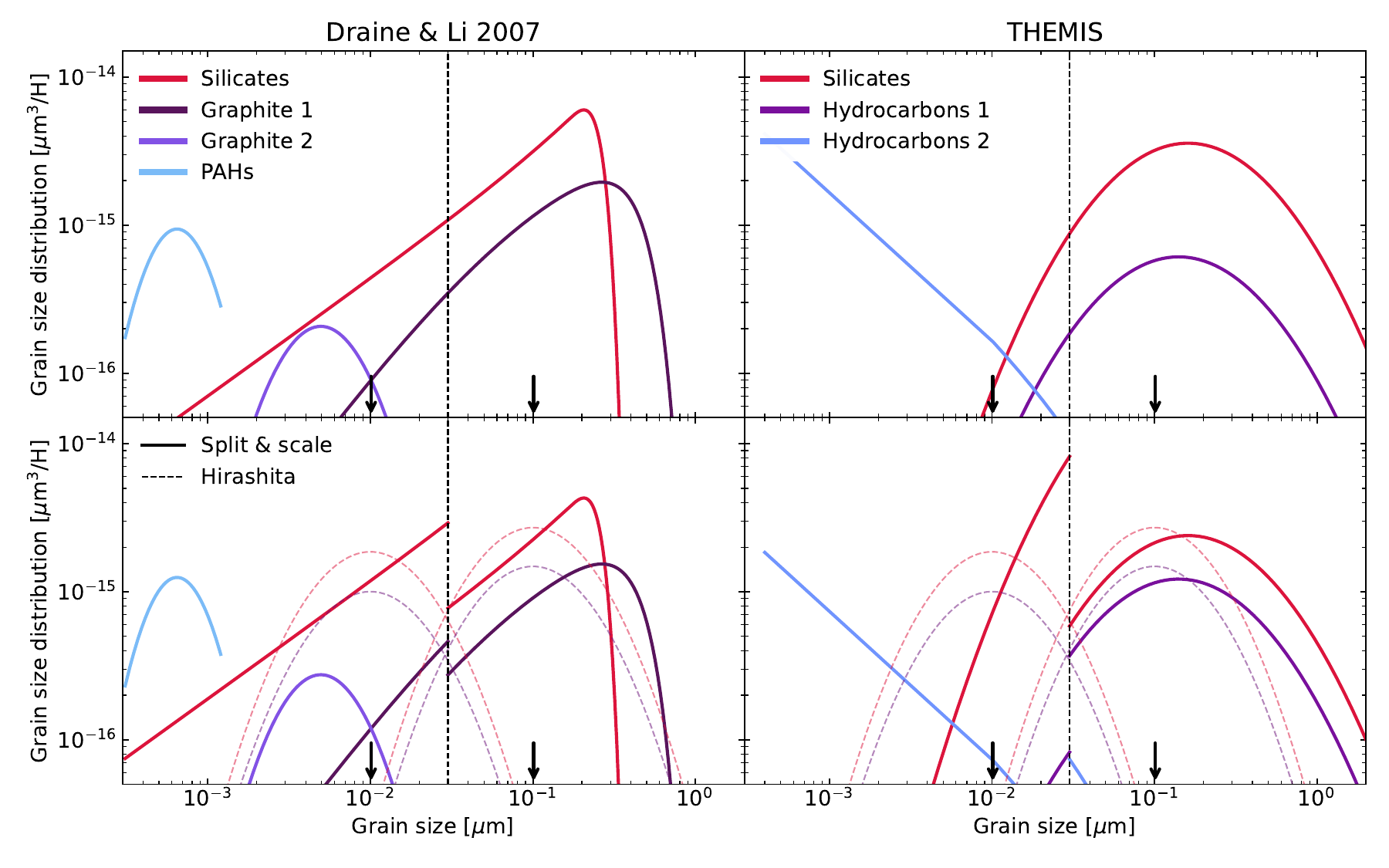}
    \caption{Dust grain size distributions of the two dust models considered. The left column shows the dust model from DL07, the right column shows the THEMIS dust model (\citealt{Jones2017}). The grain size distributions (defined as $\frac{4}{3}\pi a^3\,\mathrm{d}n(a)/\mathrm{d}\log_{10}(a)$) of the original dust models are shown in the upper panels. Arrows indicate the intrinsic COLIBRE dust grain sizes (\citealt{Trayford2026}). In the lower panels, we show modified grain size distributions according to the average redshift-zero grain species abundances in COLIBRE (see Table~\ref{tab:sizeDistributions}). In the `split \& scale' approach (solid lines), the size distributions of all components are cut into two pieces at $a=0.03\,\mu\mathrm{m}$ (indicated by the vertical dashed line in the upper panels) and scaled according to the COLIBRE grain species abundances. In the `Hirashita' approach (thin dashed lines), the size distributions are replaced by log-normal distributions centered at $0.01\,\mu\mathrm{m}$ and $0.1\,\mu\mathrm{m}$ (Eq.~\ref{eq:log-normal}), again scaled to the average COLIBRE grain species abundances. Because the grain size distributions are scaled to the local grain species abundances, each COLIBRE gas particle is assigned a unique set of grain size distributions in SKIRT.}
    \label{fig:dustModels_sizeDistributions}
\end{figure*}

\begin{figure*}
    \centering
    \includegraphics[width=\textwidth]{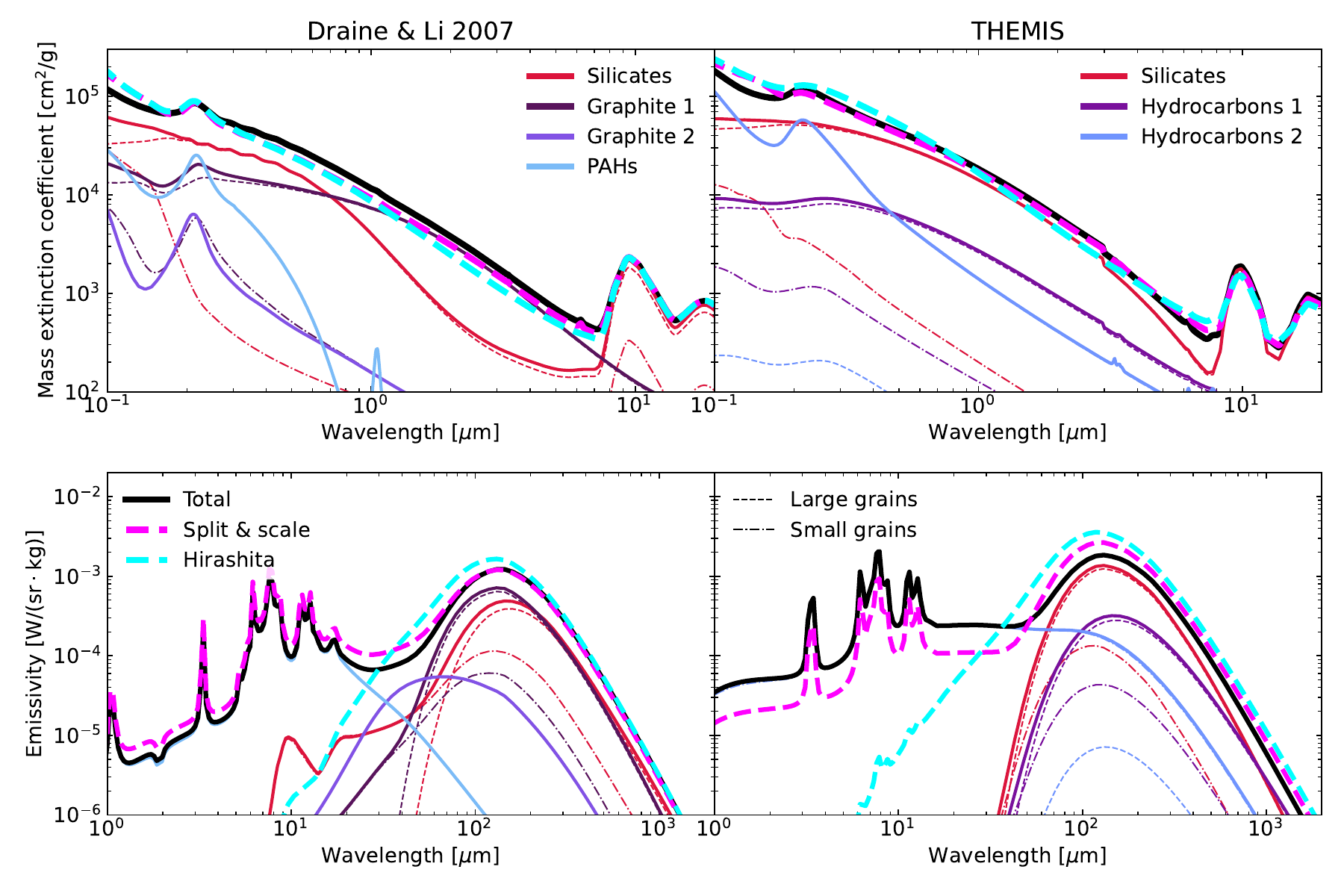}
    \caption{Optical properties of the two dust models considered. The left column shows the dust model from DL07, the right column shows the THEMIS dust model (\citealt{Jones2017}). The upper panels indicate the dust mass extinction coefficients of each individual component for the original dust models. Similarly, the emissivities of the dust model components when exposed to a \citet{Mathis1983} interstellar radiation field are shown in the bottom panels. For each component, we separately show the contribution of large ($a\geq0.03\,\mu\mathrm{m}$, dashed lines) and small ($a<0.03\,\mu\mathrm{m}$, dash-dotted lines) grains (using the original dust model). The sum of all components (corresponding to the total dust mass extinction coefficient or emissivity of the original dust model) is shown by the solid black lines. The magenta and cyan dashed lines indicate the total extinction coefficients and emissivities obtained after modifying the grain size distributions. These modifications reflect the average redshift-zero grain species abundances in COLIBRE (see Table~\ref{tab:sizeDistributions}) and correspond to the distributions shown in the lower panels of Fig.~\ref{fig:dustModels_sizeDistributions}. Because the size distributions are defined on a per-particle basis, the resulting optical properties likewise vary between gas particles.}
    \label{fig:dustModels_opticalProperties}
\end{figure*}

We show the CSED of the subsampled L200m6 box in Fig.~\ref{fig:CSEDvariation_sizeDistribution}, adopting the DL07 dust model (our fiducial choice) but varying the treatment of the grain size distribution. Differences with respect to the fiducial `split \& scale' model (described in Section~\ref{sec:Dust}) are shown in the lower panel of Fig.~\ref{fig:CSEDvariation_sizeDistribution}. A simplistic treatment where we ignore the dust species fractions predicted by COLIBRE (but we do use the predicted total dust masses) and use the original DL07 dust model equally for all gas particles (`Unmodified DL07', red line) yields only minor differences compared to our fiducial approach. This means that, at least at the level of the $z\leq0.1$ CSED, the detailed dust species information in COLIBRE (tracking two size bins and three dust species) is not necessarily required to obtain realistic results. Since the CSED is an integrated quantity, it is possible that this result no longer holds for more `resolved' quantities like color-color relations or spatially resolved observables. We defer more detailed tests of how the dust species information affects synthetic observables to future work.

We also test the two-size approximation from \citet{Hirashita2015} (`Hirashita'), where we use the following log-normal grain size distribution:

\begin{equation}\label{eq:log-normal}
    \frac{\mathrm{d}n_\mathrm{D}}{\mathrm{d}a}\propto\frac{1}{a^4}\exp\Bigl(-\frac{(\ln(a/a_0))^2}{2\sigma^2}\Bigr),
\end{equation}
where $a_0$ is the centroid and $\sigma$ the width of the size distribution. \citet{Hirashita2015} showed that grain size distributions in galaxy evolution simulations can be approximated by considering two discrete sizes (as is done in COLIBRE), and suggest Equation~\ref{eq:log-normal} to expand two discrete grain sizes into continuous distributions for dust radiative transfer calculations (see also \citealt{Dubois2024}; \citealt{Byun2025}). We test this approach using the COLIBRE grain sizes ($0.01\,\mu\mathrm{m}$ and $0.1\,\mu\mathrm{m}$) as centroids and adopt $\sigma=0.75$ following \citet{Hirashita2015}. The entire dust model is then composed of four log-normal size distributions (small silicates, large silicates, small carbonaceous grains, and large carbonaceous grains), with the normalizations of the distributions derived directly from the dust species masses (see dashed lines in Fig.~\ref{fig:dustModels_sizeDistributions}). For the optical properties of the dust grains, we adopt the DL07 silicates component for the COLIBRE silicate grains, the DL07 graphite for the large carbonaceous grains, and the DL07 PAHs for the small carbonaceous grains.

The result of this calculation is shown in Fig~\ref{fig:CSEDvariation_sizeDistribution} (`Hirashita', dark orange line). While the UV-NIR CSED is very similar to the fiducial model, the IR dust emission is dramatically different. This is due to the lack of very small (carbonaceous) grains in the log-normal size distributions, which leads to a complete absence of MIR PAH emission (as discussed in Section~\ref{sec:Dust}) despite using the optical properties of the DL07 PAH grains. To mitigate this lack of PAH emission, we also test a model where we do not use a log-normal size distribution for the small carbonaceous grains, but instead use the original DL07 size distribution for PAHs. This model (`Hirashita + PAH', light brown line) is much closer to the fiducial model, with slightly more PAH emission and less dust emission around $\lambda\approx40\,\mu\mathrm{m}$.

In Fig.~\ref{fig:CSEDvariation_sizeDistribution}, we showed different grain size distributions while keeping the optical properties of the dust grains fixed. In Fig.~\ref{fig:CSEDvariation_opticalProperties} we instead compare different optical properties (THEMIS and DL07) using our `split \& scale' approach. We find that the attenuation and FIR dust emission increase significantly when using THEMIS, significantly worsening the agreement of the COLIBRE CSED with observational data. We found that this also holds for other grain size distributions when using the optical properties from THEMIS (see Fig.~\ref{fig:CSEDvariation_THEMIS}).

To explain these differences in the CSED, we separately show the dust grain size distributions (Fig.~\ref{fig:dustModels_sizeDistributions}) and optical properties of the DL07 and THEMIS dust models (Fig.~\ref{fig:dustModels_opticalProperties}). The upper panels in Fig.~\ref{fig:dustModels_sizeDistributions} show the grain size distributions of the original dust models: The DL07 dust model consists of a silicate component, two graphite components (with the same optical properties and mass densities), and PAHs. The PAH component consists of 50\,\% neutral and 50\,\% ionized PAHs, with exactly the same size distribution and similar optical properties. Hence, we always show the total PAH component (sum of neutral and ionized PAHs) in Figs.~\ref{fig:dustModels_sizeDistributions} and~\ref{fig:dustModels_opticalProperties}. THEMIS (\citealt{Jones2017}) consists of a silicate component which is composed of 50\,\% enstatite and 50\,\% forsterite. Since the size distributions are the same and the optical properties are very similar for these two silicate grains, we only show the total silicate component in Figs.~\ref{fig:dustModels_sizeDistributions} and~\ref{fig:dustModels_opticalProperties}. THEMIS also has two hydrocarbon components which have the same optical properties, but slightly different mass densities.

In the lower panels of Fig.~\ref{fig:dustModels_sizeDistributions}, we show modified grain size distributions scaled to the average redshift-zero grain species abundances in COLIBRE (see Table~\ref{tab:sizeDistributions}). In our fiducial `Split \& scale' approach (solid lines), the original distributions are divided at $a=0.03\,\mu\mathrm{m}$ and each segment is rescaled according to these abundances. This introduces a discontinuity at the split, which is particularly pronounced for THEMIS. This reflects the fact that THEMIS differs substantially from the effective grain composition of the $z=0$ COLIBRE galaxy population: COLIBRE contains a relatively large fraction of small silicate grains ($\approx30\,\%$ of the total dust mass) and fewer small carbonaceous grains ($\approx10\,\%$). In contrast, THEMIS contains only a small fraction of silicates below $0.03,\mu\mathrm{m}$ ($\approx3\,\%$ of the total dust mass), but a substantially larger fraction of small carbonaceous grains ($\approx23\,\%$, in the form of hydrocarbons).

For comparison, we also show the commonly used prescription of \citet{Hirashita2015} (red dashed lines for silicates, purple dashed lines for carbonaceous grains), in which small and large grains are represented by separate log-normal size distributions (Eq.~\ref{eq:log-normal}). These four log-normal components (small silicates, large silicates, small carbonaceous grains, and large carbonaceous grains) are normalised using the same COLIBRE grain species abundances, making the resulting grain size distributions equal between the DL07 and THEMIS dust models.

\begin{figure*}
    \centering
    \includegraphics[width=\textwidth]{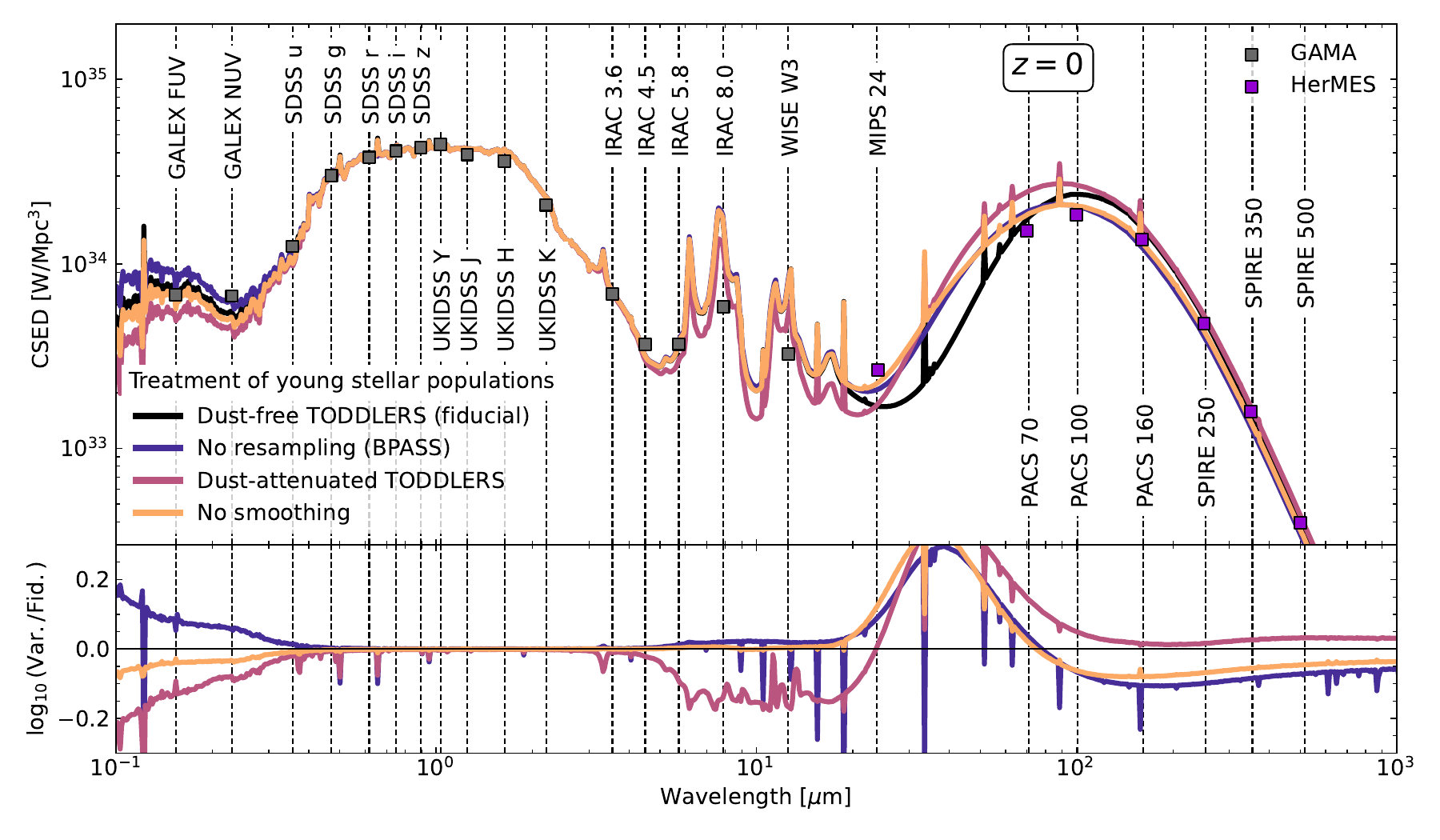}
    \caption{Same as Fig.~\ref{fig:CSEDvariation_sizeDistribution}, but varying the treatment of young stellar populations. The fiducial model, where the star-forming gas particles are used as sources without any subgrid dust attenuation, is shown in black. The `No resampling (BPASS)' CSED (blue line) uses the young star particles directly (modelled with BPASS) instead of the star-forming gas. We also show a variation including subgrid dust attenuation applied to the star-forming gas particles (`Dust-attenuated TODDLERS' CSED, purple line). Lastly, the `No smoothing' (orange line) CSED is computed by setting all smoothing lengths of star-forming gas particles to $0.001\,\mathrm{pc}$. The UV and MIR wavelength ranges are particularly sensitive to the treatment of young stellar populations.}
    \label{fig:CSEDvariation_SFregions}
\end{figure*}

Fig.~\ref{fig:dustModels_opticalProperties} illustrates how the different components and grain size distributions are reflected in the optical properties of the dust. In the upper panels of Fig.~\ref{fig:dustModels_opticalProperties}, we show how the different dust model components contribute to the total dust mass extinction coefficient of the original dust models (`Total', solid black line). We split all components into their contributions from large ($a\geq0.03\,\mu\mathrm{m}$, dashed lines) and small ($a<0.03\,\mu\mathrm{m}$, dash-dotted lines) grains, except for the DL07 PAHs and Graphite 2 components because these only consist of small grains. Since small grains exhibit steeper extinction curves than large grains (\citealt{Mathis1977}; \citealt{Draine2003}), large grains dominate extinction in the optical/NIR, while small grains contribute significantly in the UV. When adjusting the grain size distributions from the original dust models (lower panels in Fig.~\ref{fig:dustModels_sizeDistributions}) according to the COLIBRE dust species fractions in $z=0$ galaxies (see Table~\ref{tab:sizeDistributions}), the extinction curves become steeper for the DL07 dust model, both for the fiducial `split \& scale' approach (dashed pink line) and the \citet{Hirashita2015} log-normal approximation (dashed cyan line). This happens because the abundance of small silicate grains is increased by a factor of $\approx3$ for the DL07 dust model. For THEMIS, the steepening of the extinction curve is less strong, but the 2175\,\AA{} bump weakens significantly. This is due to a reduction of small carbonaceous grains by a factor of $\approx2$. The reduced extinction in the UV is overcompensated by a factor of $\approx9$x increase of the abundance of small silicate grains.

We also show the emissivities of the dust grains when exposed to a \citet{Mathis1983} interstellar radiation field (which is typical of the solar neighbourhood) in the lower panels of Fig.~\ref{fig:dustModels_opticalProperties}. Since smaller grains have lower heat capacities, they can be heated to higher temperatures and emit at shorter wavelengths (\citealt{Draine2003}). The very small carbonaceous grains which are stochastically heated give rise to the MIR PAH emission features (\citealt{Draine2001}). The \citet{Hirashita2015} log-normal approximation, which is designed to reproduce dust extinction curves, completely lacks PAH emission in the MIR. On the other hand, our fiducial `split \& scale' approach yields slightly more PAH emission in the DL07 dust model, since the small carbonaceous grains are enhanced by a factor of $\approx1.3$ relative to the original dust model. The situation is reversed for THEMIS, where small carbonaceous grains are reduced by a factor of $\approx2.2$ relative to the original dust model, leading to a reduction of the MIR PAH emission.

Altogether, the changes in the dust extinction curves and emissivities explain the differences seen in the CSEDs in Figures~\ref{fig:CSEDvariation_sizeDistribution} and~\ref{fig:CSEDvariation_opticalProperties}. Figs.~\ref{fig:dustModels_sizeDistributions} and~\ref{fig:dustModels_opticalProperties} also underscore the challenge for dust radiative transfer calculations applied to simulations with a live dust model: the live dust model gives some constraints on the grain sizes and composition, but these grains still need to be associated with optical properties. Since the optical properties of dust grains are tied to specific grain size distributions in commonly used dust models such as DL07 or THEMIS, it is not obvious how to incorporate the dust species information from the simulation without `breaking' the dust model. Because the DL07 size distributions yield relatively similar dust species fractions as $z=0$ COLIBRE galaxies and the `split \& scale' approach reproduces the original DL07 dust model well without introducing additional free parameters, this justifies our adoption of the DL07 dust model together with our `split \& scale' approach for the fiducial COLIBRE-SKIRT pipeline.

\subsection{Pipeline variations: Young stellar populations}\label{sec:Variations-YoungStars}

As outlined in Section~\ref{sec:StarFormingRegions}, modelling the emission from young stellar populations comes with challenges related to sampling and subgrid attenuation. To combat the sampling problem, we use the star-forming gas particles as sources instead of the actually formed young star particles. The impact of this on the $z=0$ L200m6 CSED is modest as shown in Fig.~\ref{fig:CSEDvariation_SFregions}, where we compare our fiducial model (`Dust-free TODDLERS', black line) with a model where we use the young star particles directly as sources (`No resampling (BPASS)', blue line). This model variation exhibits a slight increase in the UV flux and a shift of the dust emission peak to shorter wavelengths (hotter dust). The hotter dust improves the agreement with observational data, increasing the MIPS 24 flux by $\approx0.10\,\mathrm{dex}$ while reducing the PACS 100 flux by $\approx 0.07\,\mathrm{dex}$.

The increase in the UV flux in the `No resampling'-case persists, to some degree, to dust-free SEDs: Upon consideration of the dust-free CSED, we find that the `No resampling'-case predicts $\approx7\,\%$ more flux in the GALEX FUV band. This is due to the discrete time sampling\footnote{We find that when summing the stellar mass formed over the past $10\,\mathrm{Myr}$ from all 300 galaxies in our L200m6 sample the SFR is $\approx5.0\,\%$ higher than when summing all ten-Myr-averaged SFRs from the star-forming gas (accounting for the small correction due to parent gas particles of actually formed stars, see Section~\ref{sec:StarFormingRegions}). When repeating the calculation for the main L100m6 sample at $z=0$ (23\,490 galaxies), the difference between the two SFRs is $\approx6.2\,\%$, so this is not just a stochastic effect from the relatively small L200m6 sample. In addition, the discrete time sampling inherent to using formed star particles results in a UV luminosity that is enhanced relative to a smoothly sampled star-formation history over the past $10\,\mathrm{Myr}$ for the same formed stellar mass. This effect depends on metallicity and increases the UV luminosity by $\approx2.7\,\%$ at $Z=0.014$.} of the simulation, which becomes apparent for short averaging timescales like $10\,\mathrm{Myr}$. For the CSEDs shown in Fig.~\ref{fig:CSEDvariation_SFregions} which include dust attenuation, the difference in the GALEX FUV band is $\approx20\,\%$. Hence, the `No resampling'-model exhibits less efficient dust attenuation compared to the fiducial SKIRT setup.

For our fiducial postprocessing, we use the high-resolution dust-free TODDLERS templates, which do not contain any dust attenuation or nebular continuum, but do contain nebular emission lines as predicted by \texttt{Cloudy}. We show the CSED using the TODDLERS templates with birth-cloud dust attenuation (`Dust-attenuated TODDLERS, purple line) in Fig.~\ref{fig:CSEDvariation_SFregions}. Employing this subgrid dust attenuation reduces the UV and PAH emission, and shifts the dust emission to higher temperatures. The stronger attenuation in the UV drives the COLIBRE CSED away from the observed data in this wavelength range, while the PAH emission features in the MIR are lowered (broadband fluxes are lower by $\approx0.1\,\mathrm{dex}$) and hence better reproduced in this model. 

We note that more extensive testing is required to assess whether a subgrid model for dust attenuation is necessary in COLIBRE. In the present implementation of this test, dust is effectively applied twice: once through the subgrid attenuation built into the TODDLERS SED templates, and again via the resolved COLIBRE dust associated with the star-forming gas particles. For the EAGLE postprocessing, this double-counting was avoided by subtracting the subgrid dust mass from the resolved dust distribution (\citealt{Camps2016}). However, this approach is less straightforward in the context of COLIBRE and TODDLERS, since COLIBRE predicts dust properties on a per-particle basis, the dust prescriptions in COLIBRE and TODDLERS differ substantially, and the total subgrid dust masses in a galaxy can exceed those predicted by COLIBRE (see also Section~\ref{sec:NebularEmission}).

We further note that the TODDLERS library contains free parameters, namely the cloud density and star-formation efficiency. In our fiducial setup, where subgrid dust attenuation is not included, these parameters only affect nebular emission lines, which have a negligible impact on the present study. However, when subgrid dust attenuation is enabled, the results become sensitive to their choice, as both UV and MIR broadband fluxes are significantly affected (see figure 4 in \citealt{Kapoor2024}). Consequently, more detailed testing with the TODDLERS library will be required to robustly assess the role of subgrid dust attenuation in COLIBRE-SKIRT postprocessing.

In our fiducial approach, the emission from young stellar populations is spatially extended, i.e. distributed according to the quartic spline smoothing kernel, with typical smoothing lengths of $\sim1\,\mathrm{kpc}$ at m6 resolution and $z=0$. As a result, the luminosity density is exactly co-spatial with the dust density, since the dust associated with each star-forming gas particle follows the same distribution. In contrast, when using dust-attenuated TODDLERS templates, both the stellar emission and the additional subgrid dust attenuation are confined to much smaller spatial scales.

An intermediate model between our fiducial approach (no subgrid dust attenuation) and the dust-attenuated TODDLERS templates is to concentrate the stellar emission while leaving the dust distribution unchanged. This can be implemented by adopting dust-free TODDLERS templates (as in the fiducial model), but assigning much smaller smoothing lengths to the star-forming gas particles when computing the primary emission. In Fig.~\ref{fig:CSEDvariation_SFregions}, we show the resulting CSED (`No smoothing’, orange line), where smoothing lengths of $0.001\,\mathrm{pc}$ are used for the star-forming gas particles (and for the parent gas particles of young star particles).

Similar to the `No resampling’ model, this leads to hotter dust temperatures and improved agreement with observational data in the MIPS 24 and PACS 100 bands. Given that, in the fiducial model, the luminosity density of young stellar populations is distributed over $\sim1\,\mathrm{kpc}$ - far larger than the observed sizes of young star clusters and associations ($\sim1-10\,\mathrm{pc}$; \citealt{PortegiesZwart2010}) - this approach offers a promising route to improve the physical realism of the post-processing framework and the agreement with the observed low-redshift CSED. However, such a treatment is inherently subgrid, operating below the spatial resolution of COLIBRE. We therefore defer a more detailed investigation of the smoothing and subgrid treatment of young stellar populations to future work. In particular, spatially resolved diagnostics (e.g. images) are likely better suited than integrated quantities such as the CSED to constrain the appropriate smoothing scales.

\section{Discussion}\label{sec:Discussion}

\subsection{Convergence with COLIBRE resolution}

\begin{figure*}
    \centering
    \includegraphics[width=\textwidth]{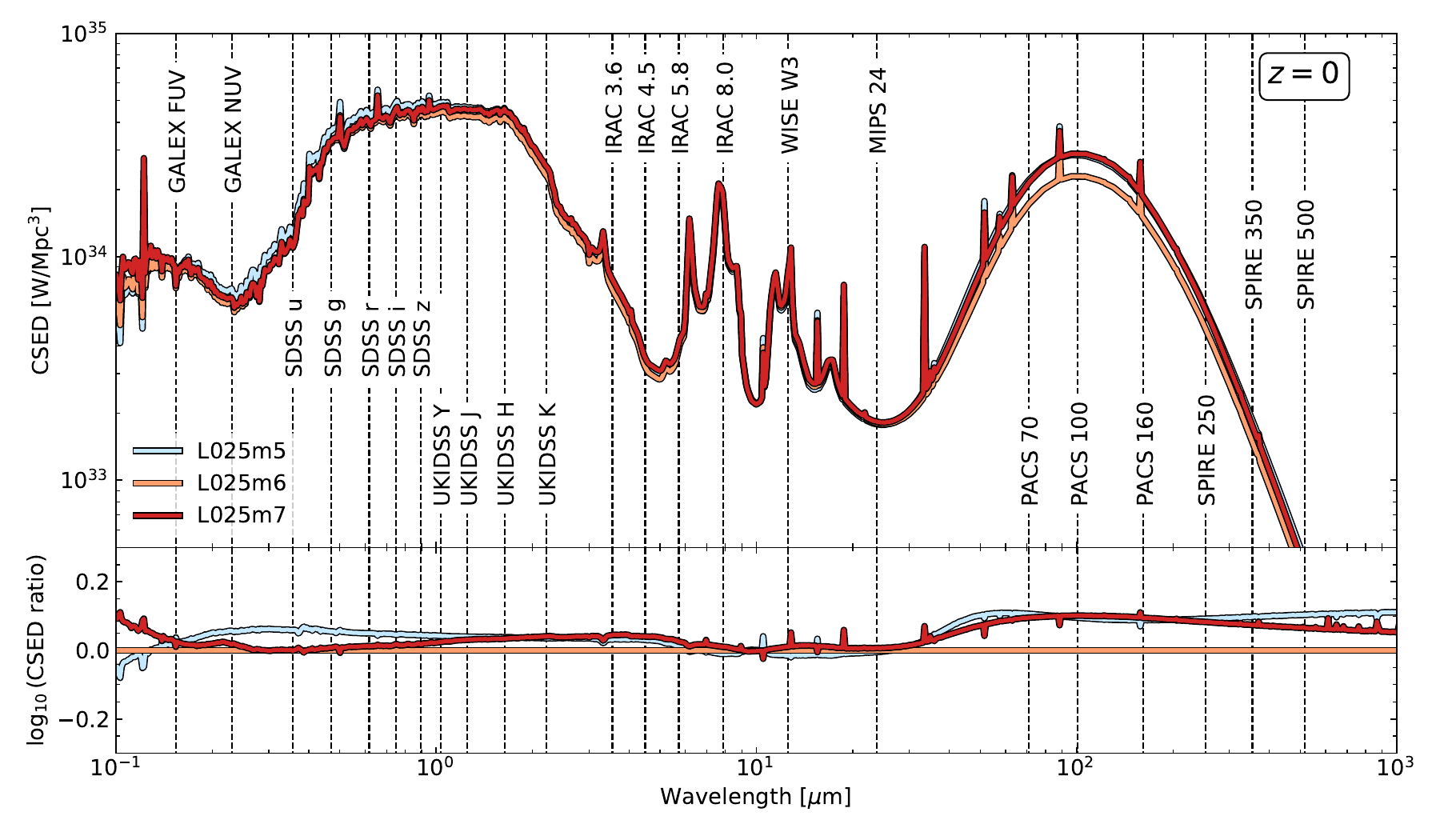}
    \caption{The CSED at $z=0$ for different COLIBRE resolution levels, using the L025m5, L025m6, and L025m7 simulations. We use our fiducial COLIBRE-SKIRT pipeline here, postprocessing all galaxies with $M_\star\geq10^7\,\mathrm{M}_\odot$ for this resolution test. The CSEDs are converged to within $\pm0.1\,\mathrm{dex}$, with the biggest differences arising in the FIR.}
    \label{fig:resolutionCSED}
\end{figure*}

\begin{table*}
    \centering
    \begin{tabular}{cccccc}
         Box & $\varepsilon_\star\,[\mathrm{M}_\odot\mathrm{Mpc}^{-3}]$ & $\varepsilon_\mathrm{dust}\,[\mathrm{M}_\odot\mathrm{Mpc}^{-3}]$ & $\varepsilon_\mathrm{SFR}\,[\mathrm{M}_\odot\mathrm{Mpc} ^{-3}\mathrm{yr}^{-1}]$ & $N_\mathrm{gal}\ (M_\star\geq10^7\,\mathrm{M}_\odot)$ & $N_\mathrm{gal}\ (M_\star\geq10^{8.5}\,\mathrm{M}_\odot)$ \\ \hline
       L025m5 & $2.85\times10^8$ & $3.70\times10^5$ & $1.35\times10^{-2}$ & 1502 & 456\\
       L025m6 & $2.76\times10^8$ & $2.69\times10^5$ & $1.16\times10^{-2}$ & 1633 & 382\\
       L025m7 & $3.03\times10^8$ & $2.66\times10^5$ & $1.28\times10^{-2}$ & 2077 & 362\\
    \end{tabular}
    \caption{Resolution comparison between the 25-Mpc COLIBRE boxes. We list the stellar mass density, dust mass density, and star-formation rate density at $z=0$ for the three COLIBRE simulations. These properties were computed by summing the corresponding quantities for all galaxies in our sample ($M_\star\geq10^7\,\mathrm{M}_\odot$) and dividing by the simulation volume. We also indicate the number of galaxies in these samples as well as the number of galaxies above a higher-mass threshold ($M_\star\geq10^{8.5}\,\mathrm{M}_\odot$).}
    \label{tab:resolution}
\end{table*}

All results shown previously were obtained at m6 resolution. Here, we discuss the convergence of our results with respect to the COLIBRE simulation resolution. To this end, we select all galaxies\footnote{\label{footnote:smoothing_lowmass}The low stellar mass threshold adopted for this sample implies that some galaxies contain very few star particles, down to a single particle at m7 resolution. In addition to the fact that three-dimensional dust radiative transfer is not meaningful for such poorly sampled systems, this also introduces a technical issue: the smoothing lengths of evolved stellar populations are defined as the distance to the 32nd nearest gravitationally bound star particle. For galaxies with fewer than 33 star particles, this definition formally yields an infinite smoothing length. To address this in the convergence tests presented in this section, we impose an upper limit on the stellar smoothing lengths of $\min(r_\star,10\,\mathrm{kpc})$, where $r_\star$ is the distance from the star particle to the galaxy centre. For gas particles (which are used both for the emission from star-forming gas and for the dust distribution), this issue does not arise, as their smoothing lengths are always finite. Nevertheless, at m6 and m7 resolution we find very large gas particle smoothing lengths (up to $\sim100\,\mathrm{kpc}$) for low-mass galaxies with $M_\star\lesssim10^8\,\mathrm{M}_\odot$.} with $M_\star\geq10^7\,\mathrm{M}_\odot$ from the 25-Mpc boxes at the three different resolution levels (m5, m6, and m7) at $z=0$. The redshift-zero CSED obtained with our fiducial COLIBRE-SKIRT pipeline is shown in Fig.~\ref{fig:resolutionCSED} for the three resolution levels. The CSED is well converged to within $\pm0.1\,\mathrm{dex}$, with the biggest differences in the FIR.

To explain the differences in the CSED, we list box-integrated physical quantities (stellar mass density, dust mass density, and instantaneous star-formation rate density) for the three simulations in Table~\ref{tab:resolution}. To compute these densities, we consider all galaxies with $M_\star\geq10^7\,\mathrm{M}_\odot$, but the numbers remain similar when using a higher stellar mass limit of $M_\star\geq10^{8.5}\,\mathrm{M}_\odot$. When comparing L025m5 and L025m6, we find that the stellar mass densities are well converged. On the other hand, the L025m5 simulation has significantly more dust ($\approx38\,\%$) and star formation ($\approx16\,\%$) compared to L025m6, explaining the relatively higher FIR flux for L025m5. The higher dust mass for L025m5 suggests that this model could use a smaller dust clumping factor; however, since this discrepancy arises primarily at $z<0.5$, it may also be driven by less effective feedback and correspondingly higher cold gas fractions (Vijayan et al. in prep.). We also note that at m7 resolution, the dust coagulation timescale was reduced by a factor of $10^{0.5}$ to improve convergence in the small-to-large grain size ratio (see Table 1 in \citealt{Schaye2026}).

L025m7 exhibits a similarly high FIR flux relative to L025m6, while the optical fluxes are well converged. The L025m7 and L025m6 simulations have comparable dust mass densities, but L025m7 has more stars ($\approx10\,\%$) and star formation ($\approx10\,\%$), Taken together, this suggests that dust attenuation (and hence dust heating) is slightly more effective at the lower m7 resolution. Potentially, this is due to the ISM being more smooth at lower resolution, leading to a higher covering fraction of the dust which promotes dust attenuation and heating.

\begin{figure*}
    \centering
    \includegraphics[width=\textwidth]{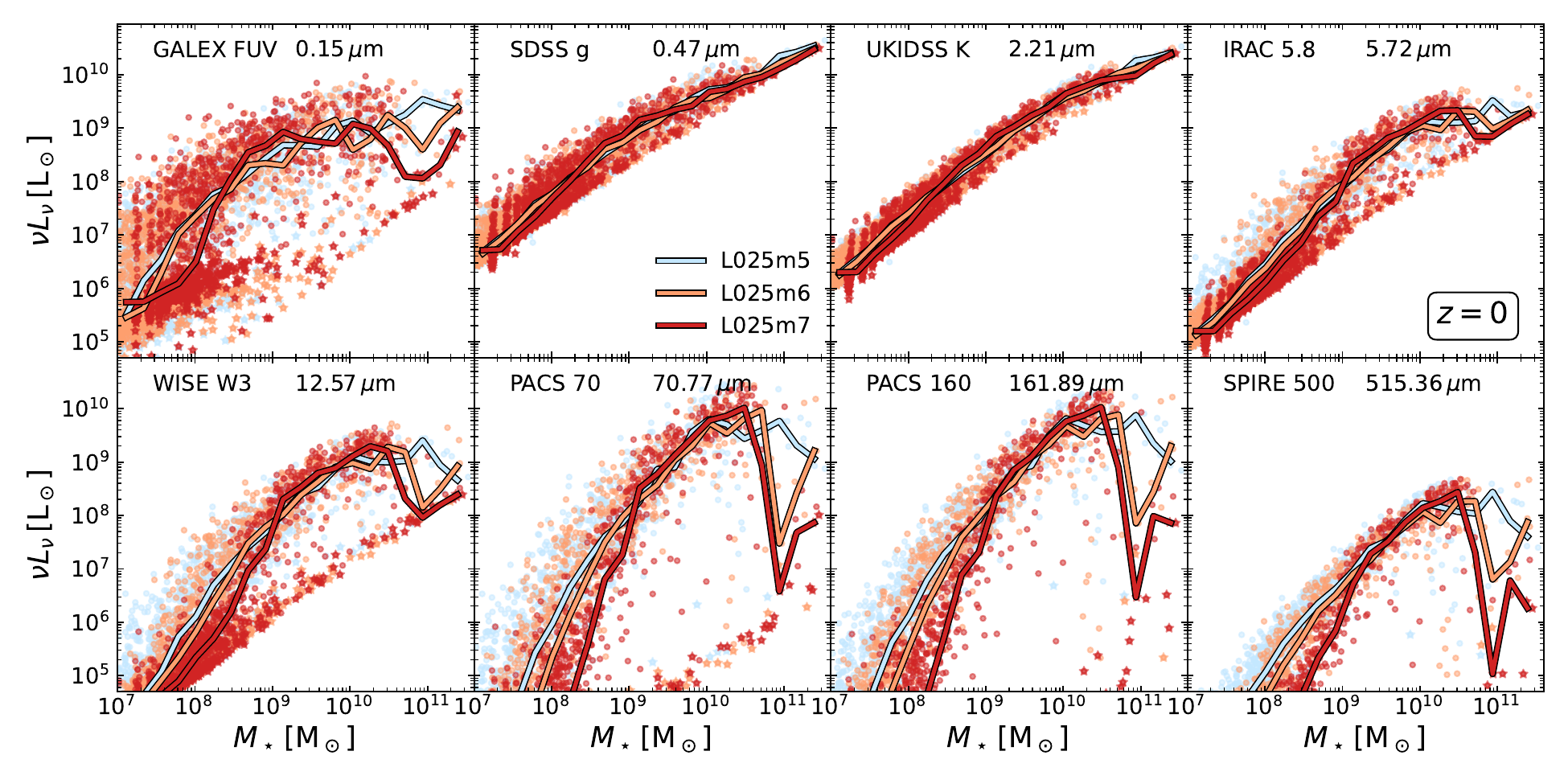}
    \caption{Broadband fluxes in various bands as a function of stellar mass, for the COLIBRE 25-Mpc simulations at $z=0$. We postprocess all galaxies with $M_\star\geq10^7\,\mathrm{M}_\odot$ for this resolution test. Lines indicate running medians in bins of stellar mass (with bin widths of $0.225\,\mathrm{dex}$. Galaxies with low specific dust masses ($M_\mathrm{dust}/M_\star\leq10^{-5}$) are indicated by star symbols. We find good convergence over the entire FUV-FIR wavelength range, down to stellar masses of $\sim10^9\,\mathrm{M}_\odot$ (for L025m7) and $\sim10^{8.5}\,\mathrm{M}_\odot$ (for L025m6).}
    \label{fig:resolutionMstar}
\end{figure*}

Since the CSED is a quantity integrated over the entire galaxy population at $z=0$, more subtle resolution effects could be washed out. We perform another resolution test by comparing the fluxes in various bands as a function of stellar mass in Fig.~\ref{fig:resolutionMstar}. In the optical-NIR range we find very good convergence, down to the lowest stellar masses ($M_\star\sim10^7\,\mathrm{M}_\odot$). In the UV and MIR-FIR wavelength ranges, fluxes are converged only for intermediate stellar masses ($\sim10^9-10^{10.5}\,\mathrm{M}_\odot$). At lower stellar masses, convergence is expected to worsen due to resolution effects and large smoothing lengths (see footnote~\ref{footnote:smoothing_lowmass}). At higher stellar masses, lower-resolution galaxies contain significantly less dust and are generally less star-forming (see Figures 18 and 22 in \citealt{Schaye2026}). We note that the galaxies with a roughly constant mass-to-light ratio that are visible in some IR bands (e.g. in the WISE W3 band) are due to galaxies that are almost or completely dust-free: We indicate galaxies with $M_\mathrm{dust}/M_\star\leq10^{-5}$ by the star markers, and find that those predominantly lie within the narrow features that are visible in those bands.

\subsection{Future improvements}\label{sec:Improvements}

We expect that the fiducial COLIBRE-SKIRT pipeline presented here will evolve over time, incorporating the most recent models for stellar emission and dust. Some features like nebular emission are limited in the current approach, while active galactic nuclei (AGN) are not included at all. Furthermore, we focus the tests and applications presented in this work on spatially integrated SEDs at low redshift. Different applications, particularly at higher redshifts, might warrant modifications to the current pipeline.

\subsubsection{Nebular emission}\label{sec:NebularEmission}

With the current COLIBRE-SKIRT pipeline, we focus on continuum emission from stars and dust. Nebular emission lines are included through the TODDLERS templates, but we do not record line luminosities specifically and we have not tested their realism. Our framework currently does not include the nebular continuum, as the TODDLERS templates only incorporate this component in the dust-attenuated version. This could be problematic particularly for high-redshift galaxies, where the UV and optical broadband fluxes can be impacted by nebular emission lines (e.g. \citealt{Schaerer2009}; \citealt{Smit2015}). Morevoer, the nebular continuum can dominate the UV flux (e.g. \citealt{Cameron2024}; \citealt{Katz2025}). The nebular continuum will be included in future versions of the unattenuated TODDLERS templates (Kapoor et al. in prep.).

For the fiducial COLIBRE-SKIRT pipeline, we refrained from using the dust-attenuated TODDLERS templates because the low-redshift UV CSED is already well reproduced without invoking subgrid dust attenuation (Fig.~\ref{fig:fiducialCSED}). Furthermore, the CLOUDY calculations that are part of TODDLERS assume a dust-to-metal ratio of $\approx40-50\,\%$. This leads to the problem that for high-redshift galaxies, which have high star-formation rates but low dust masses, the subgrid dust contained in TODDLERS exceeds the dust mass predicted by COLIBRE. To overcome this problem within the current framework, a new set of TODDLERS SED templates is necessary which either sets the dust-to-metal ratio in CLOUDY to zero or treats it as a free parameter, which can be imported from the COLIBRE data (Kapoor et al. in prep.).

Other approaches to model nebular emission from HII regions in galaxy simulations are widespread (e.g. \citealt{Olsen2021}; \citealp{RamosPadilla2021,RamosPadilla2023}; \citealt{Garg2022}; \citealt{Hirschmann2023a,Hirschmann2023b}). Nebular emission predictions from large-volume simulations suffer from the fact that HII regions are unresolved, meaning that the gas densities are not high enough compared to typical HII regions with electron densities $n_e\sim100\,\mathrm{cm}^{-3}$. Furthermore, the ionizing radiation field is not directly known from the simulations and has to be estimated. The freedom in these choices leads to large systematic uncertainties in the predicted line luminosities (e.g. \citealt{Katz2023}). While some of these issues apply for the COLIBRE simulations, the inclusion of a cold ISM phase with more realistic temperatures, the sophisticated chemical network (with non-equilibrium electron, H and He abundances), and the presence of a local interstellar radiation field (scaled with the local star formation rate surface density) hold promise to at least reduce some of the ambiguities in HII region modelling.

Our approach is particularly well-suited to self-consistently include the effects of dust on nebular emission, as the photons emitted from HII regions will be attenuated by the dusty ISM of the host galaxy. A holistic model of nebular emission will also require inclusion of the diffuse ionized gas (DIG) outside of HII regions. In the local Universe, the DIG can contribute $10-50\,\%$ of the total H$\alpha$ flux and dominate the emission of low-ionization lines like [\textsc{S\,ii}] and [\textsc{N\,ii}] (e.g. \citealt{Kewley2019}). Incorporating the DIG will necessitate larger adjustments to the current workflow, as the ionizing radiation from HII regions needs to be propagated through the host galaxy in an iterative calculation since the gas opacity depends on its ionization state. We are currently implementing this workflow in the SKIRT code (Kapoor et al. in prep.; Kapoor et al. in prep.). We note that this only accounts for photoionized gas, while it has been observed that for some galaxies a significant fraction of DIG emission originates in shock-excited gas (e.g. \citealt{Martin1997}; \citealt{Ramirez-Ballinas2014}).

\subsubsection{Active Galactic Nuclei}\label{sec:AGN}

We completely omit emission from active galactic nuclei (AGN) in the current pipeline. Incorporating AGN emission into cosmological simulations suffers from similar limitations as nebular emission: Since the AGN disk and surrounding medium are unresolved, the emission from AGN is treated through subgrid models, which comes with large systematic uncertainties. This is less of a concern for the emission from the AGN disk and the corona, which can be estimated from specific templates such as QSOSED (\citealt{Kubota2018}) or RELAGN (\citealt{Hagen2023}). Such templates predicts the emission spectrum (from X-rays to the optical) based on black hole mass, Eddington ratio, black hole spin, and inclination angle for the warm Comptonzing component.

While these templates are sufficient to estimate the unobscured emission from type-I AGN, estimating the reprocessed emission from type-I or type-II (obscured) AGN accounting for the dusty torus as well as the narrow-line region (NLR) and the broad-line region (BLR) is more challenging. Such templates would depend on the viewing angle in accordance with the AGN unification scheme (e.g. \citealt{Antonucci1993}; \citealt{Urry1995}). Observational AGN templates have the drawback that they often apply to unobscured AGN as those are easier to detect, covering emission from the AGN disk and the corona (e.g. \citealt{Richards2006}; \citealt{Jin2012}; \citealt{Gupta2024}). Some observational studies also include obscured AGN with reprocessed continuum emission (e.g. \citealt{Elvis1994}; \citealt{Assef2010}) and emission lines (e.g. \citealt{Brown2019}). To create anisotropic AGN SED templates, multiple of these observational SEDs would need to be combined and linked to physical parameters such as black hole mass and accretion rate.

An alternative route would be to use the output from radiative transfer simulations that model the reprocessed AGN emission from the dusty torus (e.g. \citealt{Fritz2006}; \citealt{Nenkova2008a,Nenkova2008b}; \citealt{Stalevski2012}; \citealt{Siebenmorgen2015}; \citealt{Hoenig2017}). These templates depend on the size and internal structure of the dust torus, and have the advantage of being inherently anisotropic. Modelling the broad- and narrow-line regions (BLR/NLR) requires specifying gas densities, column densities (i.e. whether the gas is ionisation- or density-bounded), and the degree of clumpiness. For the NLR, the \citet{Feltre2016} templates could be used, although analogous treatments of the BLR remain more limited.

More physically motivated models of these regions are increasingly informed by general-relativistic radiation-magnetohydrodynamical simulations of accretion flows and outflows (e.g. \citealt{Narayan2012}; \citealt{Liska2022}). Radiative transfer frameworks such as SIROCCO (\citealt{Matthews2025}) have begun to model the emergent radiation from these regions, but at significant computational expense. The fundamental limitation of all these approaches is that they describe predominantly sub-parsec scale processes that are not resolved in cosmological simulations. As a result, these simulation-based templates involve a large number of parameters that are not directly constrained by cosmological simulations.

\subsubsection{Dust models}

A persistent challenge for dust radiative transfer is that dust models are calibrated against observations of the Milky Way. As a result, the properties of interstellar dust grains in other galaxies remain poorly constrained, which becomes particularly problematic at high redshift. While there is growing observational evidence that dust grain properties (i.e. size distributions and compositions) evolve with redshift (e.g. \citealt{Gallerani2010}; \citealt{Markov2025}; \citealt{Shivaei2025}), the dust models in radiative transfer postprocessing campaigns are normally held fixed even when postprocessing large redshift ranges (e.g. \citealt{Camps2018}; \citealt{Vogelsberger2020b}). Since COLIBRE tracks the evolution of dust properties explicitly, and we incorporate this information into our dust model through the `split \& scale' approach, this is less of a concern here. Nevertheless, our dust optical properties are still based on dust models that are tuned to reproduce Milky Way observations.

Even though dust models are based on Milky Way observations, it is still useful to compare different models - particularly dust emissivities can differ significantly (Fig.~\ref{fig:CSEDvariation_opticalProperties}). SKIRT does not support the recent THEMIS v2.0 (\citealt{Ysard2024}) and \citet{Hensley2023} dust models, which are the `descendants' of the two dust models that we consider for this study. The results from \citet{Hensley2023} challenge the notion that there are separate silicate and carbonaceous dust grains. Instead, \citet{Hensley2023} propose a single-composition large grain (`Astrodust') and small carbonaceous grains (PAHs) in their dust model. Such a dust model would require a new approach for dust radiative transfer postprocessing, as the dust model cannot be split into the different silicate and carbonaceous components predicted by COLIBRE.

\subsubsection{Stellar templates}

For the evolved stellar populations, we adopt the BPASS v2.2.1 SED templates (\citealt{Eldridge2017}; \citealt{Stanway2018}) to be consistent with the COLIBRE early feedback model (\citealt{Benitez-Llambay2026}). The construction of SED templates is complex and requires assumptions about stellar evolution (isochrones), stellar rotation, the initial mass function, stellar atmospheres, and multiplicity (i.e. binary fractions). When fixing the initial mass function, differences between SED template libraries are relatively small, as shown in Fig.~\ref{fig:CSEDvariation_SEDtemplates}. Differences in the optical-NIR colours of the SED templates are predominantly due to uncertainties in stellar evolution, particularly from thermal pulses during the asymptotic giant branch phase (e.g. \citealt{Maraston2005}; \citealt{Maraston2006}; \citealt{Conroy2009}; \citealt{Conroy2013}).

Consistent with the COLIBRE simulation, we adopt a fixed \citet{Chabrier2003} initial mass function, spanning $0.1-100\,\mathrm{M}_\odot$. To what degree this is a good assumption is controversial, particularly at high redshift where a top-heavy IMF can explain the abundance of sub-mm galaxies at $z\approx1-4$ (e.g. \citealt{Baugh2005}; \citealt{Lacey2016}) and UV-bright galaxies in the early Universe discovered with JWST (e.g. \citealt{Inayoshi2022}; \citealt{Haslbauer2022}; \citealt{Cameron2024}; \citealt{Yung2024}; \citealt{Trinca2024}; \citealt{Lu2025}; \citealt{Hutter2025}; \citealt{Mauerhofer2025}). Self-consistent cosmological, hydrodynamical simulations with a variable IMF are scarce, to our knowledge \citet{Barber2018} is the only such simulation where the EAGLE model was adjusted to vary the IMF based on local gas pressure. The COLIBRE suite also includes a version with a variable IMF, which will be presented and analyzed in Durrant et al. (in prep.). To postprocess such simulations, SED templates with variable IMFs are required. While BPASS offers a small number of different IMFs, only FSPS (\citealt{Conroy2009}; \citealt{Conroy2010}) and pySTARBURST99 (\citealt{Hawcroft2025}) offer the required flexibility to parametrize the IMF. In the simplest parametrization, a variable IMF boils down to an extra parameter (the IMF slope) for the SED templates in addition to metallicity and age. We have already implemented such a template library using FSPS in SKIRT, and will present results using this variable IMF library in Durrant et al. (in prep.).

Another point of concern regarding stellar templates are individual element abundances. Almost all SED template libraries offer just a single elemental abundance pattern. The first step beyond this is to incorporate $\alpha$-enhancement, which is the ratio of $\alpha$-elements (C, O, Ne, Mg, Si, S, Ar, and Ca) to iron. Since $\alpha$-elements and iron-peak elements (Fe, Cr, Co, Ni, Cu, Mn) originate from different stellar production channels and hence enrich the ISM on different timescales (e.g. \citealt{Tinsley1979}; \citealt{Greggio1983}), the $\alpha$-enhancement is systematically correlated with the formation history of the galaxy (e.g. \citealp{Thomas2005,Thomas2010}; \citealt{delaRosa2011}; \citealt{Segers2016}; \citealt{Matthee2018}; \citealt{Gebek2022}). Due to the rapid enrichment with $\alpha$-elements (from type-II supernovae) and the delayed enrichment with iron-peak elements (from type-Ia supernovae), stellar populations at high redshift could be particularly $\alpha$-enhanced. First steps towards building SED templates with $\alpha$-enhancement have been taken in \citet{Vazdekis2015}, \citet{Knowles2021,Knowles2023}, and \citet{Byrne2025}. However, all these template libraries incorporate $\alpha$-enhancement at the level of the stellar atmospheres only, neglecting the impact of $\alpha$-enhancement on stellar evolution. Nevertheless, these SED template libraries could be directly used in the postprocessing pipeline since COLIBRE tracks eleven different elements (including all of the $\alpha$-elements except Ar), which would be a first step towards accounting for individual element abundances.

\subsubsection{The pipeline at higher redshifts}

We systematically perform all of our pipeline tests only at $z=0$, and present results for the CSED combining the $z=0$ and $z=0.1$ snapshots. While the pipeline can be applied to any COLIBRE snapshot without modifications, some cautionary remarks are in order. We adopt a fixed SKIRT box size of $100\,\mathrm{kpc}$, which is suitable to encompass the ISM of low-redshift galaxies, but excessively large at high redshift. While it is challenging to define SKIRT box sizes that are suitable for all galaxy types and redshifts, a simple option would be to scale the SKIRT box sizes the same way as the gravitational softening lengths in COLIBRE. COLIBRE adopts a fixed comoving softening length ($1.8\,\mathrm{ckpc}$ at m6 resolution), clipped at a maximum physical softening length ($0.7\,\mathrm{kpc}$ at m6 resolution). The switch from comoving (at high redshift) to physical (at low redshift) softening length occurs at $z=1.57$. Since COLIBRE galaxies at higher redshifts have smaller softening (and smoothing) lengths, it makes sense to also increase the resolution of the SKIRT dust grid. This could be achieved by scaling the SKIRT box size:

\begin{equation}
    L_\mathrm{box}=100\,\mathrm{kpc}\,\times\,\min(1,\frac{1+1.57}{1+z})=\min(100\,\mathrm{kpc}, 257.14\,\mathrm{ckpc}).
\end{equation}
If we scale the SKIRT box size in this manner, then the SKIRT dust grid cell sizes correspond to a fixed fraction of the COLIBRE softening lengths, independent of redshift. Alternative choices for the SKIRT box sizes (e.g. scaling with galaxy size) are also viable, with the most appropriate option depending on the intended application.

Since the ISM of high-redshift galaxies is denser, dust self-absorption can become important. \citet{Camps2018} report that infrared fluxes of high-redshift galaxies can be underestimated by up to a factor of 2.5 when neglecting dust self-absorption. SKIRT can model dust self-absorption, which necessitates iterative calculations of the dust temperature and dust emission spectrum. Due to the increased computational demand, we have not included dust self-absorption in our fiducial setup. Dust grains can also be heated by the cosmic microwave background (CMB). Due to the low CMB temperature at $z=0$ we ignored this effect, but note that it could become relevant at higher redshifts (\citealt{daCunha2013}).

\section{Summary}\label{sec:Summary}

We established a calibration-free dust radiative transfer postprocessing pipeline with the SKIRT code for the COLIBRE large-volume cosmological simulations. In the context of dust radiative transfer, the most relevant novelties of COLIBRE are its live dust model, which predicts the amount of dust in two size bins and three representative chemical compositions at the level of individual gas particles, and the direct simulation of the multiphase ISM, including the cold, molecular phase. We computed spatially integrated SEDs from the FUV to the FIR ($0.09-2000\,\mu\mathrm{m}$) for a large sample of galaxies at low redshift, and tested the impact of the choices made in the postprocessing. Our results can be summarized as follows:

\begin{itemize}
    \item To achieve an optimal balance between SKIRT simulation speed and accuracy, we scale the grid refinement criterion $\delta_\mathrm{max}$ (which controls the resolution of the adaptive dust grid) with the dust surface density of the galaxy (Eqns.~\ref{eq:refinement_m6} and~\ref{eq:refinement_m5}). With this setup, postprocessing all resolved galaxies at a single snapshot from the L100m6 simulation takes $\sim100\,000$ CPU hours (Fig.~\ref{fig:CPUhours_redshift}).
    \item We compute the cosmic SED (CSED) from the COLIBRE L100m6 simulation at $z=0$ and $z=0.1$ using all $47\,219$ galaxies with $M_\star\geq10^{8.5}\,\mathrm{M}_\odot$ (Fig.~\ref{fig:fiducialCSED}). We find excellent agreement with observational data from GAMA (\citealt{Andrews2017}) and HerMES (\citealt{Marchetti2016}) over most of the wavelength range, with some residual tension for $5\lesssim\lambda\lesssim200\,\mu\mathrm{m}$ (up to $\approx0.23\,\mathrm{dex}$ in the IRAC 8.0 band). We emphasize that our postprocessing pipeline has not been calibrated \textit{a priori} to reproduce any observational data. Nevertheless, the COLIBRE CSED matches the observational data as well as, if not better than, previous SKIRT postprocessing campaigns for EAGLE (\citealt{Camps2016}) and IllustrisTNG (\citealt{Trcka2022}), which were calibrated to observations.
    \item To compute dust emission, the two discrete grain sizes tracked by COLIBRE ($0.01\,\mu\mathrm{m}$ and $0.1\,\mu\mathrm{m}$) need to be expanded into continuous distributions. We propose a novel `split \& scale' approach that modifies a given dust model (specifying the size distributions and optical properties for all components of the dust model) by splitting it at $a=0.03\,\mu\mathrm{m}$ and scaling the different components to match the COLIBRE dust properties (Fig.~\ref{fig:dustModels_sizeDistributions}). This adjustment is done at the gas particle level, meaning that every COLIBRE gas particle has its own dust extinction curve and emissivity.
    \item We compare two different dust models (\citealt{Draine2007} and THEMIS, \citealt{Jones2017}) in terms of their CSEDs (Fig.~\ref{fig:CSEDvariation_opticalProperties}) and dust optical properties (Fig.~\ref{fig:dustModels_opticalProperties}). Since the DL07 dust model exhibits dust species fractions that align more closely with the low-redshift dust in COLIBRE (see Table~\ref{tab:sizeDistributions}), we find this dust model more suitable for postprocessing COLIBRE galaxies. When modifying the THEMIS dust model to account for the COLIBRE dust properties, dust attenuation and dust emission in the FIR is overly effective, leading to CSEDs in tension with the observational constraints (Fig.~\ref{fig:CSEDvariation_THEMIS}).
    \item To improve the sampling of the youngest stellar populations, we use the star-forming gas particles as sources, effectively averaging the recent star-formation history over the past $10\,\mathrm{Myr}$. For the first time, we do not invoke subgrid dust attenuation for star-forming regions in a large-volume cosmological simulation, as the resolved ISM dust in COLIBRE sufficiently attenuates the UV (Fig.~\ref{fig:CSEDvariation_SFregions}). We attribute this to the explicit modelling of the cold ISM in COLIBRE, leading to a clumpier dust distribution and higher optical depths around the sites of star formation compared to simulations without a cold ISM phase.
    \item We find that broadband fluxes are well converged with COLIBRE resolution, to $\approx0.1\,\mathrm{dex}$ for the CSED (Fig.~\ref{fig:resolutionCSED}). In the UV-NIR (MIR-FIR) wavelength range, we find that fluxes as a function of stellar mass converge for galaxies with $\gtrsim10$ ($\gtrsim100$) star particles (Fig.~\ref{fig:resolutionMstar}).

\end{itemize}

This pipeline can be readily applied to the COLIBRE universe to calculate various synthetic observables like broadband fluxes, high-resolution spectra, galaxy images, and IFU spectra. In upcoming work we plan to analyze luminosity functions at various redshifts (Lu et al. in prep.), the redshift evolution of dust attenuation curves (Andreadis et al. in prep.), color-magnitude relations at low redshift (Baes et al. in prep.), and the impact of the live dust model on galaxy luminosities and colours (Gebek et al. in prep.). The framework presented here is the first version of our COLIBRE-SKIRT pipeline, developed for and tested at low redshift. We plan to improve several aspects of the pipeline that are mostly relevant at high redshift, such as the inclusion of nebular continuum and diffuse ionized gas emission as well as active galactic nuclei.

\begin{acknowledgements}

We are thankful for the helpful discussions with many colleagues that improved this work. In particular, we thank Matthieu Willems for insightful discussions on dust models, Anthony Jones and Nathalie Ysard on the THEMIS dust framework, Jan Eldridge and Elizabeth Stanway on the BPASS SED templates, Rob J. McGibbon on COLIBRE data handling, and Simon P. Driver on the GAMA observational data.

\\

AG gratefully acknowledges financial support from the Fund
for Scientific Research Flanders (FWO-Vlaanderen, project FWO.3F0.2021.0030.01). NA acknowledges financial support by the Flemish Fund for Scientific Research (FWO-Vlaanderen) through the research grant G0C4723N. AUK acknowledges support from the Belgian Federal Science Policy Office (BELSPO) via the PRODEX Programme of the European Space Agency (ESA) under contract number 4000143202. CB gratefully acknowledges support from the Forrest Research Foundation. CGL acknowledges support from STFC consolidated grants ST/T000244/1 and ST/X001075/1. EC acknowledges support from STFC consolidated grant ST/X001075/1. AD is supported by an STFC doctoral studentship. FH acknowledges funding from the Netherlands Organization for Scientific Research (NWO) through research programme Athena 184.034.002. SP acknowledges support by the Austrian Science Fund (FWF) through grant-DOI: 10.55776/V982. ABL acknowledges support by the Italian Ministry for Universities (MUR) program `Dipartimenti di Eccellenza 2023-2027' within the Centro Bicocca di Cosmologia Quantitativa (BiCoQ), and support by UNIMIB's Fondo Di Ateneo Quota Competitiva (project 2024-ATEQC-0050).

\\

This work used the DiRAC@Durham facility managed by the Institute for Computational Cosmology on behalf of the STFC DiRAC HPC Facility (\url{www.dirac.ac.uk}). The equipment was funded by BEIS capital funding via STFC capital grants ST/K00042X/1, ST/P002293/1, ST/R002371/1 and ST/S002502/1, Durham University and STFC operations grant ST/R000832/1. DiRAC is part of the National e-Infrastructure.

\\

The resources and services used in this work were provided by the VSC (Flemish Supercomputer Center), funded by the Research Foundation - Flanders (FWO) and the Flemish Government.

\\

This work made use of v2.2.1 of the Binary Population and Spectral Synthesis (BPASS) models as described in \citet{Eldridge2017} and \citet{Stanway2018}.

\end{acknowledgements}

\section*{Software}

SKIRT is an open-source code, publicly available at \url{https://skirt.ugent.be}. We made extensive use of the \texttt{python} programming language, particularly the following software packages: \texttt{numpy} (\citealt{Harris2020}), \texttt{scipy} (\citealt{Virtanen2020}), \texttt{astropy} (\citealp{astropy2013,astropy2018,astropy2022}), \texttt{matplotlib} (\citealt{Hunter2007}), \texttt{swiftsimio} (\citealt{Borrow2020}), and \texttt{swiftgalaxy} (\citealt{Oman2025}). Parts of the results in this work make use of the colormaps in the \texttt{CMasher} package (\citealt{vanDerVelden2020}).

\section*{Data availability}

The observational data used in this paper (\citealt{Marchetti2016}; \citealt{Andrews2017}) are publicly available. SKIRT fluxes for the EAGLE simulation are available via the EAGLE database (\url{https://icc.dur.ac.uk/Eagle/database.php}), and those for IllustrisTNG can be accessed through the IllustrisTNG website (\url{https://www.tng-project.org/data/docs/specifications/#sec5w}).

The SKIRT fluxes for COLIBRE, computed in this work, will be made publicly available following the public release of the COLIBRE simulation suite. In the meantime, researchers wishing to access the SKIRT data or the COLIBRE simulations are invited to contact the COLIBRE team. Further information on the COLIBRE project can be found at \url{https://colibre.strw.leidenuniv.nl/}.

The scripts used to run the COLIBRE-SKIRT pipeline are publicly available at \url{https://github.com/andreagebek/colibre-skirt}. The analysis scripts required to reproduce the results and figures presented in this work will be made publicly available upon publication of the manuscript.

\bibliographystyle{aa}
\bibliography{main.bib}

\begin{appendix}

\section{Convergence with box size and subsampling}\label{sec:subsampling}

\begin{figure*}
    \centering
    \includegraphics[width=\textwidth]{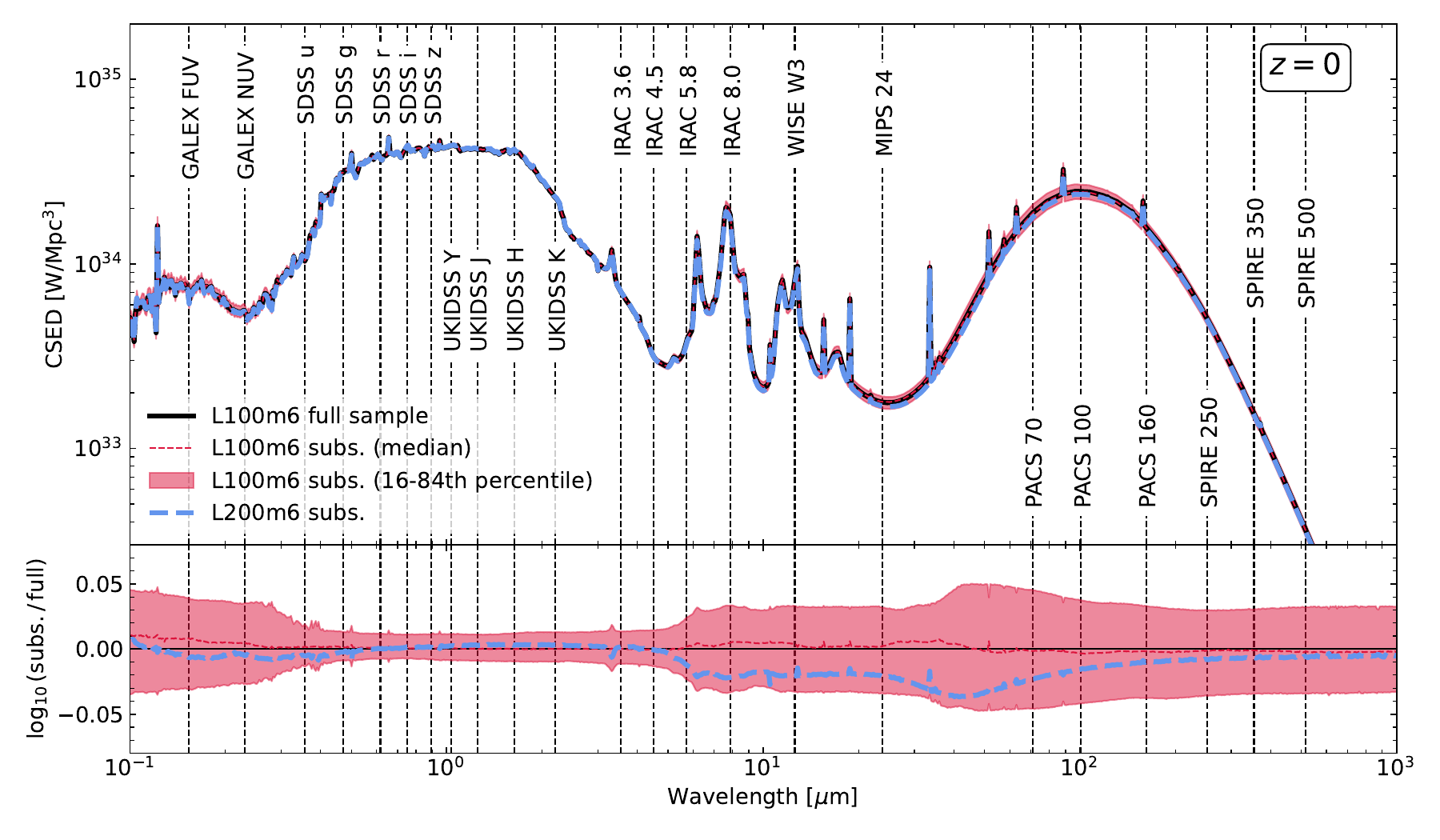}
    \caption{The CSED of COLIBRE at $z=0$, testing the convergence with subsampling and simulation box size. All results correspond to our fiducial SKIRT setup. The solid black line is the reference L100m6 CSED (`L100m6 full sample'), calculated using all 23\,490 galaxies with $M_\star\geq10^{8.5}\,\mathrm{M}_\odot$. We also compute the CSED by subsampling the L100m6 box, using 300 galaxies with a flat distribution in $\log_{10}(M_\star)$ (see text for more details). We compute this subsampled L100m6 CSED 100 times and show the median with the red dashed line and the 16th-84th percentile range with the red shaded area. The orange dashed line shows the L200m6 CSED, computed using the same subsampling strategy with 300 galaxies.}
    \label{fig:subsampling}
\end{figure*}

In Fig.~\ref{fig:subsampling}, we test the convergence of our CSED with respect to COLIBRE box size and subsampling. All of the SEDs are computed with our fiducial SKIRT setup. The reference CSED (black line in Fig.~\ref{fig:subsampling}) uses the 23\,490 galaxies with $M_\star\geq10^{8.5}\,\mathrm{M}_\odot$ from the L100m6 box at $z=0$. Our subsampling strategy for the L200m6 box uses 300 galaxies with a flat distribution in $\log_{10}(M_\star)$ (six mass bins with width $0.5\,\mathrm{dex}$). To test this subsampling strategy, we recalculate the L100m6 CSED 100 times by subsampling it with 300 galaxies. The median of these 100 subsampled L100m6 CSEDs is shown by the dashed red line in Fig.~\ref{fig:subsampling}, and the 16th-84th percentile range is indicated by the red shaded area. We find that the median subsampled CSED agrees very well with the full-box CSED, and the scatter is relatively small with a maximum of $\approx0.09\,\mathrm{dex}$ in the FIR.

To check the impact of the box size, we consider the L200m6 CSED (blue dashed line in Fig.~\ref{fig:subsampling}). This CSED is also calculated with our subsampling strategy using 300 galaxies, as it would be computationally expensive ($\sim1\,000\,000$ CPU hours) to postprocess all galaxies from the L200m6 box. The L200m6 and L100m6 CSEDs are consistent within the spread due to the subsampling, demonstrating convergence with simulation box size.

\section{Other SKIRT postprocessing variations}\label{sec:PipelineVariationsAppendix}

We presented some SKIRT postprocessing variations (different dust models and treatments of young stellar populations) in Section~\ref{sec:PostprocessingChoices}. Here, we show additional variations and discuss the convergence tests that we performed for the numerical SKIRT parameters. All of the SKIRT runs discussed here are performed for the subsampled L200m6 box at $z=0$.

\subsection{Grain size distributions with THEMIS optical properties}

\begin{figure*}
    \centering
    \includegraphics[width=\textwidth]{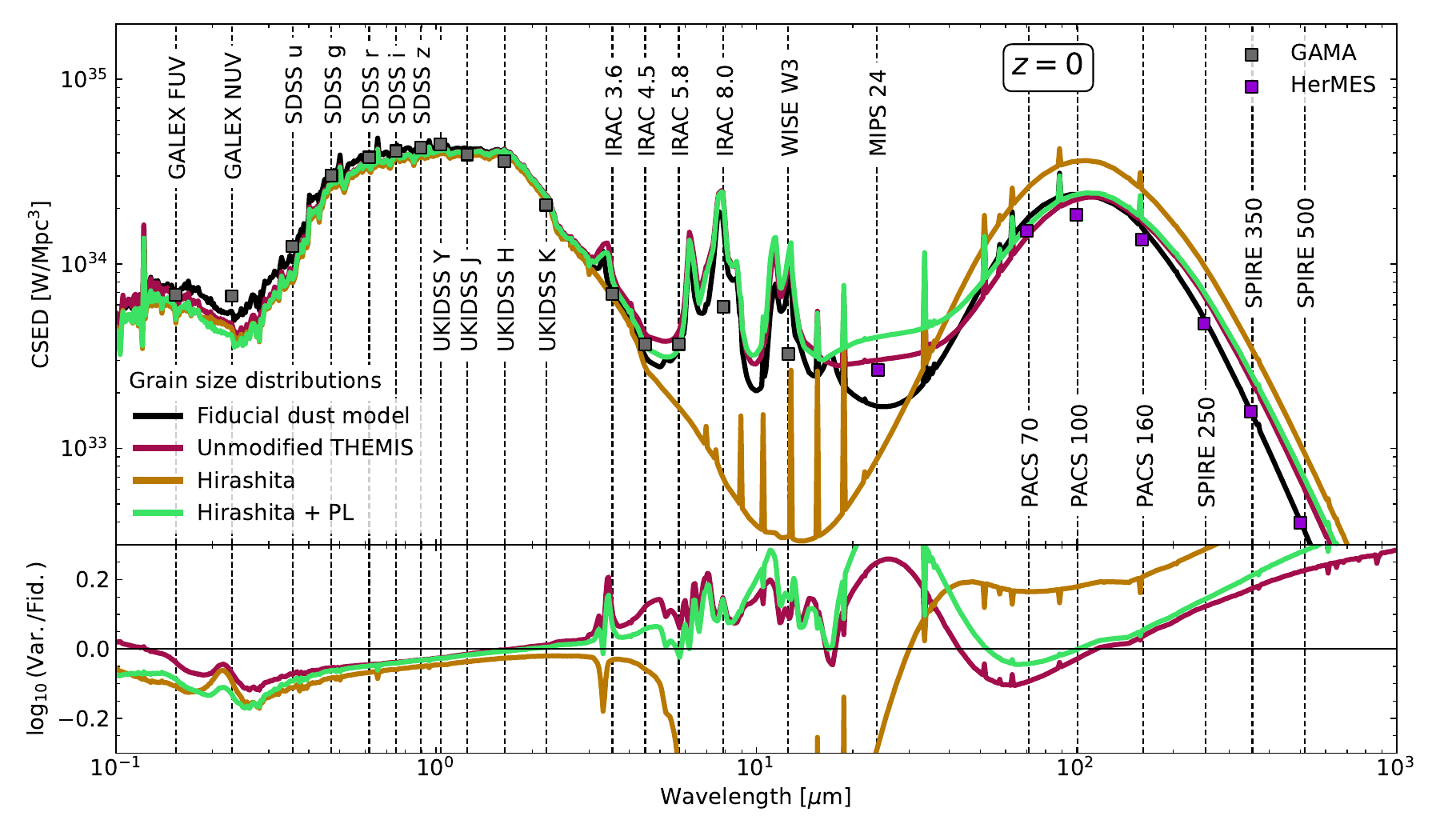}
    \caption{The COLIBRE L200m6 redshift-zero CSED, varying the treatment of the grain size distributions. The fiducial model is shown in black, where we use optical properties from DL07 and the `split \& scale' approach. All other models use optical properties from THEMIS. For the `Unmodified THEMIS' CSED (red line), we did not take the dust species fractions predicted by COLIBRE into account, and instead used the original THEMIS grain size distributions for all gas particles. The `Hirashita' CSED (orange line) employs log-normal size distributions similar to \citet{Hirashita2015}. For the `Hirashita + PL' model (green line) we distribute all small grains in COLIBRE according to the THEMIS power-law for small hydrocarbons, while the large dust grains follow log-normal size distributions. The observational data is shown by the grey (GAMA) and purple (HerMES) markers.}
    \label{fig:CSEDvariation_THEMIS}
\end{figure*}

We already showed variations in the grain size distributions when using the optical properties from DL07 in Fig.~\ref{fig:CSEDvariation_sizeDistribution}, and compared the DL07 and THEMIS optical properties when using the same `split \& scale' approach for the grain size distributions in Fig.~\ref{fig:CSEDvariation_opticalProperties}. In Fig.~\ref{fig:CSEDvariation_THEMIS}, we test different grain size distributions when using the THEMIS optical properties. We compare those CSEDs to our fiducial one, obtained with the DL07 optical properties and the `split \& scale' method for the grain size distribution.

In the `Unmodified THEMIS' model, we ignore the dust species fractions predicted by COLIBRE (but we do use the predicted total dust masses) and use the same original THEMIS dust model for all gas particles. For the `Hirashita' model, we use log-normal grain size distributions centered on the COLIBRE grain sizes of $0.01\,\mu\mathrm{m}$ and $0.1\,\mu\mathrm{m}$. Similar to Fig.~\ref{fig:CSEDvariation_sizeDistribution}, this size distribution lacks very small carbonaceous grains and hence does not produce the MIR PAH emission features. To resolve that, we assign all small grains to the THEMIS power-law size distribution in the `Hirashita + PL' model (the large grains are still described by a log-normal distribution).

In general, THEMIS leads to more attenuation and increased FIR flux compared to our fiducial dust model, which is based on DL07. The agreement with the GAMA data substantially worsens in the NUV and FIR SPIRE bands when using THEMIS, particularly for the `Hirashita' model. We find that the `Unmodified THEMIS' and `Hirashita + PL' model yield similar CSEDs. We note that the silicate grains in THEMIS effectively become carbonaceous when they are very small because they exhibit a carbonaceous mantle of thickness $5\,\mathrm{nm}$. Constraints on anomalous microwave emission precludes a significant fraction of bare silicate grains in the ISM (\citealt{Ysard2022}). The `Hirashita + PL' model effectively means that all small COLIBRE grains are treated as carbonaceous, according to the `Hydrocarbons 2' component in Fig.~\ref{fig:dustModels_sizeDistributions}. We have also experimented with a model where we only put the small carbonaceous grains in COLIBRE into the power-law component and model the small silicate grains with a log-normal size distribution. This leads to a CSED in between the `Hirashita' and `Hirashita + PL' models shown in Fig.~\ref{fig:CSEDvariation_THEMIS}.

\subsection{SED templates for evolved stellar populations}

\begin{figure*}
    \centering
    \includegraphics[width=\textwidth]{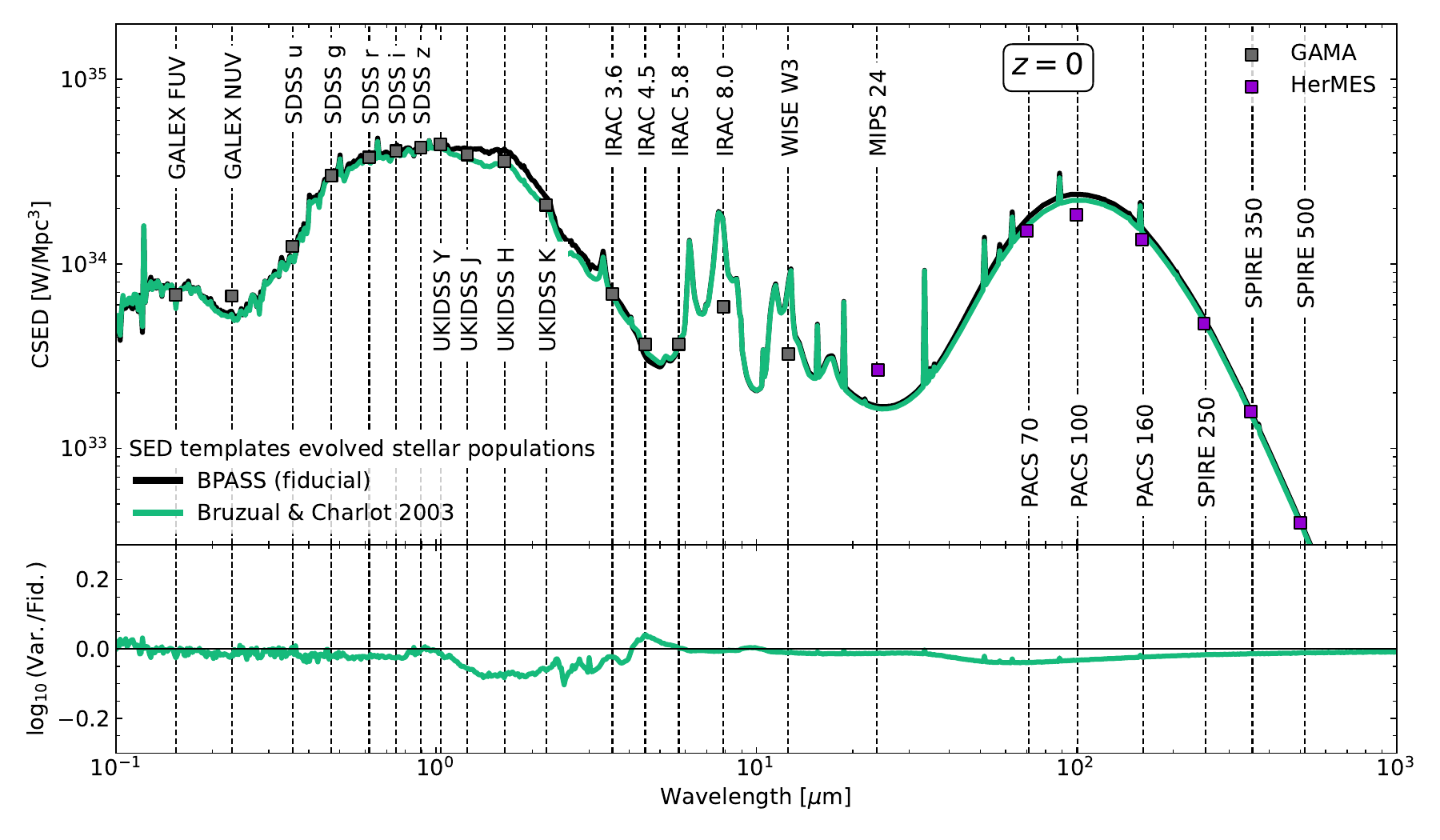}
    \caption{Same as Fig.~\ref{fig:CSEDvariation_THEMIS}, but varying the SED templates of the evolved stellar populations. We compare our fiducial choice (BPASS) to the \citet{Bruzual2003} SED templates, finding minor differences in the NIR.}
    \label{fig:CSEDvariation_SEDtemplates}
\end{figure*}

In Fig.~\ref{fig:CSEDvariation_SEDtemplates}, we compare two different SED templates for evolved stellar populations, namely BPASS v2.2.1 (our fiducial choice) and the commonly used \citet{Bruzual2003} templates. The IMFs are exactly the same for the two SED templates (\citealt{Chabrier2003} IMF within $0.1-100\,\mathrm{M}_\odot$). We find small differences between the results obtained with the two template libraries mostly in the NIR, where the CSED is lower by $\approx5\,\%$ when using the \citet{Bruzual2003} templates. We note that the difference between these two libraries is significantly greater for galaxies at cosmic noon, where $V-J$ colours differ by up to $\approx0.37\,\mathrm{mag}$ (\citealt{Gebek2025}). The different optical-NIR colours stem from varying treatments of the asymptotic giant branch phase and different stellar atmosphere models (\citealt{Stanway2018}).

\subsection{Numerical choices for SKIRT}

Lastly, we discuss the convergence with the most important numerical settings for SKIRT. Regarding the spatial grid, the choices for the maximum refinement level and grid refinement criterion (scaled with the galaxy dust surface density) are made to minimize the number of grid cells while retaining sufficient accuracy in the FUV flux, as shown in Appendix~\ref{sec:spatialGrid}. We find that the SEDs are converged when increasing the minimum refinement level from 15 to 18.

For the number of photon packets, we find that the Monte Carlo error is generally $\lesssim1\,\%$ on our broadband fluxes for our choice of $N_\mathrm{pp}=10^{7.5}$. When lowering $N_\mathrm{pp}$, the Monte Carlo error increases, but we also find systematic offsets in the MIR around $\approx40\,\mu\mathrm{m}$. Since the SKIRT runtime is relatively constant for $N_\mathrm{pp}\lesssim10^{7.5}$ and starts increasing for larger $N_\mathrm{pp}$, $N_\mathrm{pp}=10^{7.5}$ provides the optimal balance between speed and accuracy. While it would be possible to scale $N_\mathrm{pp}$ with the number of source particles (\citealt{Camps2022}), our tests did not reveal systematic trends between galaxy properties ($M_\star$, SFR, $M_\mathrm{dust}$) and convergence with $N_\mathrm{pp}$.

We discretize all grain size distributions into ten size bins. When increasing this to fifteen size bins, we find that the fluxes are well converged. The largest differences occur for low-mass galaxies ($8.5\leq\log_{10}(M_\star/\mathrm{M}_\odot)<9$) in the SPIRE 500 band, with median differences of $\approx0.50\,\%$ (only considering dust-containing galaxies).

To store the primary radiation field (from evolved stellar populations and star-forming regions) and to compute the secondary emission (from dust), the wavelength ranges are discretized into a number of wavelength bins. We use 25 grid points to store the primary radiation field (for $0.09\,\mu\mathrm{m}<\lambda<20\,\mu\mathrm{m}$), and 200 grid points to compute the dust emission (100 points for $0.09\,\mu\mathrm{m}<\lambda<2000\,\mu\mathrm{m}$, and 100 additional points for $1\,\mu\mathrm{m}<\lambda<30\,\mu\mathrm{m}$). To test how the resolution of these wavelength grids affects the SEDs, we have increased all numbers of grid points by a factor of two. We find that this does systematically affect the dust emission SEDs. Median differences in the SEDs (only considering dust-containing galaxies) are highest in the MIR bands and reach up to $2.9\,\%$ (IRAC 3.6, $9.5\leq\log_{10}(M_\star/\mathrm{M}_\odot)<10$) and down to $-4.1\,\%$ (MIPS 24, $8.5\leq\log_{10}(M_\star/\mathrm{M}_\odot)<9$).

Lastly, we test the impact of the primary source wavelength range. By  default, we set this to $0.09\,\mu\mathrm{m}<\lambda<2000\,\mu\mathrm{m}$. We perform a test increasing the lower wavelength to the Lyman limit ($\lambda=0.09113\,\mu\mathrm{m}$), equivalent to assuming that all ionizing photons are absorbed by the gas-phase ISM and do not contribute to dust heating. We find that this slightly reduces dust emission, with the biggest impact in the MIPS 24 band ($-1.4\,\%$ for $8.5\leq\log_{10}(M_\star/\mathrm{M}_\odot)<9$, considering only dust-containing galaxies).

\section{Aperture impact on the CSED}\label{sec:aperture}

\begin{figure*}
    \centering
    \includegraphics[width=\textwidth]{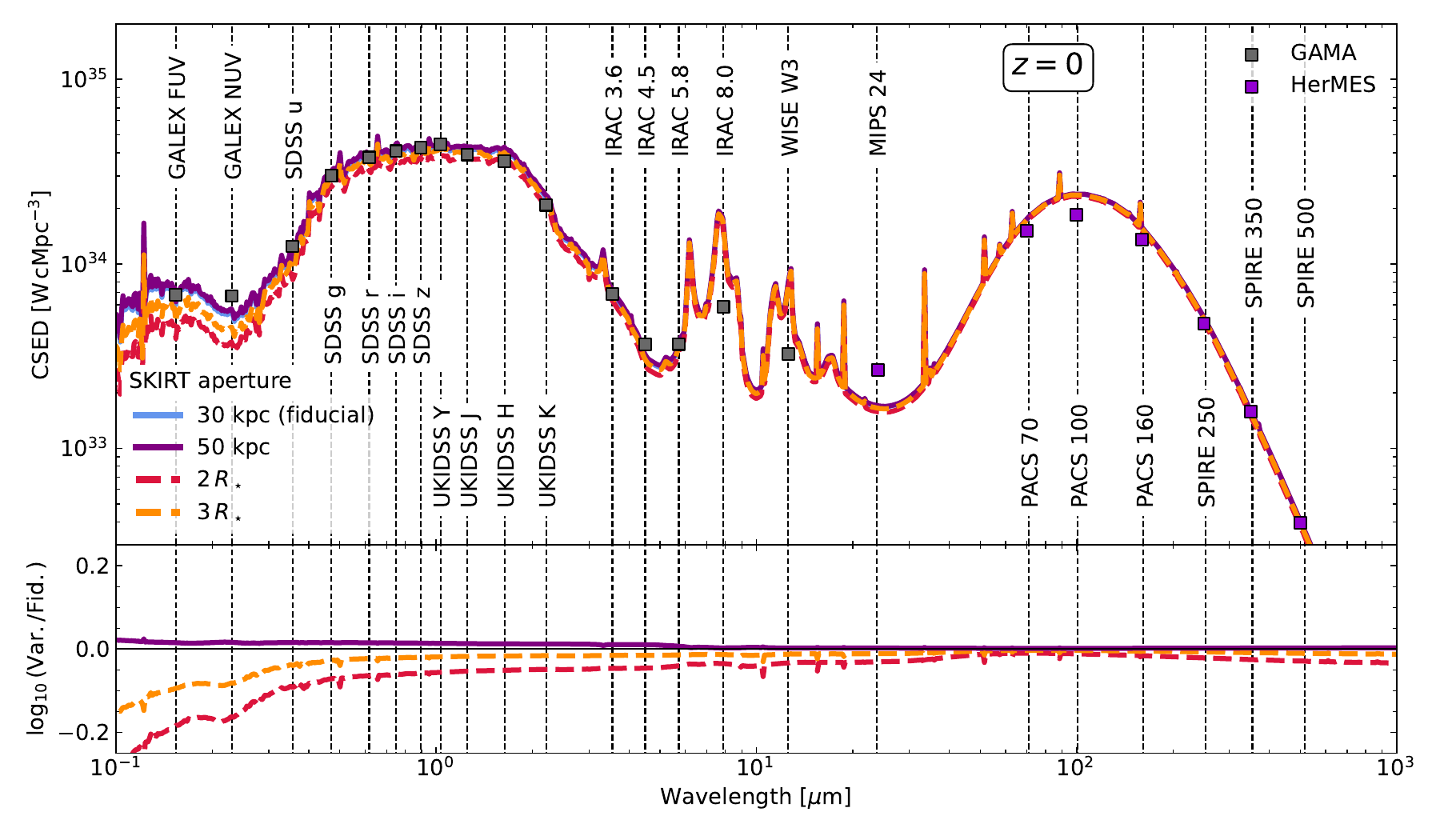}
    \caption{The impact of the SKIRT aperture on the COLIBRE CSED, using the L200m6 sample. The observational data is shown by the grey (GAMA) and purple (HerMES) markers. The fiducial aperture radius ($30\,\mathrm{kpc}$) is shown in blue. We also test apertures of $50\,\mathrm{kpc}$ (purple line) and two or three stellar half-mass radii (red and orange dashed lines, respectively).}
    \label{fig:apertures}
\end{figure*}

Throughout this study, we have adopted fixed SKIRT apertures with radii of $30\,\mathrm{kpc}$. When importing the particles, we normally use all gravitationally bound star and star-forming gas particles as primary sources, and all gas within a cube of side length $100\,\mathrm{kpc}$ for the dust medium. Other aperture choices are conceivable, here we test their impact on the low-redshift CSED using our main L100m6 sample (Fig.~\ref{fig:apertures}). As discussed in Section~\ref{sec:SKIRTsettings}, the usage of a 100-kpc box for the dust leads to an inconsistency in the treatment of primary sources inside or outside the box. Hence, for the aperture comparison here, we use a model variation (using the L200m6 sample of 300 galaxies) where we first filter out all particles (including primary sources) that lie outside the 100-kpc box.

We find that the CSED yields consistent results for both 30-kpc (our fiducial choice) and 50-kpc apertures (widely adopted for COLIBRE; \citealt{DeGraaff2022}; \citealt{Schaye2026}; \citealt{Chaikin2026b}). In contrast, scaling the aperture with the stellar half-mass radius reduces the CSED, particularly in the UV. We note that, because all particles outside the 100-kpc box are excluded in this comparison, apertures scaled with $R_{\star,1/2}$ are effectively capped at radii of $50\,\mathrm{kpc}$. For the most extended systems (the largest stellar half-mass radius in the L100m6 sample is $R_{\star,1/2}\approx237\,\mathrm{kpc}$), substantially larger dust grids would be required to capture their extended emission. However, the vast majority of galaxies in the L100m6 sample have much smaller sizes, with $\approx 98.8\,\%$ having $R_{\star,1/2}\leq10\,\mathrm{kpc}$.

\section{SKIRT spatial grid}\label{sec:spatialGrid}

SKIRT uses a spatial grid to discretize the transfer medium, in our case the dust in the ISM. We find that for our setup, the SKIRT runtime correlates strongly with $N_\mathrm{cells}$, with a Spearman rank correlation coefficient of $\rho=0.974$ for our main L100m6 sample. While tracing photon packets through the grid might be expected to scale approximately with $N_\mathrm{cells}^{1/3}$, our simulations perform a significant amount of work that scales with $N_\mathrm{cells}$. Specifically, the runtime of the secondary emission phase is dominated by the calculation of the dust temperature distribution and emission spectrum for each grain size bin and dust species in each grid cell (see \citealt{Camps2015b}). Dust-free galaxies ($\approx17\,\%$ of the main L100m6 sample), which do not require a spatial grid, are completely negligible in terms of their runtime as they require only $\approx0.03\,\%$ of the total CPU hours. Hence, it is of paramount importance to optimize the spatial grid such that the number of grid cells is as small as possible while retaining converged results.

SKIRT offers various structured and unstructured adaptive grids (\citealt{Camps2013}; \citealp{Saftly2013,Saftly2014}; \citealt{Lauwers2024}), which refine cells in denser regions according to specific refinement criteria. While the canonical choice for EAGLE and IllustrisTNG SKIRT postprocessing was the octree grid (\citealt{Camps2016}; \citealt{Trcka2022}), our tests showed that a $k$-d tree is slightly more efficient (as also advocated by \citealt{Saftly2014}). Hence, we adopt a $k$-d tree for our SKIRT simulations. This adaptive grid has four parameters: box size, minimum refinement level, maximum refinement level, and grid refinement criterion ($\delta_\mathrm{max}$). $\delta_\mathrm{max}$ sets the maximum allowed fraction of dust mass in a SKIRT grid cell relative to the total dust mass in the grid: if that fraction exceeds $\delta_\mathrm{max}$ then the cell is subdivided (within the limits set by the minimum and maximum refinement levels). We fix the box size to $100\,\mathrm{kpc}$ (proper) and the minimum refinement level to 15 for all our SKIRT runs.

\begin{table}
    \centering
    \begin{tabular}{cccc}
         $\log_{10}(\delta_\mathrm{max})$ & \shortstack{$N_\mathrm{cells}$ (max.\\level 33)} & \shortstack{$N_\mathrm{cells}$ (max.\\level 36)} & \shortstack{$N_\mathrm{cells}$ (max.\\level 39)} \\ \hline
         -4 & $4.7\times10^4$ & $4.7\times10^4$ & $4.7\times10^4$ \\
         -4.5 & $7.8\times10^4$ & $7.8\times10^4$ & $7.8\times10^4$ \\
         -5 & $1.7\times10^5$ & $1.8\times10^5$ & $1.8\times10^5$ \\
         -5.5 & $4.3\times10^5$ & $4.8\times10^5$ & $4.9\times10^5$ \\
         -6 & $1.1\times10^6$ & $1.4\times10^6$ & $1.5\times10^6$ \\
         -6.5 & $2.6\times10^6$ & $3.9\times10^6$ & $4.5\times10^6$ \\
         -7 & $5.9\times10^6$ & $1.0\times10^7$ & $1.4\times10^7$
    \end{tabular}
    \caption{Median number of cells in the SKIRT grid for the 150 galaxies from the L200m6 spatial grid sample. We quote $N_\mathrm{cells}$ as a function of the grid refinement criterion ($\delta_\mathrm{max}$), for three different maximum refinement levels. $N_\mathrm{cells}$ is mostly driven by $\delta_\mathrm{max}$, the maximum refinement level only has a secondary impact (and only for $\delta_\mathrm{max}\lesssim10^{-5.5}$).}
    \label{tab:Ncells}
\end{table}

The choice of maximum refinement level and $\delta_\mathrm{max}$ is an optimisation problem: we aim to minimise the number of grid cells while maintaining sufficient accuracy in the radiative transfer calculations. We explore this using a dedicated sample of dust-containing galaxies drawn from L025m5 (176 galaxies) and L200m6 (150 galaxies), as described in Section~\ref{sec:spatialGridSample}. For these systems, we run SKIRT 21 times, varying $\delta_\mathrm{max}$ across seven log-spaced values between $10^{-4}$ and $10^{-7}$ (in steps of $0.5\,\mathrm{dex}$) and adopting three maximum refinement levels (33, 36, and 39). At fixed parameter combinations, both samples yield similar numbers of grid cells per galaxy. We therefore report the median $N_\mathrm{cells}$ for the 150 L200m6 galaxies (with L025m5 showing consistent behaviour) in Table~\ref{tab:Ncells}.

We find that $N_\mathrm{cells}$ is primarily controlled by $\delta_\mathrm{max}$, with an approximate scaling $N_\mathrm{cells}\sim1/\delta_\mathrm{max}$ for maximum refinement levels of 36 and 39. The maximum refinement level has only a minor impact on $N_\mathrm{cells}$ for $\delta_\mathrm{max}\gtrsim10^{-5.5}$, and remains subdominant even at lower values of $\delta_\mathrm{max}$.

Our full panchromatic tests show that the FUV is the most sensitive band to spatial discretisation effects. We therefore assess radiative transfer accuracy using the total (infinite-aperture) flux at $\lambda = 0.15,\mu\mathrm{m}$. To reduce the Monte Carlo noise, we perform oligochromatic simulations at this wavelength and increase the number of photon packets to $N_\mathrm{pp}=10^8$.

To determine an appropriate maximum refinement level, we first examine the relative FUV flux errors $\varepsilon_\mathrm{FUV}$ for a fixed $\delta_\mathrm{max}=10^{-6}$ (Table~\ref{tab:FUVerror}). We adopt results obtained with a maximum refinement level of 39 as the reference solution and compare them to levels 33 and 36. While increasing the refinement level from 33 to 36 leads to a clear improvement in accuracy, the number of grid cells changes only marginally. We therefore adopt a maximum refinement level of 36. We note that this choice could in principle be adjusted with simulation resolution; however, at this refinement level the largest errors (84th percentile in Table~\ref{tab:FUVerror}) remain comparable between L025m5 ($2.4\times10^{-4}$) and L200m6 ($1.8\times10^{-4}$).

\begin{table}
    \centering
    \begin{tabular}{cccc}
         Simulation & Percentile & \shortstack{$\varepsilon_\mathrm{FUV}$ (max.\\level 33)} & \shortstack{$\varepsilon_\mathrm{FUV}$ (max.\\level 36)} \\ \hline
         \multirow{3}{*}{L025m5} & $15.87\,\%$ & $5.3\times10^{-5}$ & $9.3\times10^{-6}$ \\
         & $50\,\%$ & $5.1\times10^{-4}$ & $9.4\times10^{-5}$ \\
         & $84.14\,\%$ & $1.6\times10^{-3}$ & $2.4\times10^{-4}$ \\ \hline
         \multirow{3}{*}{L200m6} & $15.87\,\%$ & $9.8\times10^{-6}$ & $5.8\times10^{-6}$ \\
         & $50\,\%$ & $1.1\times10^{-4}$ & $3.8\times10^{-5}$ \\
         & $84.14\,\%$ & $5.5\times10^{-4}$ & $1.8\times10^{-4}$
    \end{tabular}
    \caption{Relative error of the FUV flux, for different maximum refinement levels. We use a fixed $\delta_\mathrm{max}=10^{-6}$ here. The reference solution is computed using a maximum grid refinement level of 39.}
    \label{tab:FUVerror}
\end{table}

We next select $\delta_\mathrm{max}$. It is common practice to adopt a single global value (typically $\delta_\mathrm{max}\sim10^{-6}$; e.g. \citealt{Camps2016}; \citealt{Kapoor2021}; \citealt{Trcka2022}), which yields a similar number of grid cells for all galaxies. However, this approach is not necessarily optimal: galaxies with higher dust optical depths require finer spatial resolution to capture more rapidly varying radiation fields.

\begin{figure*}
    \centering
    \includegraphics[width=\textwidth]{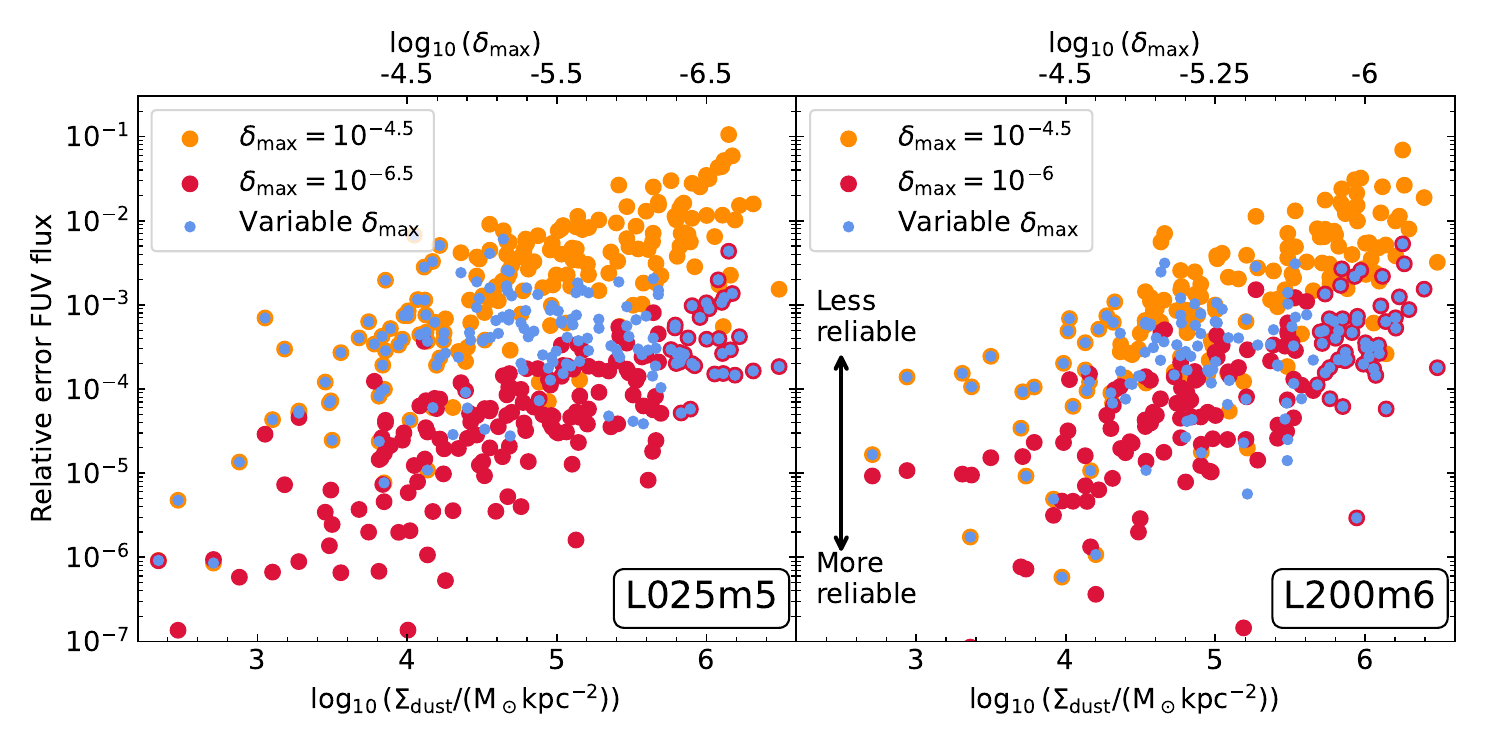}
    \caption{Relative error in the FUV fluxes as a function of galaxy dust surface densities. The maximum grid refinement level is fixed to 36 here. We show results from both L025m5 (176 galaxies) and L200m6 (150 galaxies) using the respective `spatial grid subsamples'. The reference FUV flux is computed using a very stringent grid refinement ($\delta_\mathrm{max}=10^{-7}$). The orange and red points indicate the relative error in the FUV flux with respect to the reference flux when using higher $\delta_\mathrm{max}$. The upper y-axes indicate the scaling of $\delta_\mathrm{max}$ that we use for the COLIBRE-SKIRT pipeline (Eqns.~\ref{eq:refinement_m6} and~\ref{eq:refinement_m5}). We also show the relative errors using these variable $\delta_\mathrm{max}$ values (blue points), which achieve a much more balanced error distribution with errors consistently below $\sim1\,\%$.}
    \label{fig:gridRefinement}
\end{figure*}

To investigate this, we introduce the dust surface density $\Sigma_\mathrm{dust}$ as a proxy for optical depth and study its relation to $\varepsilon_\mathrm{FUV}$. We require an orientation-independent definition, as the grid construction should not depend on the placement of the SKIRT instruments. After testing several options, we adopt $\Sigma_\mathrm{dust}=M_\mathrm{dust}/(\pi R_{\star,1/2}^2)$ where $M_\mathrm{dust}$ is measured within a 50-kpc exclusive-sphere aperture and $R_{\star,1/2}$ is the 3D stellar half-mass radius computed from all gravitationally bound star particles. Both quantities are taken from the SOAP catalogues.

Figure~\ref{fig:gridRefinement} shows the correlation between $\Sigma_\mathrm{dust}$ and $\varepsilon_\mathrm{FUV}$ (using a reference solution with $\delta_\mathrm{max}=10^{-7}$). At fixed $\delta_\mathrm{max}$, we find strong correlations, with Spearman coefficients of $\rho \approx 0.75$. This implies a non-uniform error distribution: galaxies with higher dust surface densities (and thus larger optical depths) systematically exhibit larger discretisation errors when using a constant $\delta_\mathrm{max}$.

To mitigate this, we introduce a heuristic scaling of $\delta_\mathrm{max}$ with $\Sigma_\mathrm{dust}$ (Eqns.~\ref{eq:refinement_m6} and~\ref{eq:refinement_m5}). Since FUV errors are slightly larger in the higher-resolution COLIBRE simulation, we adopt a more stringent lower bound on $\delta_\mathrm{max}$ for L025m5 than for L200m6. This adaptive prescription significantly homogenises the error distribution (blue points in Fig.~\ref{fig:gridRefinement}). Importantly, it maintains errors below $\sim 1\,\%$ even for the most dust-rich galaxies, while simultaneously reducing $N_\mathrm{cells}$ for low-optical-depth systems. We discuss the corresponding impact on SKIRT runtimes in Appendix~\ref{sec:runtimes}.

\begin{figure}
    \centering
    \includegraphics[width=\columnwidth]{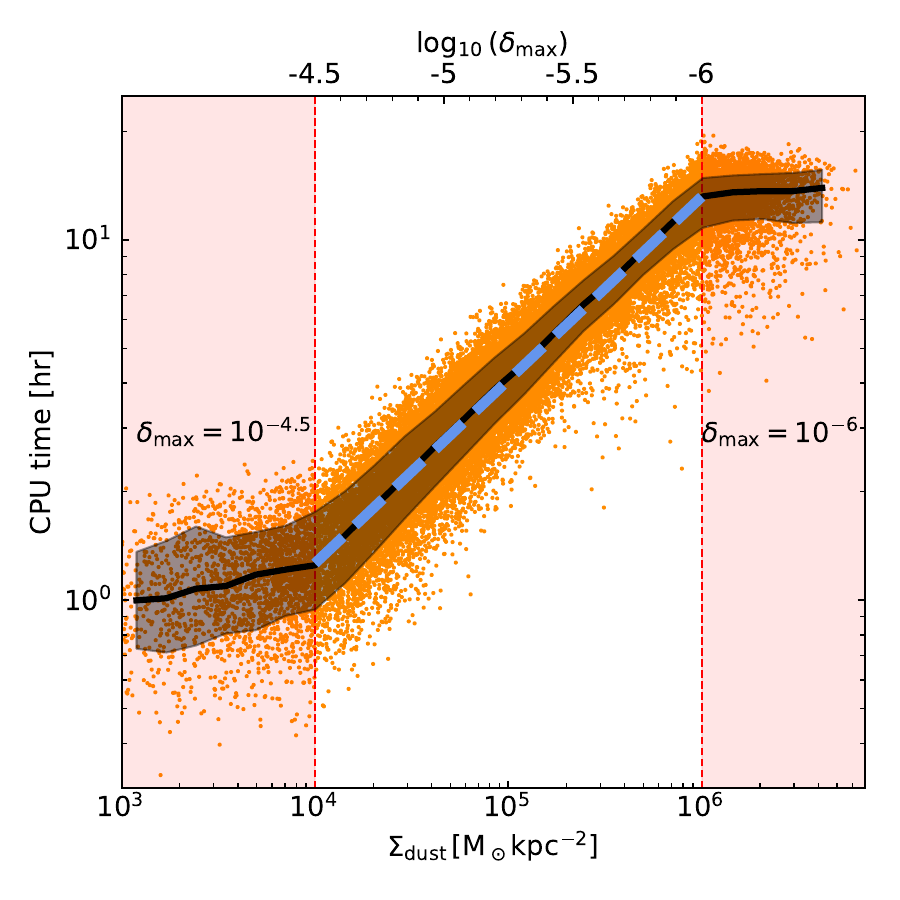}
    \caption{SKIRT runtimes for the main L100m6 sample as a function of dust surface density. Orange points indicate the runtimes of individual galaxies. Since the grid refinement criterion $\delta_\mathrm{max}$ is scaled with $\Sigma_\mathrm{dust}$ (see upper $x$-axis), a tight correlation exists between $\Sigma_\mathrm{dust}$ and the required number of CPU hours. The solid black line indicates the running median, the shaded area shows the 16th-84th percentile range. The median line can be well fit with a linear relation in log-log space, as shown by the dashed blue line.}
    \label{fig:CPUhours_SigmaDust}
\end{figure}

\section{SKIRT runtimes}\label{sec:runtimes}

Our scaling of $\delta_\mathrm{max}$ with $\Sigma_\mathrm{dust}$ means that galaxies with higher dust surface densities have more grid cells, and hence require more CPU hours. We show this for the main L100m6 sample in Fig.~\ref{fig:CPUhours_SigmaDust}, where we find a tight relation between $\Sigma_\mathrm{dust}$ and CPU time ($t_\mathrm{CPU}$). We fit a linear relation between these two quantities in log-log space:

\begin{equation}\label{eq:CPUfit}
\begin{aligned}
     &\log_{10}(t_\mathrm{CPU}/\mathrm{hr})=\\
    &\begin{cases}
      1.12 & \text{if $\Sigma_\mathrm{dust}\geq10^6\,\mathrm{M}_\odot\mathrm{kpc}^{-2}$}\\
      0.509\log_{10}\Bigr(\Sigma_\mathrm{dust}/(\mathrm{M}_\odot\mathrm{kpc}^{-2})\Bigl)-1.934 & \text{else}\\
      0.10 & \text{if $\Sigma_\mathrm{dust}\leq10^4\,\mathrm{M}_\odot\mathrm{kpc}^{-2}$}
    \end{cases}
\end{aligned}
\end{equation}
We find CPU times between $1.26\,\mathrm{hr}$ ($\Sigma_\mathrm{dust}\leq10^4\,\mathrm{M}_\odot\mathrm{kpc}^{-2}$) and $13.2\,\mathrm{hr}$ ($\Sigma_\mathrm{dust}\geq10^6\,\mathrm{M}_\odot\mathrm{kpc}^{-2}$). Since Eq.~\ref{eq:CPUfit} provides a link between galaxy properties (which are directly available from the \texttt{SOAP} catalogues) and SKIRT runtime, it is possible to make postprocessing runtime estimates for entire COLIBRE boxes. For instance, we find an estimated SKIRT runtime of $\approx238\,000$ CPU hours to postprocess all galaxies with $M_\star\geq10^{8.5}\,\mathrm{M}_\odot$ in the L100m6 box at $z=0$ and $z=0.1$, close to the true SKIRT runtime for this sample ($241\,000$ CPU hours). According to Eq.~\ref{eq:CPUfit}, a fixed $\delta_\mathrm{max}=10^{-6}$ would require a runtime of $\approx518\,000$ CPU hours, $\approx2.1$ times longer than with our variable $\delta_\mathrm{max}$ approach. A cautionary remark is in order: The CPU times discussed here are obtained for a single hardware system (the Tier-1 high-performance computing facility of the Flemish Supercomputing Center), running SKIRT with four cores. For other computing infrastructures, the runtimes will differ from the ones found here.

\begin{figure}
    \centering
    \includegraphics[width=\columnwidth]{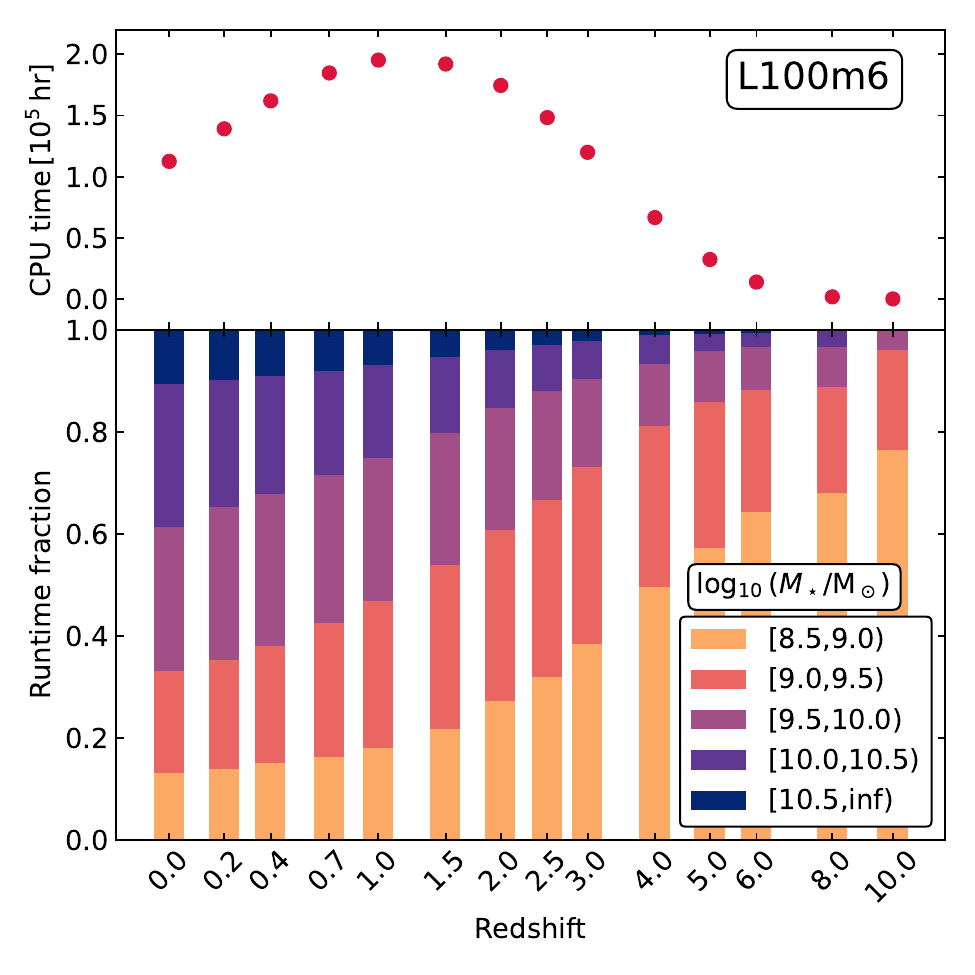}
    \caption{SKIRT runtime predictions for the COLIBRE L100m6 simulation at different redshifts. The runtimes are calculated from the linear fit to galaxy dust surface densities (see Fig.~\ref{fig:CPUhours_SigmaDust}). The upper panel shows the total runtimes when postprocessing all galaxies with $M_\star>10^{8.5}\,\mathrm{M}_\odot$. In the lower panel, the fractional contributions due to different stellar mass bins are shown. The SKIRT runtimes peak around $z\approx1$ at $\approx200\,000$ CPU hours.}
    \label{fig:CPUhours_redshift}
\end{figure}

In Fig.~\ref{fig:CPUhours_redshift}, we show runtime projections for the L100m6 simulation up to $z=10$. These are computed based on the galaxy dust surface densities and Eq.~\ref{eq:CPUfit}. The SKIRT runtimes first rise to $\approx200\,000$ CPU hours at $z\approx1$ and decrease for higher redshifts. We also show the contribution of different stellar mass ranges to the total SKIRT runtime, finding balanced contributions in terms of $\log_{10}(M_\star/\mathrm{M}_\odot)$ at $z=0$. Similar runtime projections can be made for different COLIBRE simulations and/or SKIRT setups, which is useful to design (sub-)sampling strategies when postprocessing large galaxy samples.

\end{appendix}

\end{document}